\documentclass[10pt,preprint2]{aastex}

\newcommand{\be}{\begin{eqnarray}} 
\newcommand{\ee}{\end{eqnarray}} 
\newcommand{\Msol}{\mbox{$M_{\odot}\;$}}
\newcommand{\Msun}{\mbox{$M_{\odot}\;$}}

\newcommand\simgreater{\buildrel > \over \sim}
\newcommand\simless{\buildrel < \over \sim}
\newcommand\appropto{\buildrel \sim \over \propto}
\begin{document}

\title{Minimal Cooling of Neutron Stars: A New Paradigm}

\author{Dany Page}
\affil{Departamento de Astrof\'{\i}sica Te\'orica, 
       Instituto de Astronom\'{\i}a, UNAM, 
       04510 Mexico D.F., Mexico}
\email{page@astroscu.unam.mx}
\author{James M. Lattimer}
\affil{Department of Physics and Astronomy, State University of New York 
at Stony Brook, Stony Brook, NY-11794-3800, USA}
\email{lattimer@mail.astro.sunysb.edu}
\author{Madappa Prakash}
\affil{Department of Physics and Astronomy, State University of New York 
at Stony Brook, Stony Brook, NY-11794-3800, USA}
\email{prakash@snare.physics.sunysb.edu}
\author{Andrew W. Steiner}
\affil{School of Physics and Astronomy, University of Minnesota, 
Minneapolis, MN 55455, USA}
\email{stein@physics.umn.edu}
\shorttitle{Cooling of Neutron Stars}
\shortauthors{PAGE, LATTIMER, PRAKASH, \& STEINER}

\begin{abstract}
A new classification of neutron star cooling scenarios, involving
either ``minimal'' cooling or ``enhanced'' cooling is proposed.  The
minimal cooling scenario replaces and extends the so-called standard
cooling scenario to include neutrino emission from the Cooper pair
breaking and formation process.  This emission dominates that due to
the modified Urca process for temperatures close to the critical
temperature for superfluid pairing.  Minimal cooling is distinguished
from enhanced cooling by the absence of neutrino emission from any
direct Urca process, due either to nucleons or to exotica such as
hyperons, Bose condensates or deconfined quarks.  Within the minimal
cooling scenario, theoretical cooling models can be considered to be a
four parameter family involving the equation of state (including
various compositional possibilities) of dense matter, superfluid
properties of dense matter, the composition of the neutron star
envelope, and the mass of the neutron star.  The consequences of
minimal cooling are explored through extensive variations of these
parameters.  The results are compared with the inferred properties of
thermally-emitting neutron stars in order to ascertain if enhanced
cooling occurs in any of them.  

All stars for which thermal emissions have been clearly detected are
at least marginally consistent with the lack of enhanced cooling,
given the combined uncertainties in ages and temperatures or
luminosities.  The two pulsars PSR 0833-45 (Vela) and PSR
1706-44 would require enhanced cooling in case their ages and/or
temperatures are on the lower side of their estimated values whereas
the four stars PSR 0656+14, PSR 1055-52, Geminga, and RX J0720.4-3125
may require some source of internal heating in case their age and/or
luminosity are on the upper side of their estimated values.  The new
upper limits on the thermal luminosity of PSR J0205+6449 (in the
supernova remnant 3C58) and RX J0007.0+7302 (in CTA 1) are 
indicative of the occurrence of some enhanced neutrino emission beyond
the minimal scenario.  

\end{abstract}


\newpage


\section{INTRODUCTION}

Within the last several years, several candidates for
thermally-emitting neutron stars have been discovered (see, e.g.,
\cite{PZ02} for a short review).  These stars are presumably cooling
through the combination of neutrino emission from the interior and
photon cooling from the surface, the latter responsible for their
observed thermal emissions.  Their temperatures have been deduced by
fitting atmosphere models to their spectra whereas ages can be inferred
from kinematics or from the associated pulsar spin-down timescale.
Neutron star cooling depends upon the equation of state (EOS) of dense
matter as well as the neutron star mass and their envelope
composition.  It has been hoped that comparing theoretical cooling
curves, i.e., the temperature-age or luminosity-age relation for
neutron stars, with observations could yield information about their
internal properties.

A goal of this paper to explore how observations of thermal
emission from neutron stars might be able to constrain the equation of
state of dense matter.  
We will treat theoretical neutron star cooling trajectories as a four 
parameter series of models.  
The parameters are: \\

\noindent (1) the equation of state (including various compositional
possibilities), \\
\noindent (2) superfluid properties of the relevant components, \\ 
\noindent (3) the envelope composition, and \\
(4) \noindent the stellar mass.  \\ 

In nature, only one equation of state and one set of superfluid
properties is realized, but at the present time, the theoretical range
of superfluid properties for a given equation of state is so broad
that these must be treated as an independent parameter.  In the
future, as more observations become available, it should be possible
to eliminate some combinations of EOS and superfluid parameter sets,
if not entire families of possibilities.  It is also not known if a
neutron star's envelope composition is unique (at least for a given
mass) or if it varies from star to star or as a function of time.  The
neutron star mass is constrained to lie between the maximum mass (a
function of the equation of state) and a minimum mass of about 1.1
M$_\odot$ (set by theoretical considerations of neutron star birth
(see, e.g., \cite{BL86}).

Historically, theoretical neutron star cooling models have fallen into
two categories, ``standard'' cooling or enhanced cooling.  The
so-called ``standard cooling'' scenario has no ``enhanced cooling''
which could result from any of the direct Urca processes involving
nucleons, hyperons, meson condensates or quark matter 
(see, e.g. \cite{P92} and \cite{Pr98}).  Until recently, ``standard''
cooling has been treated as being dominated by the modified Urca process
\citep{FM79}.  However, in the
presence of superconductivity or superfluidity in the neutron star
interior, an additional source of neutrino emission, Cooper
pair breaking and formation, occurs  \citep{FRS76,VS87}.
For temperatures near the associated gap energies, Cooper pairs, in
fact, dominate the neutrino emissivities.  Although the magnitude of
the superfluid gap energies as a function of density is somewhat
uncertain at present, it is generally accepted that superfluidity
occurs in neutron star matter.  For this reason, we embark on the
``minimal cooling'' scenario in which ``standard cooling'' is extended
to include the effects of superfluidity, including Cooper
pair breaking and formation.

The purpose of this paper is to explore in as complete a fashion as
possible the consequences of this minimal cooling paradigm, employing
the four kinds of parameters described above, and to compare our
results with the inferred properties of cooling neutron stars.  In
this way, it will become apparent to what extent one or more of the
so-called enhanced cooling mechanisms might be necessary to understand
the observations.  A future paper will explore in a similar fashion
the consequences of enhanced cooling.

One consequence of the minimal cooling paradigm, that enhanced cooling
will not occur, is a restriction upon the equation of state involving
the symmetry energy.  It is well-known (see, e.g., \cite{LPPH91}) that
the density dependence of the nuclear symmetry energy controls the
charge fraction in uniform beta-equilibrium matter.  Since the direct
Urca process occurs in uniform beta-equilibrium matter when the charge
fraction exceeds 1/9 (in the absence of muons or hyperons), minimal
cooling thus restricts the density dependence of the nuclear symmetry
energy: the critical density for the onset of the direct Urca process
must remain above the star's central density.  When muons are
considered, this critical density, for a given equation of state, is
slightly lowered.  The appearance of hyperons can also trigger other
direct Urca processes~\citep{PPLP92}.  Thus, to the extent that
minimal cooling can explain existing observations, a constraint on the
equation of state could be inferred.

In \S~\ref{Sec:Data}, observations of cooling neutron stars are
reviewed.  The input physics, including the equation of state,
superfluid properties, and neutrino emissivities are discussed in
\S~\ref{Sec:Physics}.  The influence of the neutron star envelope is
briefly discussed in \S~\ref{Sec:Envelope}.  The results of cooling
calculations for minimal cooling models are extensively discussed in
\S~\ref{Sec:General}. The coldest stars possible within the minimal
cooling scenario are identified in \S~\ref{Sec:Try-the best} and
\S~\ref{Sec:Minimal-Data} contains a summary of the confrontation of
the minimal cooling paradigm with existing data.  A comparison with
other studies is performed in \S~\ref{Sec:Other}.  Conclusions are
offered in \S~\ref{Sec:Conclusion}.
\section{DATA ON COOLING NEUTRON STARS}
         \label{Sec:Data}

Observations of neutron stars whose thermal emission has been
unambiguously detected give rise to the information summarized in
Tables~\ref{htable} and \ref{btable} whereas Table~\ref{utable}
contains results about objects for which only upper limits have been
set.  Tables~\ref{htable} and \ref{btable} display four inferred
quantities: the total thermal luminosity $L_\infty$, the surface
temperature $T_\infty$, the distance $d$ and the age $t$, whereas
Table~\ref{utable} omits $T_\infty$.  The subscript $\infty$ refers to
quantities observed at the Earth which are redshifted relative to
their values at the stellar surface.  The data in these tables are
taken from references that are detailed in Appendix~\ref{App:data}.
In cases where a range of estimates is presented in these references,
the particular parameters selected for inclusion in these tables are
elaborated in Appendix~\ref{App:data}.  Table ~\ref{htable} presents
properties as inferred from models incorporating atmospheres dominated
by hydrogen, whereas Table~\ref{btable} presents properties inferred
from blackbody or heavy-element dominated atmospheres.  The stars
displayed in Table~\ref{htable} are a subset of those in
Table~\ref{btable} because the inferred radii of the excluded stars
are far above theoretically plausible values for neutron star radii
(see discussion below).  The objects listed in Table~\ref{utable} are
fainter than those listed in the first two tables and the upper
limits listed have been obtained only very recently thanks to
the extended capabilities of  {\em XMM-Newton} and {\em Chandra}.
As a result, much less information has been obtained and the data
analysis has not been as detailed as for the stars in
Table~\ref{htable} and \ref{btable}.  Moreover, in four cases no
compact object has been detected. Indicated by a ``?'' in 
Table~\ref{utable}, these compact remnants may contain 
isolated black holes instead of  neutron stars.

For the purposes of restricting cooling models, we will not use two of
these sources. RBS 1223 is suspected of being a magnetar, judging from
its inferred extremely high magnetic field.  In addition, its period
derivative is highly uncertain.  RX J0720.4-3125 has a very uncertain
age estimate, and both objects possibly have additional heating
sources compared to the other sources.  RX J0720.4-3125 will be
included in the relevant figures, however, for purposes of comparison.

\begin{deluxetable}{cccccc}
\tablecaption{Neutron Star Properties with Hydrogen Atmospheres\label{htable}}
\tablehead{
\colhead{Star}&
\colhead{\hspace{-.7cm}$\begin{array}{c}\log_{10}t_{sd}\\\hspace{0.8cm}{\rm yr}\end{array}$}&
\colhead{$\begin{array}{c}\log_{10}t_{kin}\\{\rm yr}\end{array}$}&
\colhead{$\begin{array}{c}\log_{10}T_\infty\\{\rm K}\end{array}$}&
\colhead{$\begin{array}{c}d\\{\rm kpc}\end{array}$}&
\colhead{$\begin{array}{c}\log_{10}L_\infty\\ {\rm erg/s}\end{array}$}
}
\startdata
RX J0822-4247&3.90&$3.57^{+0.04}_{-0.04}$&$6.24^{+0.04}_{-0.04}$&
$1.9-2.5$ & $33.85-34.00$\\
1E 1207.4-5209&$5.53^{+0.44}_{-0.19}$&$3.85^{+0.48}_{-0.48}$& 
$6.21^{+0.07}_{-0.07}$&$1.3-3.9$&$33.27-33.74$\\
RX J0002+6246&--& $3.96^{+0.08}_{-0.08}$ & $6.03^{+0.03}_{-0.03}$&
$2.5-3.5$&$33.08-33.33$\\
%
PSR 0833-45 (Vela) & 4.05 & $4.26^{+0.17}_{-0.31}$ &
 $5.83^{+0.02}_{-0.02}$&$0.22-0.28$&$32.41-32.70$\\
PSR 1706-44        & 4.24 &    --                  &
 $5.8^{+0.13}_{-0.13}$ & $1.4-2.3$ &$31.81-32.93$ \\
PSR 0538+2817      & 4.47 &    --                  &
 $6.05^{+0.10}_{-0.10}$&   $1.2$   &$32.6 - 33.6$ \\
%
\enddata\\
\vskip .1in
References are cited in Appendix~\ref{App:data}
\end{deluxetable}

\begin{deluxetable}{ccccccc}
\tablecaption{Neutron Star Properties with Blackbody Atmospheres\label{btable}}
\tablehead{
\colhead{\hspace{-1.cm}Star}&
\colhead{\hspace{-0.7cm}$\begin{array}{c}\log_{10}t_{sd}\\\hspace{0.8cm}{\rm yr}\end{array}$}&
\colhead{$\begin{array}{c}\log_{10}t_{kin}\\{\rm yr}\end{array}$}&
\colhead{$\begin{array}{c}\log_{10}T_\infty\\{\rm K}\end{array}$}&
\colhead{$\begin{array}{c}R_\infty\\{\rm km}\end{array}$}&
\colhead{$\begin{array}{c}d\\{\rm kpc}\end{array}$}&
\colhead{$\begin{array}{c}\log_{10}L_\infty\\{\rm erg/s}\end{array}$}
}
\startdata
\hspace{-1.cm}RX J0822-4247&3.90&$3.57^{+0.04}_{-0.04}$&$6.65^{+0.04}_{-0.04}$&
$1-1.6$&$1.9-2.5$&$33.60-33.90$\\
\hspace{-1.cm}1E 1207.4-5209&$5.53^{+0.44}_{-0.19}$&$3.85^{+0.48}_{-0.48}$& 
$6.48^{+0.01}_{-0.01}$&$1.0-3.7$&$1.3-3.9$&$32.70-33.88$\\
\hspace{-1.cm}RX J0002+6246&--& $3.96^{+0.08}_{-0.08}$ & 
$6.15^{+0.11}_{-0.11}$&
$2.1-5.3$&$2.5-3.5$&$32.18-32.81$\\
\hspace{-1.cm}PSR 0833-45 (Vela) & 4.05 & $4.26^{+0.17}_{-0.31}$ &
 $6.18^{+0.02}_{-0.02}$&$1.7-2.5$&$0.22-0.28$&$32.04-32.32$\\
\hspace{-1.cm}PSR 1706-44 & 4.24 & -- & $6.22^{+0.04}_{-0.04}$&
 $1.9-5.8$&$1.8-3.2$&$32.48-33.08$\\
\hspace{-1.cm}PSR 0656+14 & 5.04 & -- & $5.71^{+0.03}_{-0.04}$&
 $7.0-8.5$&$0.26-0.32$&$32.18-32.97$\\
\hspace{-1.cm}PSR 0633+1748 (Geminga) & 5.53 & -- & $5.75^{+0.04}_{-0.05}$&
 $2.7-8.7$&$0.123-0.216$&$30.85-31.51$\\
\hspace{-1.cm}PSR 1055-52 & 5.43 & -- & $5.92^{+0.02}_{-0.02}$&
 $6.5-19.5$&$0.5-1.5$&$32.07-33.19$\\
\hspace{-1.cm}RX J1856.5-3754 &    --       & $5.70^{+0.05}_{-0.25}$ 
& $5.6 - 5.9$&  $> 16$&$0.105-0.129$&$31.44-31.68$\\
\hspace{-1.cm}RX J0720.4-3125 & $6.0\pm0.2$ & -- & 
$5.55 - 5.95 $ & 5.0 -- 15.0 & 0.1 -- 0.3 & $31.3-32.5$
\enddata\\
 \vskip .1in
\noindent References are cited in Appendix~\ref{App:data}
\end{deluxetable}

\begin{deluxetable}{ccccc}
\tablecaption{Properties of Barely Detected or Undetected Objects\label{utable}}
\tablehead{
\colhead{Star (SNR)}&
\colhead{\hspace{-.7cm}$\begin{array}{c}\log_{10}t_{sd}\\\hspace{0.8cm}{\rm yr}\end{array}$}&
\colhead{$\begin{array}{c}\log_{10}t_{kin}\\{\rm yr}\end{array}$}&
\colhead{$\begin{array}{c}d\\{\rm kpc}\end{array}$}&
\colhead{$\begin{array}{c}\log_{10}L_\infty\\{\rm erg/s}\end{array}$}
}
\startdata
CXO J232327.8+584842 (Cas A) &   -  &     2.51     & $3.3-3.6$ & $<34.5$ \\
J0205+6449 (3C 58)           & 3.74 &     2.91     & $2.6-3.2$ & $<33.0$ \\
PSR J1124-5916 (G292.0+1.8)  & 3.45 & $3.15-3.30$  &  $5 - 6$  & $<33.3$ \\
RX J0007.0+7302 (CTA 1)      &   -  & $4.0 - 4.2$  & $1.1-1.7$ & $<32.3$ \\
? (G084.2-0.8)               &   -  & $3.5 - 4.0$  &$\approx4 .5$& $<30.68 - 31.45$ \\ 
? (G093.3-6.9)               &   -  & $3.3 - 4.0$  & $2.1-2.9$ & $<30.65 - 31.55$ \\
? (G127.1+0.5)               &   -  & $3.3 - 3.9$  & $1.2-1.3$ & $<29.6  - 30.75$ \\
? (G315.4-2.3)               &   -  & $3.5 - 4.17$ & $2.4-3.2$ & $<30.65 - 31.80$ \\
PSR J0154+61                 & 5.29 &      -       & $1.7-2.2$ & $<32.14$\\
\enddata\\
\vskip .1in
References are cited in Appendix~\ref{App:data}
\end{deluxetable}

\subsection{Temperatures \label{sec:data-Temp}}

The estimation of $L_\infty$ and $T_\infty$ from the observed spectral
fluxes requires atmospheric modeling in which three additional factors
are involved: the composition of the atmosphere, the column density
of x-ray absorbing material  between the star and the Earth, and
the surface gravitational redshift (the surface gravity does not play
a major role in fitting broad spectral flux distributions.  The
column density is important because the bulk of the emitted flux from
neutron stars is absorbed (mostly by interstellar hydrogen)
before it reaches the Earth.  The surface gravitational redshift,
although not a factor in blackbody models, can  influence
heavy-element atmosphere models.  In many references, the
gravitational redshift was not optimized, but was set to the canonical
value 0.304 implied by $M=1.4$ M$_\odot$ and $R=10$ km.

Since narrow spectral lines are not observed in any of the stars in
Tables~\ref{htable} and \ref{btable}, the atmospheric composition of
these neutron stars is unknown.  However, some information can be
deduced from the shape of the spectral distribution.  Broadly
speaking, neutron star atmospheres can be described as being either
light-element (i.e., H or He) or heavy-element dominated.
Heavy-element atmospheres have spectral distributions more closely
resembling the blackbody distribution than do light-element
atmospheres \citep{R87}.  This is due to the higher opacities of heavy
elements, and seems to be the case even in the presence of strong
magnetic fields.  Since the wavelength range of available x-ray
spectra is relatively small, it is possible to fit x-ray spectra with
both kinds of atmosphere models.  
In general, an x-ray spectrum that is fit with a light-element atmosphere 
will predict the star to have a lower temperature and a larger angular size 
than will be the case if a heavy-element atmosphere or blackbody is assumed.  
If the distance is known, the neutron star radius can be inferred (see
more on this in \S~\ref{sec:Data-DL}).  In some cases, fitting a star
with a light-element atmosphere results in a predicted neutron star
radius much larger than the canonical range of 10--15 km. In other
cases, fitting a star with a heavy-element atmosphere could result in
an inferred radius that is too small.

\begin{figure}
\plotone{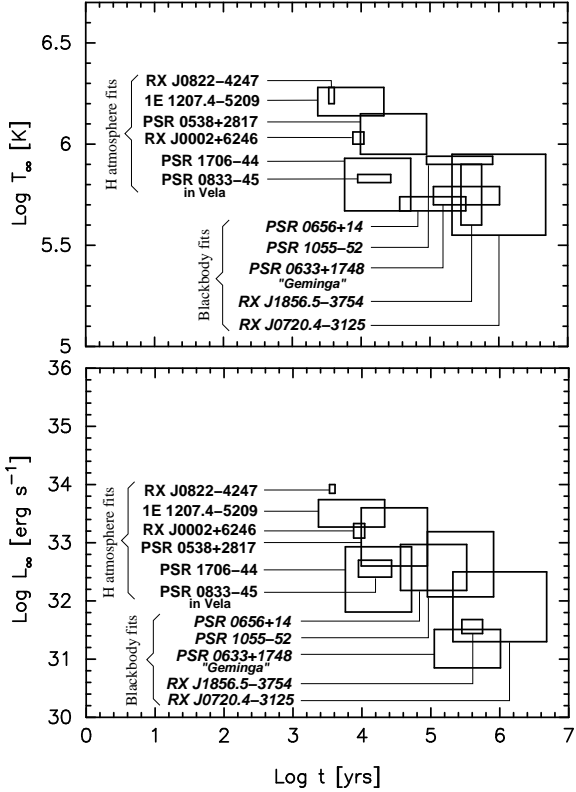}
\caption{Inferred temperature $T_\infty$ (top) and luminosity
$L_\infty$ (bottom) versus age for neutron stars with thermal emission.
Data from Table~\ref{htable} are marked as ``H atmosphere fits'' and
data from Table~\ref{btable} as ``Blackbody fits''.}
\label{Fig:TL}
\end{figure}

\cite{CB03a,CB03b} have discussed a trend
observed from atmospheric modeling of thermal neutron star spectra
\citep{P00}:
the inferred neutron star radii for stars younger than about $10^5$
years are consistent with canonical values only if they are modeled
with light-element atmospheres (magnetized or non-magnetized).  
Stars older than about $10^5$ years, on the other hand, have inferred 
radii close to the canonical range only when modeled with heavy-element
atmospheres.  
For this reason, Table~\ref{htable} is limited to stars with ages
less than about $10^5$ years, and it displays results inferred from
modeling them with H atmospheres.  
Table~\ref{btable}, on the other hand, lists all stars, and displays 
properties deduced from blackbody (or heavy-element dominated) models.
The temperatures and luminosities are plotted in Figure~\ref{Fig:TL}, 
and are also selected according to this trend and our desire that the 
inferred stellar radius lies in a theoretically plausible range.  
The temperature and luminosity are taken from Table~\ref{btable} unless 
values for them appear in Table~\ref{htable}.

The above trend implies that the atmospheric composition of a neutron star
evolves from light to heavy elements with a timescale of about $10^5$
years.  
This possible evolution is considered in more detail in 
\S~\ref{Sec:Decay-Env}.

\begin{figure}
\plotone{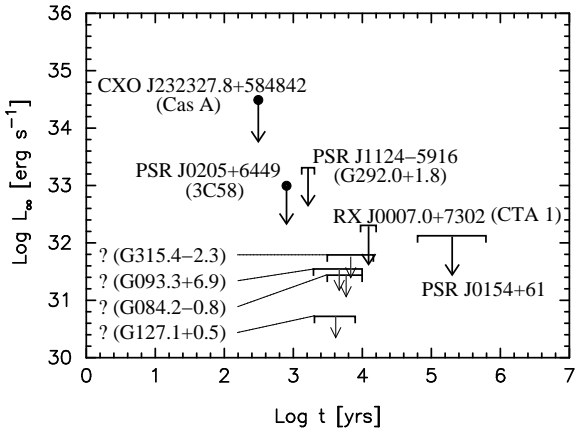}
\caption{Inferred upper limits on thermal luminosity $L_\infty$, versus age,
for the compact objects listed in Table~\ref{utable}.}
\label{Fig:UL}
\end{figure}

\subsection{Ages \label{sec:Data-age}}

The precise ages of observed cooling neutron stars are not always
known.  Most stars listed in Tables 1 through 3 are known radio and/or
x-ray pulsars and their ages can be estimated from the oberved
spin-down rate using $t_{sd} \equiv P/2\dot{P}$, where $P$ and $\dot
P$ are the period and its time derivative,respectively.  In some
cases, kinematic information is available and ages can be inferred by
relating pulsar transverse velocities to the distances from the
presumed sites of origin as, e.g., the geometric center of the
associated supernova remnant or a nearby cluster of massive OB stars.
In the case of an association with a supernova remnant, the age can also
be estimated by the general properties of the remnant and, in the best
cases, by association with historical supernovae.  We will generically
refer to these various alternatives to $t_{sd}$ as the ``kinematic
age'' $t_{kin}$.  Both ages, where available, are shown in
Tables~\ref{htable}, \ref{btable} and \ref{utable}.  The data in these
tables show that, in general, there is a large discrepancy between the
spin-down age and the kinematic age in cases where both are given.  In
most cases, the spin-down age is longer, but in the case of the Vela
pulsar, it is shorter.  Typical discrepancies are of order 3 or
larger.  For this reason, we have used the kinematic age in
Figures~\ref{Fig:TL} and \ref{Fig:UL} where available, and otherwise
have assigned an uncertainty of a factor of 3 in each direction to the
spin-down ages.

\subsection{Distances and Luminosities \label{sec:Data-DL}}

The distances are estimated from pulsar dispersion measures, estimated
distances to the related supernova remnants, or observations of
interstellar absorption to other stars in proximity.  In three cases,
parallax estimates are available.  The details are discussed in
Appendix~\ref{App:data}.  Uncertainties in the distances are in many
cases rather large.  Since the inferred luminosities of the stars are
proportional to the square of the assumed distances, it is usually the
case that the inferred stellar luminosity has greater relative error
bars than the inferred stellar temperature.  However, in cases in
which the composition of the stellar atmosphere is uncertain, but the
distance to the source is accurately known, the inferred stellar
luminosity might be more accurately estimated.   A
consistency check of the measurements of $T_\infty$, $L_\infty$ and
the distance $d$ is that the relation 
\be
L_\infty = 4 \pi R_\infty^2 \cdot \sigma_{\scriptscriptstyle SB} T_\infty^4 
\ee
should give a radius at infinity $R_\infty$ comparable to the radius
of a neutron star.  This is the case for the measurements listed in
Table~\ref{htable}, whereas for the measurements based on BB spectral
fits listed in Table~\ref{btable} only 1055-52 has a possibly
acceptable $R_\infty$, but with very large errors due to the
uncertainty in $d$.  However, BB models are overly simplistic.
Non-magnetic heavy-element dominated atmospheres tend to have values
of $R_\infty$ factors of 2 to 3 larger than a BB \citep{R87}, so that
essentially all the sources listed in Table~\ref{btable} satisfy this
consistency check.  For the objects listed in Table~\ref{utable}, this
consistency test is only marginally possible for J0205+6449 (3C58).

Theoretical cooling calculations also give an effective temperature
$T_e^\infty$ and a luminosity $L^\infty$ which are related to each
other by the equation~(see also equation (\ref{eq:L_inf})) 
\be
L^{\infty} \; \equiv \; e^{2\Phi(R)} L(R) \; = \;
4 \pi R^{\infty \; 2} \cdot \sigma_{\scriptscriptstyle SB} T_e^{\infty \; 4}\,,
\label{eq:L_inf0}
\ee
where we have used the superscript $\infty$ to
denote the theoretical values and subscript $\infty$ for the observed
values at infinity in order to emphasize the difference.  It is only
for a star for which the measured $T_\infty$ and $L_\infty$ satisfy
equation~(\ref{eq:L_inf0}), and for which an accurate measurement of
$d$ exists (implying a small error bar on $L_\infty$), that comparison
of cooling curves with data in terms of $T$ or $L$ are
equivalent.  For the stars listed in Table~\ref{btable} which do not
pass the above consistency test, the measured $T_\infty$ is thus
{\em not} an effective temperature and cannot be directly compared
with the calculated $T_e^\infty$.  In these cases, the luminosity
$L_\infty$ is more representative of  thermal emission and should
be used for comparison with $L^\infty$.  For this reason, we have
chosen to tabulate luminosities as well as temperatures in Tables~\ref{htable}
and \ref{btable}, and have plotted both temperature and luminosity in
Figure~\ref{Fig:TL}.
In Table~\ref{utable}, we have reported only upper limits to $L_\infty$,
which is the quantity that observation can usefully constrain, and
plotted them separately in Figure~\ref{Fig:UL}.

One feature notable in Figures~\ref{Fig:TL} and \ref{Fig:UL}
is the sizes of the error boxes, particularly in the age dimension.  
These uncertainties represent an inherent difficulty in using these 
observations to firmly constrain the details of neutron star cooling.  
For this reason, instead of attempting to detail properties of the 
equation of state, superconductivity, and/or neutrino emissivities 
from the observations, our approach will be to model a reasonably broad 
range of acceptable physical inputs in order to determine ranges of 
parameters that might be excluded by the present data.


\section{INPUT PHYSICS}
         \label{Sec:Physics}

The standard general relativistic equations determining the structure
and thermal evolution of a neutron star are briefly summarized in
Appendix~\ref{Sec:Equ}.  Given an equation of state (EOS), described
in \S~\ref{Sec:EOS}, we solve numerically the TOV equations of
hydrostatic equilibrium, 
and build our stars.
The equations of energy conservation, equation~(\ref{Eq:dLdr}), and 
energy transport, equation~(\ref{Eq:dTdr}), with their corresponding 
boundary conditions are then solved numerically with a fully general
relativistic Henyey-type
stellar evolution code specially developed for neutron stars \citep{P89}.
The required physics input are described in the
next section.
The outer boundary condition, equation~(\ref{Eq:Tb-Te}), is implemented 
in terms of an envelope, described in \S~\ref{Sec:Envelope}.

\subsection{The Equation of State}
\label{Sec:EOS}

The gross properties of a neutron star (such as its mass and radius)
and its interior composition (which influences the thermal evolution)
chiefly depend on the nature of strong interactions in dense matter.
Investigations of dense matter can be conveniently grouped into three broad
categories: nonrelativistic potential models, effective field
theoretical (EFT) models, and relativistic Dirac-Brueckner-Hartree-Fock
(DBHF) models.  In addition to nucleons, the presence of softening components
such as hyperons, Bose condensates or quark matter, can be
incorporated in each of these approaches.  Some general attributes,
including references and typical compositions, of equations of state
(EOS's) in each of these approaches have recently been summarized by
\citet{LP01}.

In this work, we employ four EOS's in the category of nonrelativistic
potential models in which only nucleonic degrees of freedom are
considered. Two of these are taken from the calculations of the Argonne
and Urbana groups. The EOS labeled WFF3, from \cite {WFF88}, is based on the 
variational calculations using UV14+TNI potential and that labeled APR,
from \cite{APR98},
utilizes the AV18 potential plus the UIX potential plus the $\delta
v_b$ boost.  APR represents the most complete study to date of Akmal
\& Pandharipande (1997), in which many-body and special relativistic
corrections are progressively incorporated into prior models including
that of WFF3.  

For isospin symmetric matter, the equilibrium densities of the WFF3
and APR models are $n_0= 0.163~{\rm fm}^{-3}$ and $0.16~{\rm
fm}^{-3}$, respectively, with corresponding compression modulii of $269$
MeV and $274$ MeV, respectively.  In isospin asymmetric matter, the
density dependent symmetry energy $S(n_b,x)$ is defined by the relation
\be
E(n_b,x) &=& E(n_b,1/2) + S(n_b,x) \,, 
\ee
where $E$ is the energy per particle, $n_b=n_n+n_p$ is the baryon number
density, and $x=n_p/n_b$ is the proton fraction.  In practice, $S(n_b,x)$
can be expanded as
\be 
S(n_b,x) = S_2(n_b)(1-2x)^2 + S_4(n_b)(1-2x)^4\cdots \,,
\ee
where the term involving $S_4$ is generally very small.  $S(n_b,x)$
plays a crucial role in a neutron star's thermal evolution insofar as
it determines the equilibrium proton fraction, which in turn
determines whether or not the direct Urca process, $n \rightarrow
p+e^-+\bar \nu_e$, is permitted to occur in charge neutral
beta-equilibrated matter \citep{LPPH91}.  

The equilibrium proton fraction is determined from the condition
\be
&\mu_e& = \hat \mu = \mu_n - \mu_p = 
-  (\partial E/\partial x)\\
&&= 4 (1-2x) [S_2(n_b) + 2S_4(n_b)(1-2x)^2 + \cdots]\,,\nonumber 
\ee
where $\mu_i~(i=e,n,p)$ are the chemical potentials.  For
ultrarelativistic and degenerate electrons, $\mu_e= \hbar
c(3\pi^2n_bx)^{1/3}$, since due to charge neutrality $n_e=n_p$ in
matter in which the only leptons are electrons.

When the electron Fermi energy is large enough (i.e., greater
than the muon mass), it is energetically favorable for the electrons
to convert to muons through $e^- \rightarrow \mu^- + {\overline
\nu}_\mu + \nu_e \,$.  Denoting the muon chemical potential by
$\mu_\mu$, the chemical equilibrium established by the above process
and its inverse is given by $\mu_\mu = \mu_e$.  At the threshold for
muons to appear, $\mu_\mu = m_\mu c^2 \cong 105$ MeV.  Noting that the
proton fraction at nuclear density is small, one has the approximate
relationship $4S_2(u)/m_\mu c^2 \sim 1$, where $u=n_b/n_0$. Using a
typical value $S_2(u=1)\simeq 30$ MeV, one may expect muons to appear
roughly at nuclear density $n_0 = 0.16~{\rm fm}^{-3}$.  Above the
threshold density,
\be
\mu_e=\hat\mu=\mu_\mu=
{\sqrt {(\hbar c)^2(3\pi^2n_bx_\mu)^{2/3} + m_\mu^2c^4}} \,,
\label{chememu}
\ee
where $x_\mu = n_\mu/n_b$ is the muon fraction in matter. The charge
neutrality condition now takes the form
$n_e + n_\mu = n_p$, 
which, together with the relation of chemical equilibrium in
equation (\ref{chememu}), establishes the lepton and proton fractions in
matter.  The appearance of muons has the consequence that the electron
fraction $x_e=n_e/n_b$ is lower than its value without the presence of muons.

If $S(n_b,x)$ does not rise sufficiently rapidly with density, the
equilibrium proton fraction will remain below a critical value (of
order $11\%$ in matter with $e^-$ only and $14\%$ in matter with both
$e^-$ and $\mu^-$) required for the direct Urca process.  The critical
proton fraction is determined by requiring simultaneous conservation
of energy and momentum among the participating fermions.  In this
case, cooling occurs via the modified Urca process, $n+n \rightarrow
n+p+e^-+\bar \nu_e$, and several other similar processes (see
\S~\ref{Sec:nu}),
modulated by effects of possible nucleon
superfluidity.  Because of the additional fermions involved, the
emissivity of the modified Urca process is several orders of magnitude
lower than the direct Urca process.

The symmetry energies $S_2(n_0)$ of the WFF3 and APR models at their
respective equilibrium densities $n_0$ have the values 
$29.5$ MeV and $32.6$ MeV, respectively.
In the WFF3 model, the symmetry energy $S_2(n_b)$ rises slowly with density
and $x$ never reaches the critical value for the direct Urca process.
However, in the case of the APR model, 
the direct Urca process becomes possible at $n_B > 0.78 \; {\rm fm}^{-3}$,
which corresponds to a neutron star mass of $M_{cr} = 1.97$
\Msun.  For this model, we will therefore consider only stars with
masses below this threshold.

We also consider two EOS's from the phenomenological non-relativistic
potential model of \citet{Prak97} which is designed to reproduce the
results of more microscopic calculations at zero temperature, and
which allows extensions to finite temperature. The EOS's chosen are
labeled BPAL21 and BPAL31, which have bulk nuclear matter
incompressibilities $K_s=180$ or 240 MeV, respectively.    In both
cases, the symmetry energy, at the empirical symmetric matter
equilibrium density of $n_0=0.16~{\rm fm}^{-3}$, was chosen to be 30
MeV.  Furthermore, the potential part of $S(n_b,x)$ varies approximately
as ${\sqrt {n_b/n_0}}$ in both cases, which is close to the behavior
exhibited in the EOS of APR.

\begin{figure}
\plotone{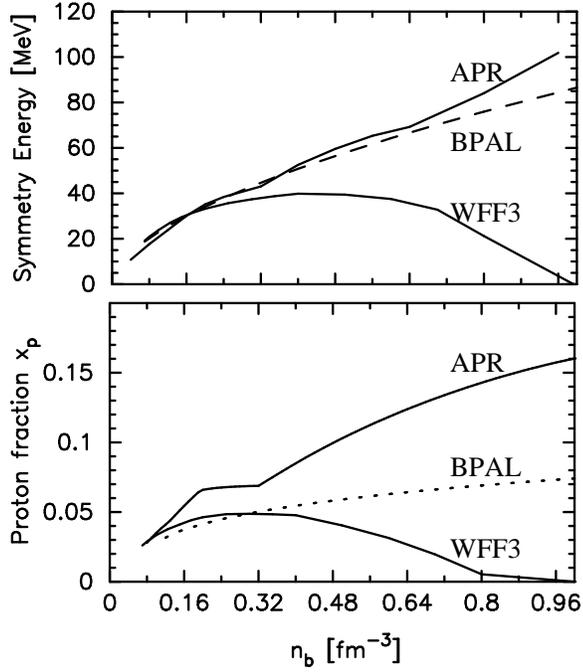}
\caption{Symmetry energy (top panel) and proton fraction (bottom panel) for
         the four EOS's used in this work. 
         \label{Fig:Sym-Xp}}
\end{figure}

\begin{figure}
\plotone{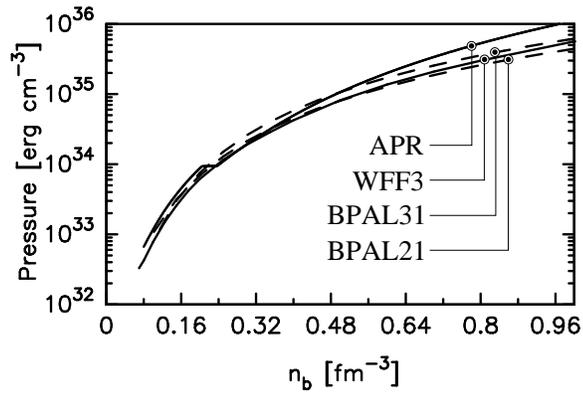}
\caption{Pressure vs baryon density for the four EOS's employed in this work. 
         \label{Fig:P-rho}}
\end{figure}

In Figure~\ref{Fig:Sym-Xp}, the symmetry energies (top panel) and
corresponding proton fractions (bottom panel) in charge-neutral
beta-stable neutron star matter are shown.  In Figure~\ref{Fig:P-rho},
the pressure of neutron star matter is shown as a function of baryon
density for the EOS's considered in this work.  The differences in the
high-density behavior of these two EOS's are largely attributed to
differences in the underlying three-body interactions.

The reason why we do not consider EOS's based on effective
field-theoretical (EFT) and relativistic Dirac-Brueckner-Hartree-Fock
(DBHF) models in this work merits some discussion. In EFT approaches based on
the prototype Walecka model, interactions between nucleons are
mediated by the exchange of $\sigma-$, $\omega-$, and $\rho-$
mesons. At the mean field level, the symmetry energy in this approach
is given by (\cite{HP01})
\be
S_2(n_b) = \frac{k_F^2}{6 {\sqrt {k_F^2+M^{*^2} } }} + \frac{ n_b }
{8 \left( \frac{g_{\rho}^2}{m_{\rho}^2} + 2 f (\sigma_0, \omega_0)  \right)} 
\,,
\ee
where $\sigma_0$, $\omega_0$, and $\rho_0$ are the mean-field
expectation values of the fields, $g_\sigma$ and $g_\rho$ are the
$\sigma-$ and $\rho-$ meson couplings to the nucleon, and
$M^*=M-g_\sigma\sigma_0$ is the nucleon's Dirac effective mass. The
quantity $f(\sigma_0, \omega_0)$ summarizes effects of density
dependent nonlinear interactions arising from $\sigma-$, $\omega-$,
and $\rho-$ mixings and have recently been employed to explore
deviations from the linear behavior of the second term with density in
the case $f=0$. When the symmetry energy rises linearly with density,
the critical proton fraction for the direct Urca process is reached at
$2-3~n_0$, which is well within the central densities of both
1.4M$_\odot$ and maximum mass stars obtained with these EOS's.
It is possible, however, to forbid the direct Urca process with a
suitable choice of $f \neq 0$~\citep{SPLE04}. These cases, however, 
resemble the potential models considered above.

All DBHF calculations reported thus far in the literature
(e.g. \cite{MPA87,Eng94}) find that proton fractions favorable
for the direct Urca process to occur are reached in stars whose masses
are larger than $\sim1.3{\rm M}_\odot$.  Since our intention here is
to explore the extent to which model predictions can account for
observations without invoking the direct Urca process and its variants
involving hyperons, Bose condensation or quarks, we defer a discussion
of these models to a separate work.

\begin{figure}
\plotone{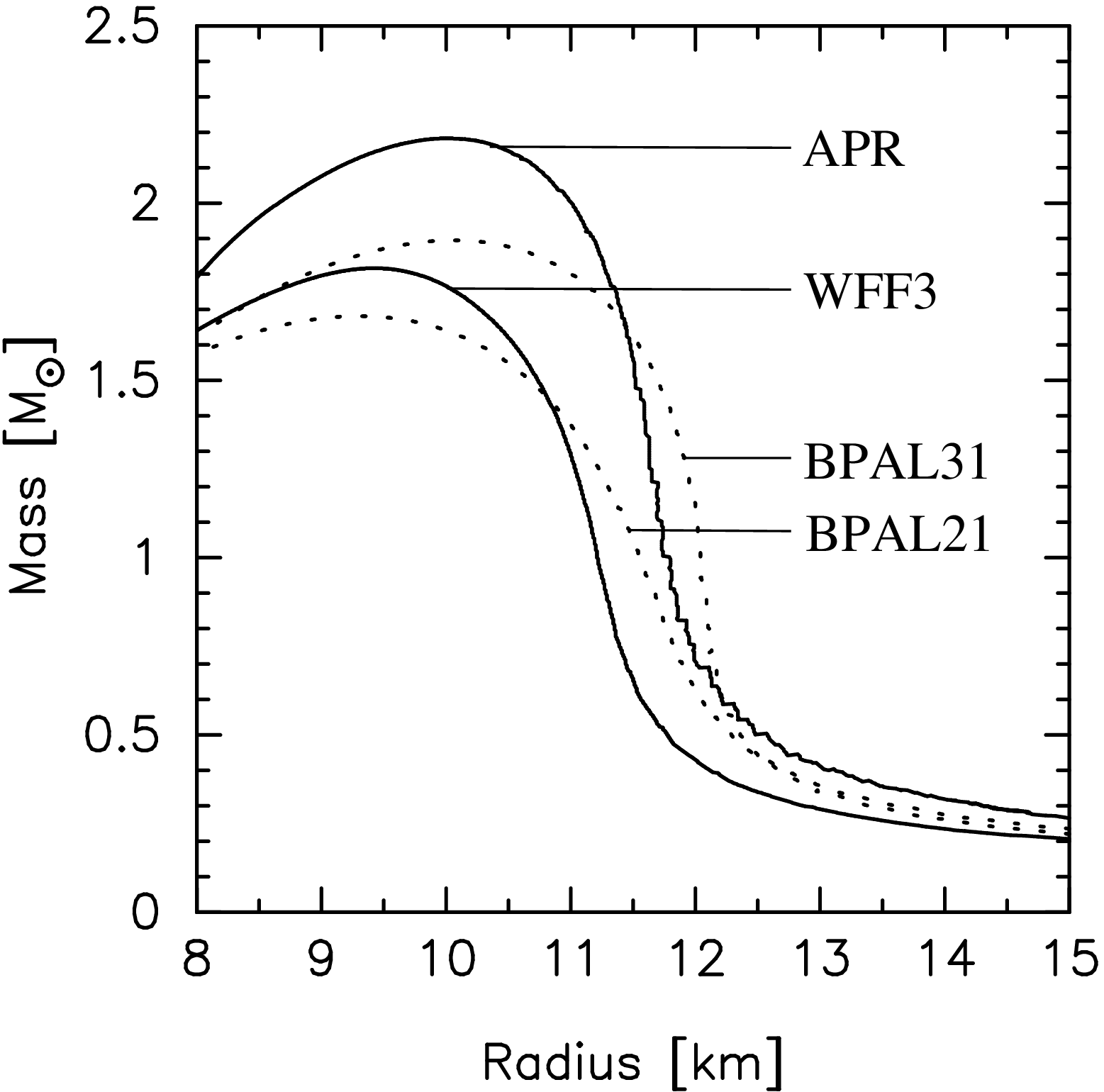}
\plotone{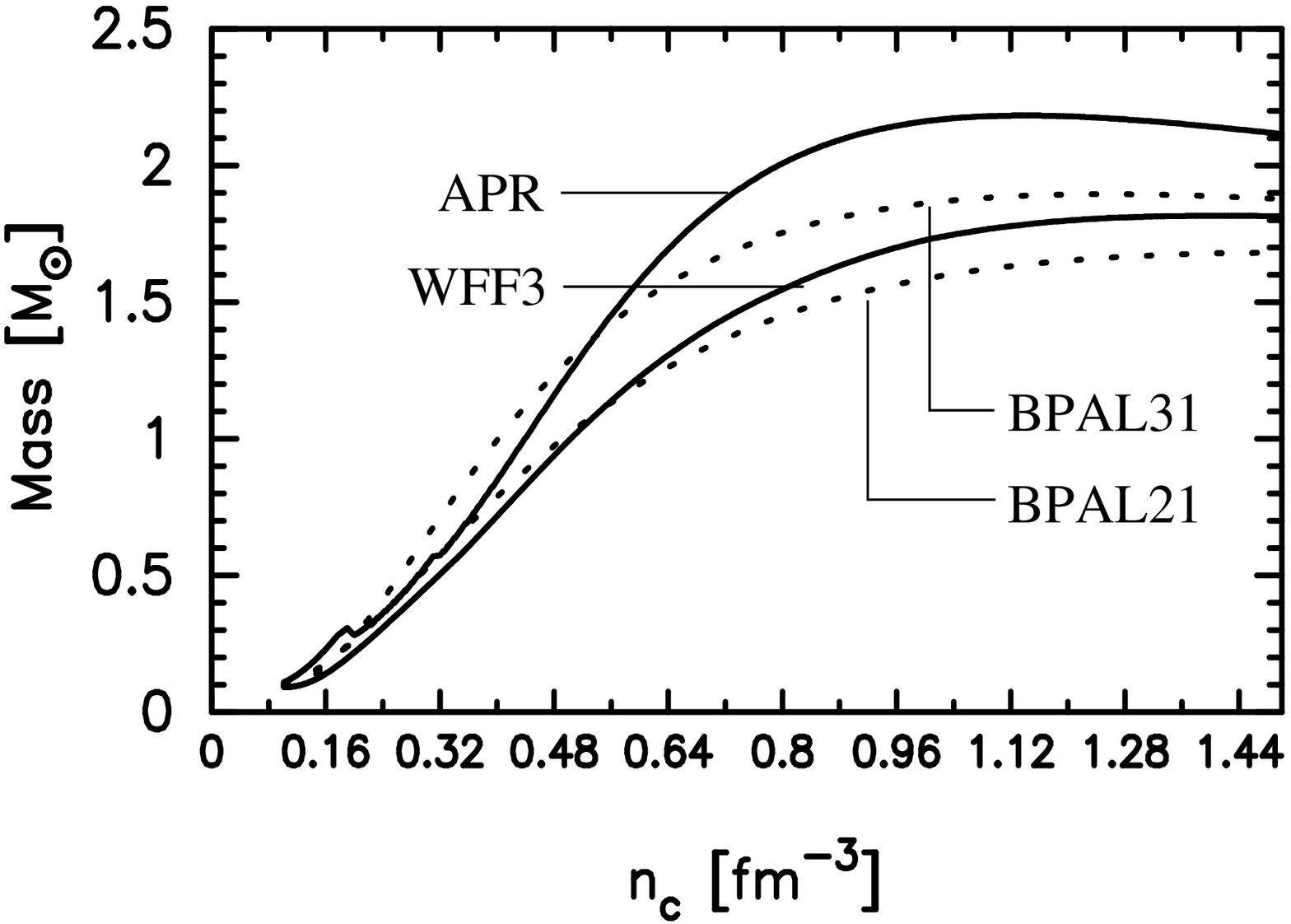}
\caption{Mass versus radius (top panel) and versus central baryon
density (bottom panel) for the four EOS's employed in this work.
\label{Fig:M-R}}
\end{figure}

In Figure~\ref{Fig:M-R}, we show the mass versus radius and versus central 
density curves for the four EOS's chosen. 
Features of relevance to the discussion of cooling
to note are:
\begin{itemize}  
\item The radii of maximum mass configurations ($1.7 < M_{max}/{\rm
M}_\odot < 2.2$)  are confined to the narrow range 9--10 km.

\item The radii of $1.4{\rm M}_\odot$ stars lie in the narrow range
$11-12$ km. 
\end{itemize}
Significant deviations from such a tight clustering of radii occur
only in those cases in which 
\begin{itemize}
\item normal nucleonic matter is described through the use of EFT or
DBHF models.  \citet{LP01} showed that the neutron star radius is
proportional to the density derivative of the symmetry energy in the
vicinity of nuclear matter density.  Therefore, in this case,
relatively large radii for both 1.4${\rm M}_\odot$ and maximum mass
configurations occur.  Similarly, a relatively large density
dependence of the symmetry energy also permits the
direct Urca process to occur in this case.
\item extreme softening is induced by the presence of additional
components such as hyperons, Bose condensates or quarks: In this case,
significantly smaller radii are possible. Such components also lead to
relatively rapid cooling.   
\end{itemize}
As mentioned earlier, both of these cases fall outside the ``Minimal
Cooling Scenario'' and will be investigated separately. 

For completeness, we note that in all four cases considered, the
crust-core transition occurs at $n_b \sim 0.1 \; {\rm fm}^{-3}$ or
equivalently $\rho \sim 1.6 \times 10^{14}$ g cm$^{-3}$.  For
the EOS in the crust region, we employ the EOS of \cite{NV73}
above neutron drip and that of \cite{HZD89} below
neutron drip.

\begin{figure}
\plotone{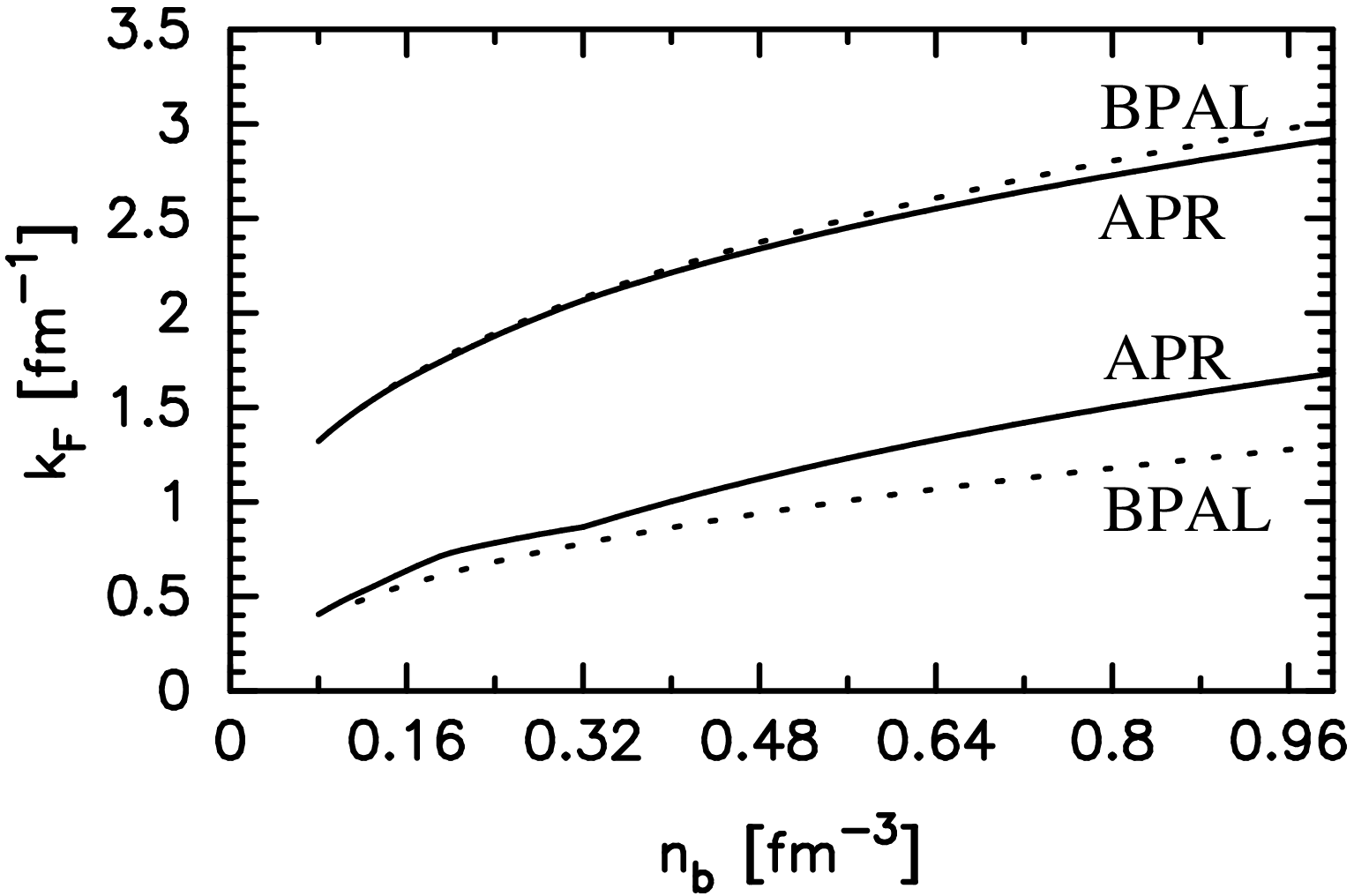}
\caption{Fermi momenta of neutrons (upper curves) and protons (lower curves)
         versus baryon density for the APR and BPAL EOS's 
         (BPAL21 and BPAL31 give nearly the same values).
         \label{Fig:kf-n}}
\end{figure}

\subsection{Nucleon Effective Masses}
\label{Sec:m*}

Under degenerate conditions ($T \ll \mu$) and in the absence of
collective excitations close to the Fermi surface, physical quantities
such as the specific heat, entropy, and superfluid gaps, and processes
such as neutrino emission from particles with energies in the
neighborhood of the Fermi energy, depend sensitively on the so-called
Landau effective masses of particles.  Formally, the Landau effective
mass $m^*$ of any degenerate fermion ($n$, $p$, $e$, and $\mu$ in our
case) is defined by
\be {m^*} \; \equiv \; p_F \left [\left. \frac{\partial
e(p)}{\partial p} \right|_{p=p_F} \right]^{-1} \,, 
\label{Eq:effm}
\ee 
where
$e(p)$ is the single-particle energy of the particle with
momentum $p$, and the derivative is evaluated at the Fermi momentum
$p_F$. 
For nonrelativistic interacting nucleons, the single particle energies
of the neutron and proton can be written as
\be
e_n (p)&=& \frac {p^2}{2m} + U_n(n_b,p) \nonumber \\
e_p (p) &=& \frac {p^2}{2m} + U_p(n_b,p) 
\ee
where $m$ is the nucleon mass in vacuum, and $U_n$ and
$U_p$ are the neutron and proton 
single-particle momentum-dependent potentials which are
obtained by appropriate functional differentiations of the potential
energy density.

\begin{figure}
\plotone{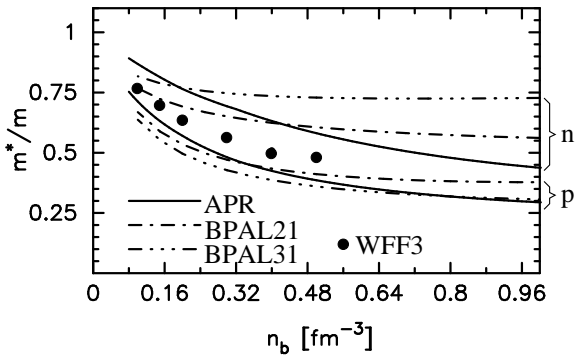}
\caption{Ratio of the Landau effective mass $m^*$ to the vacuum mass $m$ for
         neutrons and protons vs. density for the four EOS's. 
         \label{Fig:mstar}}
\end{figure}

From the Hamiltonian density in Appendix A of APR, 
the neutron and proton Landau effective masses are
\be
\frac {m^*}{m} &=& 
\left[ 1 + \frac {2m}{\hbar^2} (p_3+zp_5) n_b e^{-p_4n_b}  \right]^{-1} 
\ee
where $z=(1-x)$ for neutrons and $z=x$ for protons, and 
$p_3=89.8~{\rm MeV~fm^{5}},
~p_4=0.457~{\rm fm}^{3}$, and $p_5=-59~{\rm MeV~fm^{5}}$.  
The solid curves in 
Figure \ref{Fig:mstar} show the variation of the effective masses
with density for both neutrons and protons in charge-neutral 
beta-stable matter corresponding to the EOS of APR.

To date, single-particle energies for the WFF3 model are available
only for symmetric nuclear matter up to $n_b = 0.5~{\rm fm^{-3}}$
(\cite{W88}). Using the parameterization from equation~(7) 
of Wiringa (1988), we find that 
\be   
\frac {m^*}{m} = \left[ 1 - \frac {2m}{\hbar^2
\Lambda^2} \beta 
\left( 1 + \frac {k_F^2}{\Lambda^2}  \right)^{-2}  \right]^{-1}  
\ee
where the density-dependent parameters $\beta$ and $\Lambda$ are
tabulated in Table I of Wiringa (1988).  The filled circles in 
Figure~\ref{Fig:mstar} show the symmetric matter Landau effective masses for
this case.  Lacking further input, and encouraged by the fact that the
APR results for neutron and proton effective masses in beta-stable
isospin asymmetric matter bracket the WFF results in isospin
symmetric matter, we take the results of the APR model to apply for
the WFF model as well. As we will show, our final results
are not significantly affected by this approximation.

The neutron and proton Landau effective masses for the 
BPAL21 and BPAL31 models can be obtained straightforwardly from the
single-particle potentials given in \citet{Prak97}. Explicitly, 
\be
\frac {m^*}{m} = \left[ 1 - 
\sum_{i=1,2} (\beta_i + \gamma_i z) u \left[ 1 + \frac {(2zu)^{2/3} }{R_i^2} \right]^{-2} 
\right]^{-1}  \nonumber \\
\beta_i = \frac 25 \frac {(2C_i+4Z_I)}{E_F^{0}R_i^2} \qquad
\gamma_i =\frac 25 \frac {(C_i-8Z_I)}{E_F^{0}R_i^2} \qquad 
\ee
where $E_F^{0}$ is the
Fermi energy at the equilibrium density $n_0$ and the
parameters $C_i, Z_i$ and $R_i$ can be found in Tables 1 and  2
of \citet{Prak97}. The dash dotted curves in Figure \ref{Fig:mstar} 
show the effective masses for the BPAL models. 

The general trends to note in Figure \ref{Fig:mstar}  are 
\begin{itemize} 
\item the steady decrease of $m_n^*$ and $m_p^*$ 
with density, and 
\item $m_n^* > m_p^*$ in charge-neutral beta-stable matter.
\end{itemize}
It is interesting to observe that there is more spread in the
model predictions for $m_n^*$, particularly with increasing density,
than for $m_p^*$.

For noninteracting relativistic particles, such as $e$ and $\mu$, the
Landau effective masses are given by 
\be
m^* c^2 = \sqrt{m^2 c^4 + p_F^2 c^2} \,
\ee
where $m$ denotes the appropriate vacuum mass.

\subsection{Pairing}

The Fermi surface of a degenerate system of fermions, as are $e$'s,
$\mu$'s, $p$'s and $n$'s in a catalyzed neutron star, becomes unstable in the
presence of an attractive interaction between the particles whose
momenta lie close to the Fermi momentum: this is the Cooper theorem
\citep{Cooper56}.  As a result of this instability, the ground state
of the system is reorganized with a gap $\Delta$ in the
energy spectrum around the value of the Fermi energy; no particle can
have an energy between $E_F - \Delta$ and $E_F + \Delta$.

This instability usually appears as a second order phase transition,
the gap acting as the order parameter of the transition which has
a corresponding critical temperature $T_c$.  For $T > T_c$, the system
behaves as a normal Fermi liquid whereas for $T \simless T_c$ the gap
grows in magnitude which results in superfluidity or superconductivity.  The
precise value of $T_c$ depends on the nature of the pairing
interaction.

There is no obvious attractive interaction between $e'$s and/or $\mu'$s in
a neutron star and thus they are not expected to become paired at
temperatures relevant for our concerns here \citep{BP75}.

For nucleons, the strong interaction provides several channels in
which pairing is possible with $T_c$'s of order MeV.  Nucleon-nucleon
scattering data in vacuum indicate that at low momentum pairing should
occur through Cooper pairs with zero angular momentum $L$ in a
spin-singlet state, $^1S_0$, whereas for  larger momenta an $L =1$
spin-triplet ($J = L+S = 2$), $^3P_2$ pairing becomes favorable.  
Starting from a knowledge of the nucleon-nucleon interaction in
vacuum, the
difficulty of obtaining reliable values of the gap in a medium 
 is illustrated by considering the result for the
solution of the gap equation in the so called {\em BCS weak coupling
approximation} \citep{BCS56}:
\be
\Delta \; \sim \; E_F \; {\rm e}^{-1/V N(0)} \,,
\ee
where $V$ is the in-medium pairing interaction in the corresponding
channel and $N(0) \equiv m^* p_F/\pi^2 \hbar^3$ is the density of
states at the Fermi surface.  Small variations of $m^*$ and medium
effects on $V$ affect $\Delta$ in an exponential manner.

\begin{figure}
\plotone{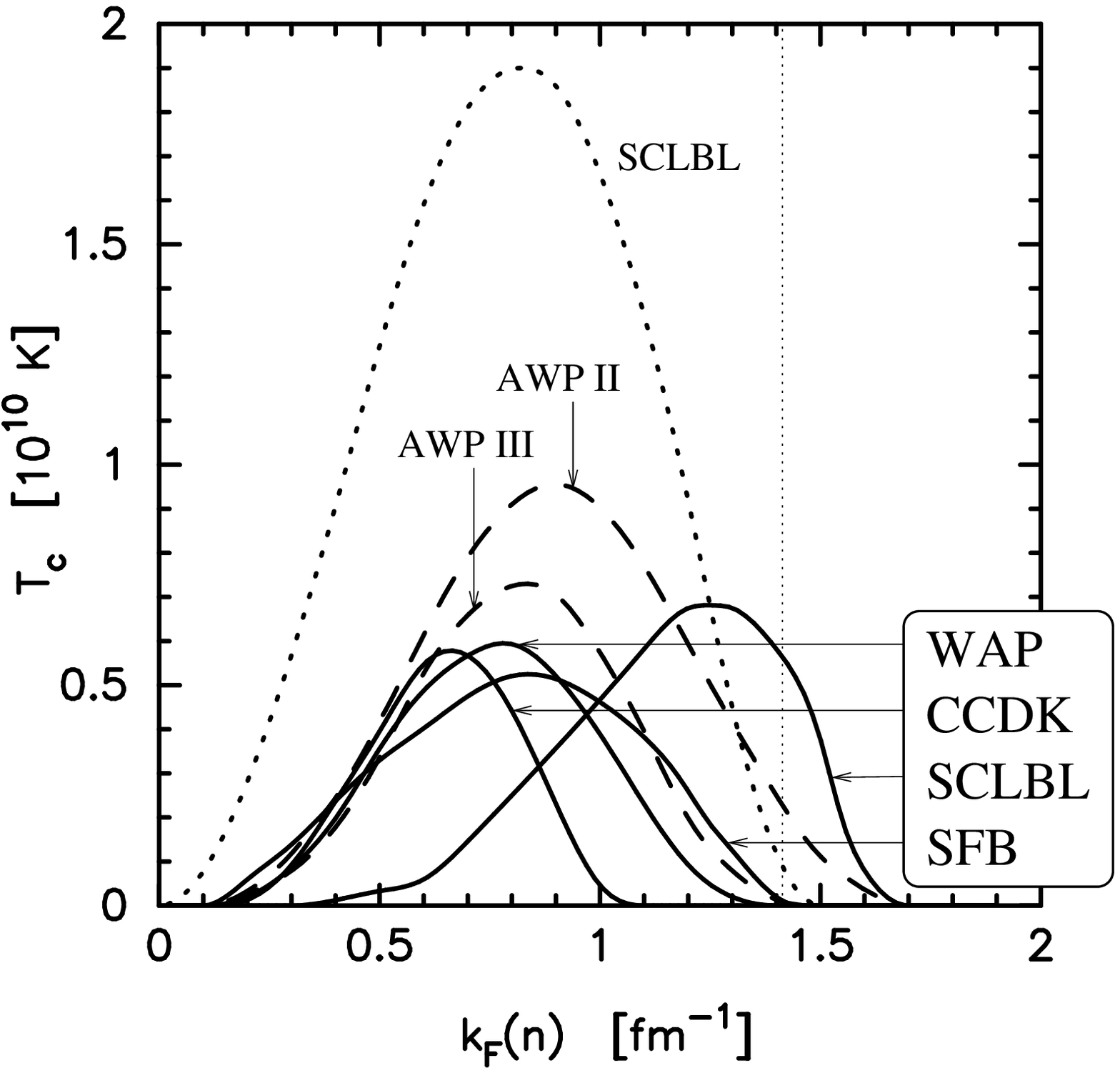}
\caption{Neutron $^1S_0$ pairing critical temperature $T_c$ vs neutron
fermi momentum $k_F$ from the calculations of \cite{AWP89} (labeled
as ``AWP II'' \& ``AWP III'': two slightly different results),
\citet{WAP93} (``WAP''), \citet{CCDK93} (``CCDK''), \cite{SCLBL96}
(``SCLBL''), and \cite{SFB03} (``SFB'').  The dotted curve shows the
results of \cite{SCLBL96} in the case where medium polarization is not
included. Medium polarization effects reduce the
$^1S_0$ gap by about a factor three.  The vertical dotted line shows the
location of the crust-core boundary.  
\label{Fig:n1S0}}
\end{figure}

The best-studied case is the $n$ $^1S_0$ gap in pure neutron
matter.  The gap appears at densities of order $n_0$ or lower, where
$m^*_n$ is well determined and the pairing interaction in vacuum for
the $^1S_0$ channel is accurately known.  Simple arguments \citep{P71}
show that medium polarization, the dominant medium effect on $V$,
should induce a reduction of the $^1S_0$ gap \citep{CKYC76} from its
value without medium polarization.  Much effort has been dedicated to
take into account medium polarization at various levels of
approximation.  With time and improving many-body techniques, the
results are beginning to show a convergence for the maximum value of
$T_c$, which is in the range $\sim 0.5~{\rm to}~0.7 \times 10^{10}$ K,
as can be seen in Figure~\ref{Fig:n1S0}.  The density range in which 
this gap is non-zero is still somewhat uncertain and corresponds to
the inner part of the crust and, possibly, the outermost part of the
core.

Since the results shown in Figure~\ref{Fig:n1S0} are for uniform pure
neutron matter, they will be altered by the presence of a small
fraction of protons in the outer core and the nonuniformity of neutron
density due to nuclei (or nuclear clusters) in the inner crust. This
latter effect has been studied recently by \citet{BBEV97} who show
that it does not alter significantly the results, at least at the
level of accuracy required for the study in the present paper.

\begin{figure}
\plotone{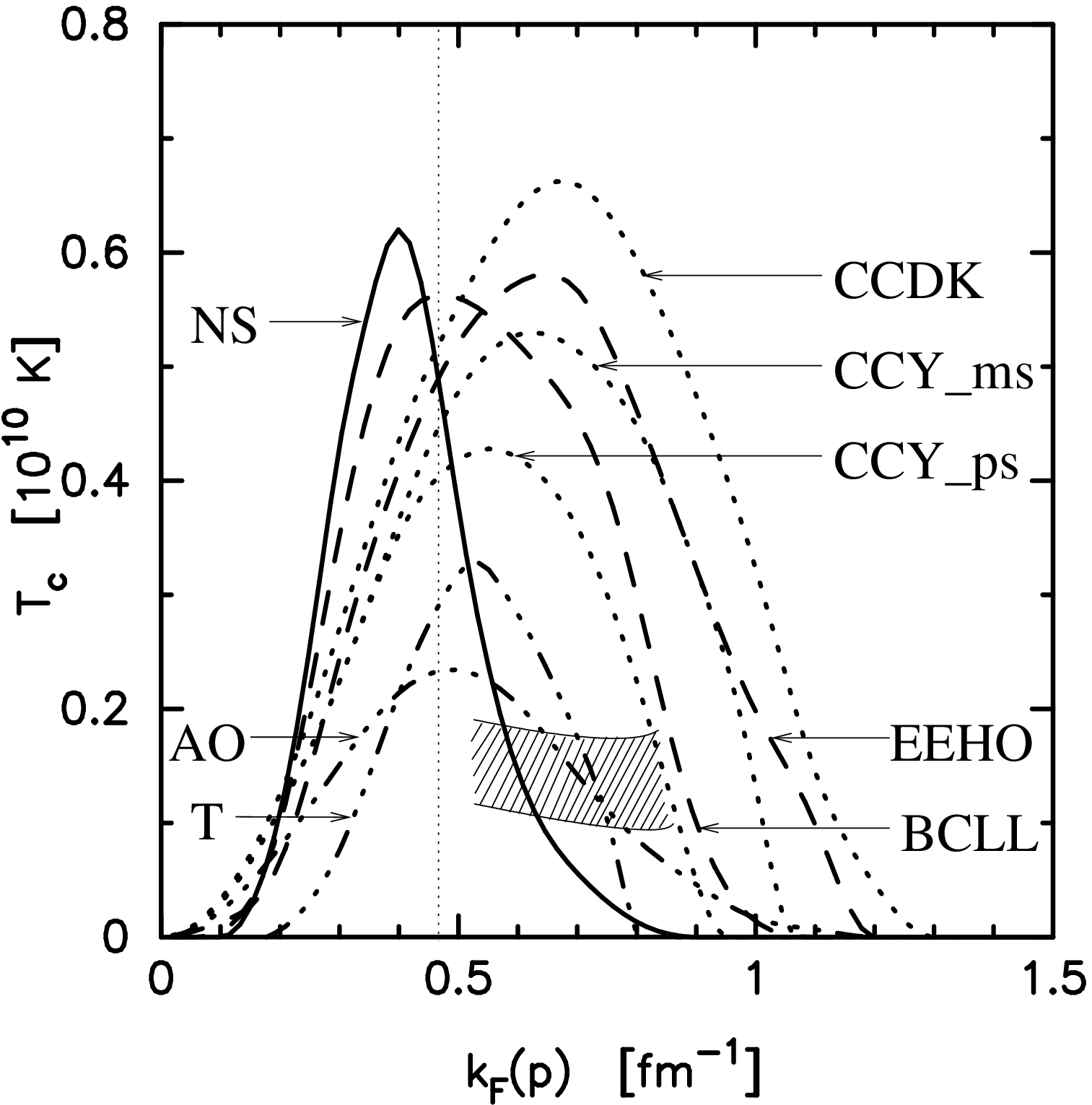}
\caption{Proton $^1S_0$ pairing critical temperature $T_c$ vs proton
fermi momentum $k_F$ from the calculations of \citet{T73} (labeled as
``T''), \citet{AO85a} (``AO''), \citet{CCY72} (``CCY\_ms'' \&
``CCY\_ps'': two slightly different results), \citet{CCDK93}
(``CCDK''), \citet{BCLL92} (``BCLL''), \citet{EEHO96} (``EEHO''),
\citet{NS81} (``NS'').  Only the calculation of \citet{NS81} include
medium polarization. The shaded region shows the estimates of
\citet{AWP91} of the range of values in which $T_c$ should lie due to
medium polarization. The vertical dotted line shows the location of
the crust-core boundary.  
\label{Fig:p1S0}}
\end{figure}

The $p$ $^1S_0$ gap is similar to the $n$ $^1S_0$ gap and
occurs at similar Fermi momenta $k_F$, but since protons represent
only a small fraction of the nucleons, this translates to high
densities which allows the gap to persist in much deeper regions of the
core than the $n$ $^1S_0$ gap.  The values of $T_c$ from several
calculations are shown in Figure~\ref{Fig:p1S0}.  An essential immediate
difference compared to the $n$ $^1S_0$ gap is that the $p$
$^1S_0$ gap is much smaller, $m^*$ being smaller for protons than for
neutrons (see Figure~\ref{Fig:mstar}).
It should be noted that all calculations shown in this figure have
employed values of $m^*_p$ larger than the values we report in
Figure~\ref{Fig:mstar}.  Insofar as the results of Figure~\ref{Fig:mstar}
for APR are indicative of the likely magnitudes of $m_p^*$, the values
of $T_c$ in Figure~\ref{Fig:p1S0} are likely overestimated, particularly
at large $k_F$.  Moreover, medium polarization effects on $V$ are much
more difficult to take into account for the $p$ $^1S_0$ gap than
for the neutron gap.  Such effects are expected to reduce the size of
the gap and, to date, only two works have attempted to include them
\citep{NS81,AWP91}.  The estimates of \citet{AWP91} show that medium
polarization reduces the $^1S_0$ gap roughly by a factor of three in
the stellar core.  It is important to notice that {\em all} these
calculations find that $T_c$ vanishes for $k_F > 1.5$ fm$^{-1}$ and in
most cases for $k_F > 1$ fm$^{-1}$.

\begin{figure}
\plotone{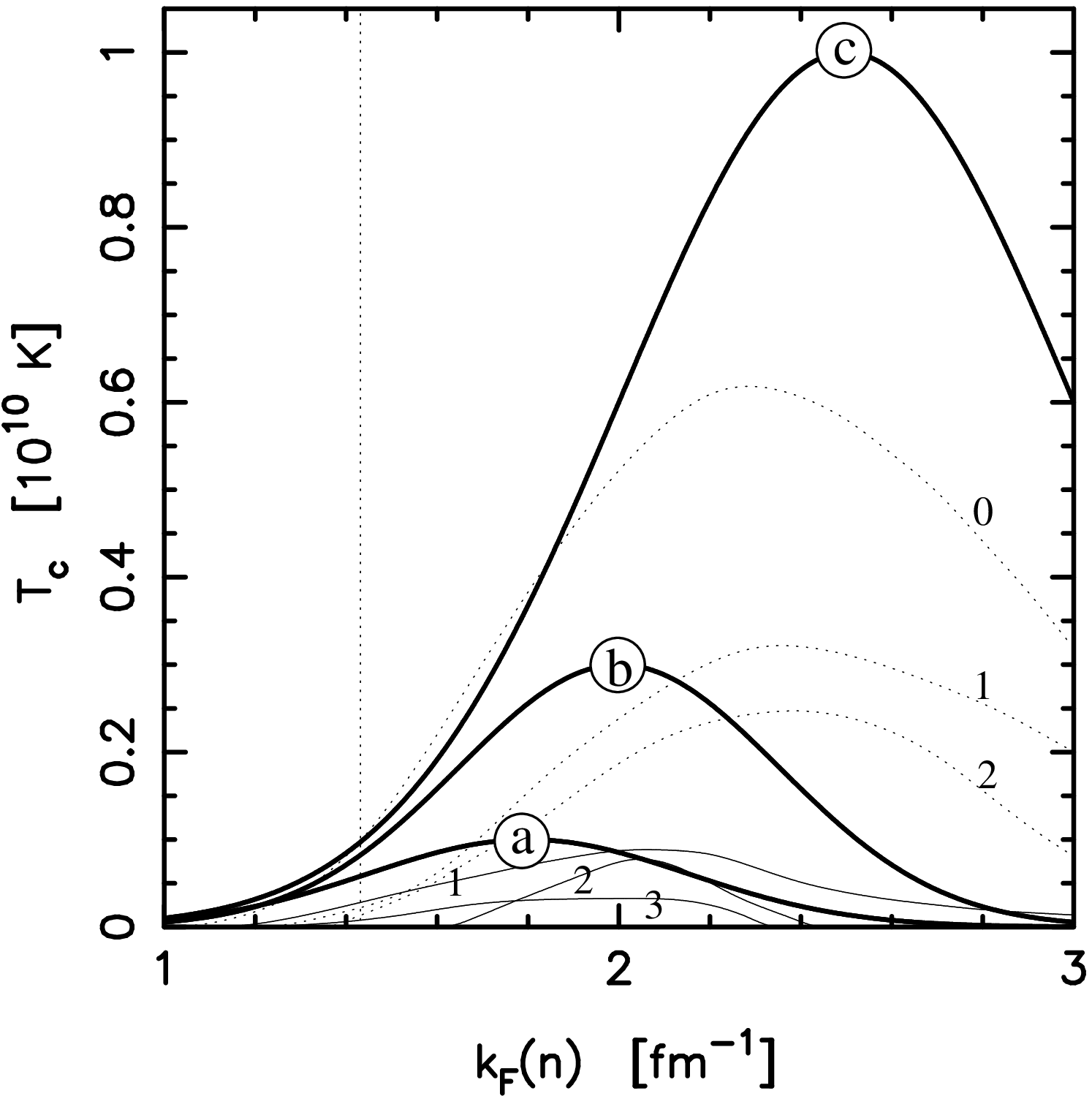}
\caption{Neutron ${^3P_2} - {^3F_2}$ pairing critical temperature
$T_c$ vs neutron fermi momentum $k_F$.  The three thin continuous
lines show the results of: (1) \citet{AO85b}, (2) \citet{T72b}, and
(3) \citet{EEHO96b}.  The three dotted lines show results obtained
assuming $m^*_n = m_n$ to illustrate the strong reduction  due
to small $m^*_n$: (0) \citet{HGRR70}, (1) \citet{AO85b}, and (2)
\citet{T72b}.  The three thick continuous lines, (a), (b), and (c),
bracket the results of \citet{BEEHS98}.  The vertical dotted line
shows the location of the crust-core boundary.
         \label{Fig:n3P2}}
\end{figure}

The so-called $n$ $^3P_2$ gap actually occurs in the ${^3P_2} -
{^3F_2}$ channel, since the tensor interaction couples
channels with $\delta L = 2$.  This coupling with the $^3F_2$ channel
increases the gap \citep{T72a}.  We present an illustrative sample of
published $T_c$ curves in Figure~\ref{Fig:n3P2}.  The large
differences among these curves points to the
inherent difficulty in pinning down the magnitude of
this gap.  The gap possibly extends to high densities where $m^*_n$ is
uncertain. In addition, the presence of a small fraction of protons
has generally been ignored except in the work of \citet{EEHO96b}, who
found a reduction of the gap by a factor $\sim 3$ when considering
neutron-proton matter in $\beta$-equilibrium.  

A fundamental problem, emphasized recently by \citet{BEEHS98}, is that
even the best modern models of the nucleon-nucleon interaction ({\em
in vacuum}) fail to reproduce the experimental phase shift in the
$^3P_2$ channel at laboratory energies above 300 MeV (corresponding to
the pion-production threshold).  Translating this energy into an
equivalent density implies that the bare pairing interaction is not
understood at densities $\simgreater 1.7 n_0$.  Moreover, medium
polarization effects have not been included in any of the calculations
displayed in Figure~\ref{Fig:n3P2}.  Estimates of such effects had
shown that they should strongly enhance the $^3P_2 - {^3F_2}$ gap
\citep{JKMS82}, but the recent results of \cite{SF04} indicate that
the medium-induced spin-orbit interaction strongly suppresses this
gap.  For this reason we will also consider the possibility that the
$n$ $^3P_2$ gap is vanishingly small.

The $J=2$ with $L=1$ or $3$ angular momentum and the tensor coupling
make the gap equation a system of coupled integral equations, one for
each value of the magnetic quantum number $m_J$ and the gap is not
isotropic: $\Delta = \Delta(\theta,\phi;T)$, where $\theta$ and $\phi$
are the polar angles of the momentum $\vec{k}$.  Until recently most
works had looked for single component solutions with $|m_J| = 0$, 1 or
2, but \citet{ZCK03} have shown that when considering multicomponent
solutions there are {\em at least} 13 distinct phases: for 7 of them
$\Delta(\theta,\phi;T)$ vanishes on the Fermi surface at some values
of $(\theta,\phi)$ whereas the other 6 are nodeless.  Nodeless gaps are
energetically favored over gaps with nodes, but by a small amount
[see, e.g., \citet{AO85b}], and may become disfavored in the case of
very fast rotation of the star or in the presence of an ultra-strong
magnetic field \citep{MSS80}.  We will, in this work, assume that the
$^3P_2 - {^3F_2}$ gaps are nodeless, since this maximizes the effect of
pairing on cooling and seems energetically favored, and more
specifically assume its angular dependence to be
$\Delta(\theta,\phi;T)\propto (1+3 \cos^2 \theta)^{1/2}$, corresponding to
the pure $m_J = 0$ phase.

For both the isotropic $^1S_0$ and the pure $m_J = 0$ phase of the 
${^3P_2} - {^3F_2}$ gap, the critical temperature $T_c$ and the $T=0$ gap
are related by
\be
k_B T_c \approx 0.57 \Delta_0
\label{Eq:T_c-Gap}
\ee
where $\Delta_0$ is $\Delta(T=0)$ for $^1S_0$ and the
angle averaged value of $\Delta(\theta,\phi;T=0)^2$ over the Fermi sphere for 
$^3P_2 - {^3F_2}$ \citep{BCLL92}. 
 The temperature dependence of $\Delta$ for these two cases
has been calculated and fitted by simple analytical expressions by
\citet{LY94a}.

Once the energy gap $\Delta$ is given, the quasi- particle energy
spectrum near the vicinity of the Fermi surface can be expressed as
\be
e({\bf k}) = E_F - {\sqrt {\Delta({\bf k})^2 + \eta^2}} \qquad {\rm for~~} k < k_F
\nonumber \\ 
=  E_F + {\sqrt {\Delta({\bf k})^2 + \eta^2}} \qquad {\rm for~~} k > k_F 
\label{Eq:egap0}
\ee
where the quantity $\eta$ is given by
\be
\eta  =  \frac {k^2}{2m^*} -  \frac {k_F^2}{2m^*}
 \nonumber \\
   \cong \left(\frac{k_F}{m^*}\right) \cdot (k-k_F) 
   \equiv v_F \cdot (k- k_F)
\ee
The effect of this change in the quasi-particle spectrum is discussed
below in \S~\ref{Sec:Cv} and \S~\ref{Sec:nu}.

Finally, it is instructive to consider $T_c$ for the $p$ $^1S_0$
gap and the $n$ $^3P_2 - {^3F_2}$ gap in terms of density and also
in terms of the volume in the stellar core, once an EOS and a stellar
mass has been chosen.  (This is presented for a few cases in
Figures~\ref{Fig:Coop_n3P2} and \ref{Fig:Coop_p1S0} in a 1.4 \Msol star
built with the APR EOS.)  The $p$ $^1S_0$ gap vanishes or is very
small in the inner core for most calculations whereas the $n$ $^3P_2
- {^3F_2}$ gap is more likely to reach the stellar center with high
values.

\subsection{The Specific Heat}
\label{Sec:Cv}

The total specific heat (per unit volume) at constant volume, $c_v$, receives
contributions from all of the constituents inside the star. In the
homogeneous phase above nuclear density, 
\be
c_v = \sum_{i=e^-,\mu^-,n,p} c_{i,v}
\ee
For a given species of unpaired spin $\frac 12$ fermions, $c_v$ is given by 
\be
c_v &=& \frac {1}{n_b} \frac {\partial \epsilon}{\partial T} \nonumber \\ 
  &=& 2 \int \frac{d^3k}{(2\pi)^3} (e-\mu) \frac {\partial
f}{\partial T} - 
T \frac {\left (2  \int \frac{d^3k}{(2\pi)^3} \frac {\partial
f} {\partial T}\right)^2} { 2  \int \frac{d^3k}{(2\pi)^3}  
\frac {\partial f} {\partial \mu}  } \nonumber \\  
\ee
where $e(k)$ is the single-particle spectrum, $f$ is the Fermi-Dirac
distribution function, and $\mu$ and $T$ are the chemical potential
and temperature, respectively. Under the degenerate conditions of
interest here, $T \ll \mu$, the contribution from the second term
above can be safely neglected with the result
\be
c_{i,v} = N_{i}(0) \frac{\pi^2}{3} k_B^2 T 
          = \frac{m_i^* n_i}{p_{i,F}^2}  \pi^2 k_B^2 T 
\label{Eq:Cv_deg}
\ee
In writing the rightmost relation above, the density of states at the
Fermi surface $N_i(0)$ has been expressed in terms of the Landau
effective mass $m_i^*$ (see equation (\ref{Eq:effm})) through the relations 
\be
N_{i}(0) = \frac {3n_i}{k_Fv_F}\,, \qquad v_F = \left. 
\frac {\partial e}{\partial k}\right |_{k_F} 
\ee
where $v_F$ denotes the velocity at the Fermi surface. 

The effect of pairing interactions on the specific heat depends on the
disposition of $T$ with respect to $T_c$.  When $T$ reaches $T_c$,
there is a sharp increase in the specific heat due to the large
fluctuations characteristic of a second order phase transition.
Subsequently, when $T \ll T_c$, a Boltzmann-like suppression occurs
due to the presence of a gap in the energy spectrum,
equation~(\ref{Eq:egap0}).  In practice, these effects are taken into account
by using control functions that would multiply the unpaired values of $c_v$; 
these functions have been calculated for nucleon pairing in
both the $^1S_0$ and $^3P_2$ channels by \cite{LY94a} and are
displayed in the upper panel of Figure~ \ref{Fig:All_coeff}.

\begin{figure}
\plotone{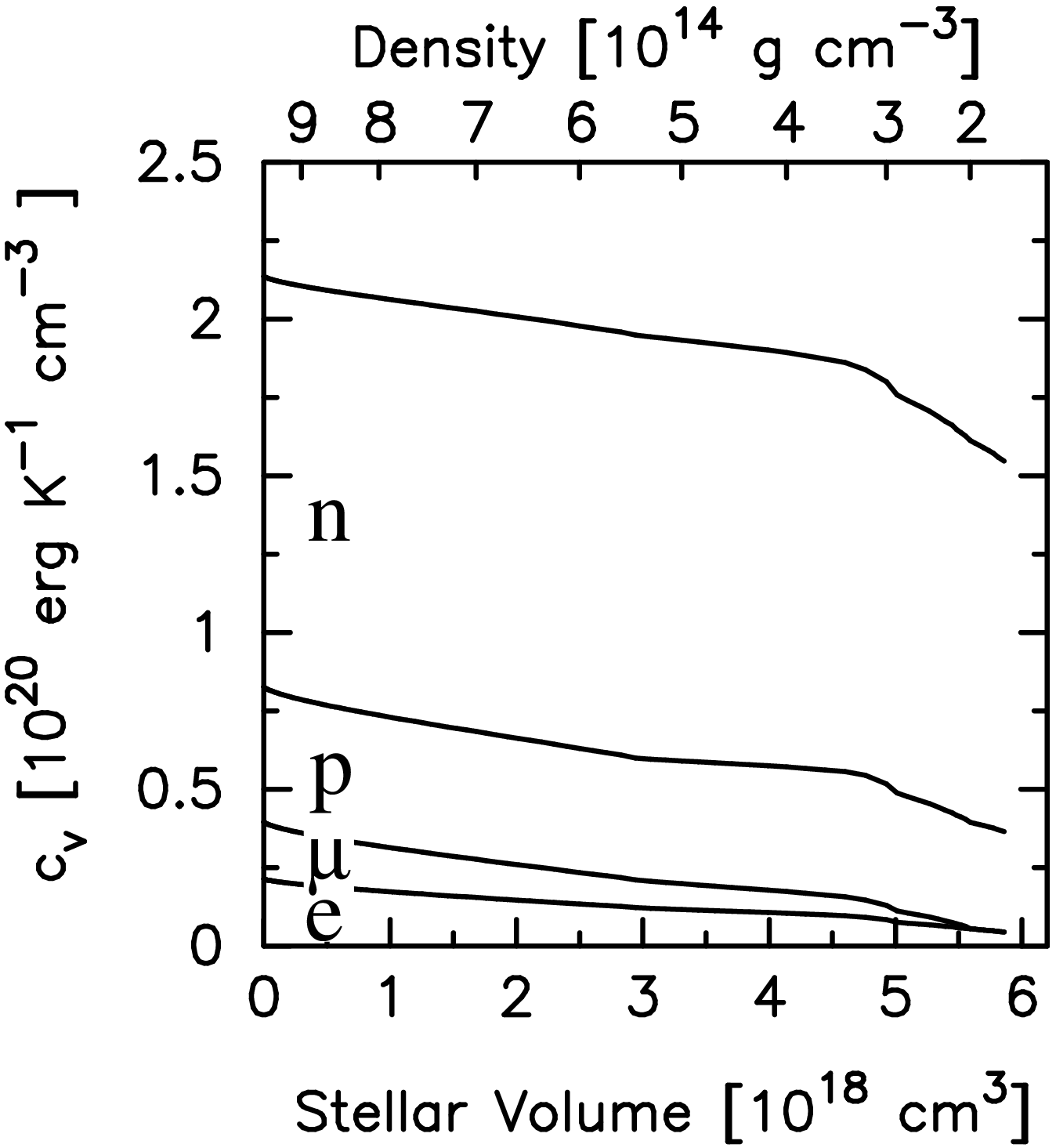}
\caption{Cumulative specific heats of $e$, $\mu$, $p$, and $n$
vs. stellar volume in the core of a 1.4 \Msun star built using the APR
EOS, at temperature $T=10^9$ K. Nucleons are assumed to be unpaired.
\label{Fig:Cv}}
\end{figure}

The cumulative contributions to $c_v$ are presented in
Figure~\ref{Fig:Cv}. 
Once a nucleonic component becomes paired,
its contribution will ultimately be suppressed when $T \ll T_c$, but
in all cases the lepton contribution will always remain.  For different
temperatures the various contributions all scale as $T$ in the absence
of pairing and their relative importance hence remains the same as
illustrated in this figure.  

\begin{figure}
\plotone{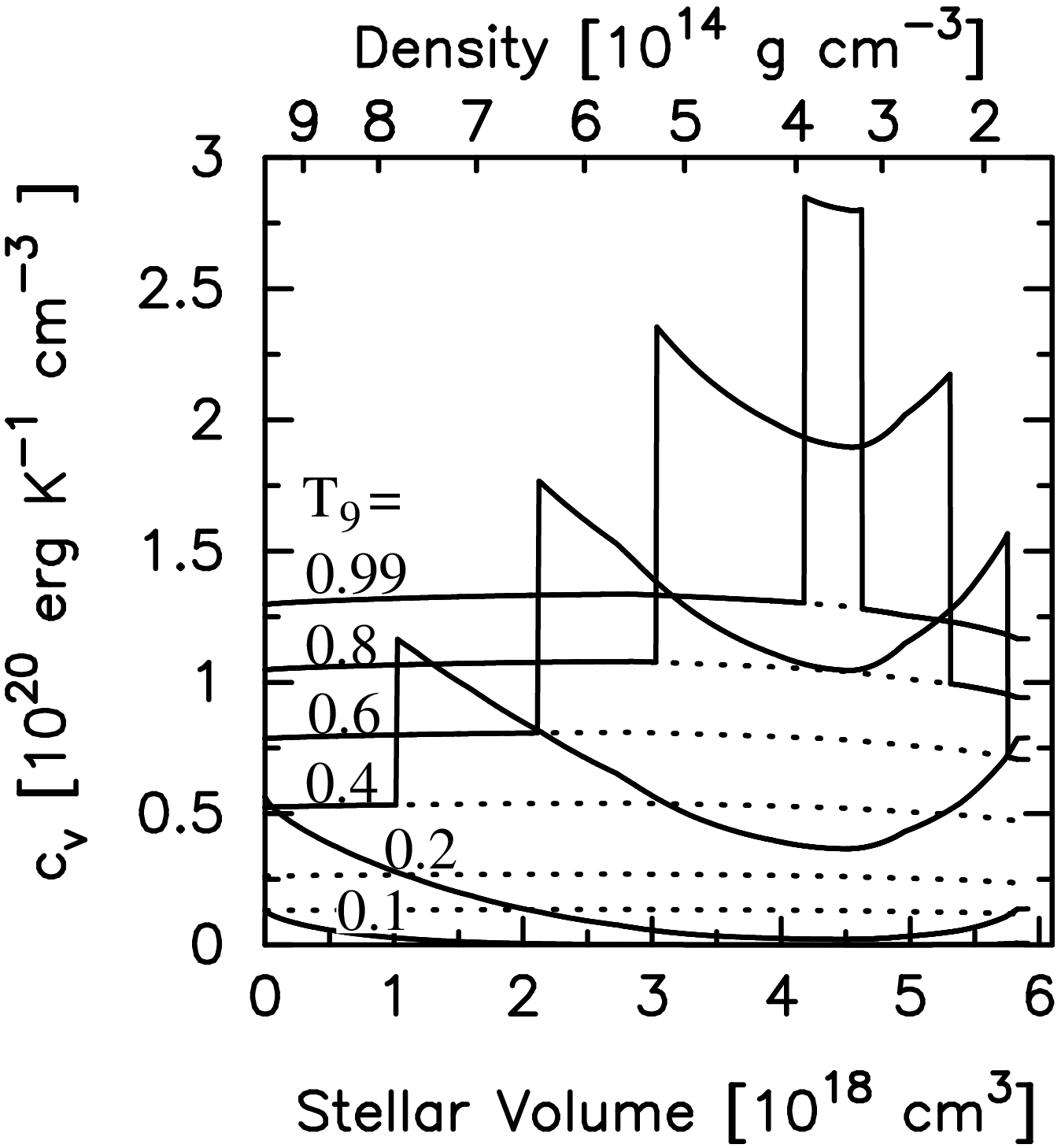}
\caption{Specific heat of neutrons in the core of a 1.4 \Msun star
built using the APR EOS, at six different temperatures, from $0.99$
down to $0.1$ times $10^9$ K illustrating the effects of pairing.  No
gap is present for results shown by the dotted lines.  The $n$ $^3P_2$
gap ''a'' is assumed for continuous curves.  This gap has a maximum
$T_c$ of $1 \times 10^9$ K at $\rho = 3.61 \times 10^{14}$ gm
cm$^{-3}$. the continuous curves show the jump of $c_v$ by a factor
2.188 (see Figure~\protect\ref{Fig:All_coeff}) at the two zones where
$T = T_c$ and its progressive suppression when $T \ll T_c$ in the
layers in between.
         \label{Fig:Cv-nnn}}
\end{figure}

The suppression by pairing is illustrated in Figure~\ref{Fig:Cv-nnn}
for the case of neutrons at various temperatures beginning with $10^9$
K as in Figure~\ref{Fig:Cv}. The results in this figure, in agreement
with those in Figure~\ref{Fig:All_coeff}, show that beyond the initial
strong enhancement at $T \simless T_c$, $c_v$ will be almost
completely suppressed only when $T < 0.1~{\rm to}~0.2 \times T_c$
everywhere within the core.  

In the crustal region, contributions to the specific heat arise from
the neutron gas in the inner crust, the degenerate electron gas and
the nuclear lattice.  The contribution of the neutron gas is strongly
suppressed by the $n$ $^1S_0$ gap.  The specific heat of the crust is
smaller than that of the leptons in the core and is hence not very
important, but included in all our calculations.


\subsection{Neutrino Emissivities}
           \label{Sec:nu}

Until the time that photon emission takes over the cooling of the
star, the thermal energy of the star is lost from both the crustal
layers and the core of the star chiefly by neutrino emission. The various
neutrino emission processes that are included in our study are
summarized below.

In the crust of the star, we include neutrino pair emission from
plasmons according to \cite{HRW94}, or, from the practically
equivalent results of \cite{IHNK96}, and from electron-ion
bremsstrahlung according to \cite{KPPTY99}.  These two processes are
not affected by nucleon pairing.  We also consider neutrino pair
emission from neutron-neutron bremsstrahlung and its suppression by
neutron pairing. The treatment of pairing suppressions is outlined
below in the context of similar suppressions in the core. In addition,
we consider neutrino pair emissions from the formation and breaking of
$n$ $^1S_0$ Cooper pairs, 
also described further below.  
In the presence of a magnetic field,
synchrotron neutrino pair emission from electrons also occurs
\citep{BHKY97}, but contributions from this process are negligible.
Photo-neutrino emission and neutrino pairs from $e^+e^-$ pair
annihilation are effective only at low density and high temperature
and are not relevant here.

In the core of the star, we include (1) the modified Urca processes
and the similar nucleon bremsstrahlung processes with their
corresponding suppressions by nucleon pairing, and (2) neutrino pair
emission from the formation and breaking of Cooper pairs.  The
emissivity from the neutron branch of the modified Urca process
\be
\begin{array}{rcl}
n + n'     & \rightarrow & p + n' + l + \bar{\nu}_l \\
p + n' + l & \rightarrow & n + n' + \nu_l \;\;\;\;\;\;\; ,
\end{array}
\ee
where $l$ is either an electron or a muon and $\nu_l$ or $\bar{\nu}_l$
is the associated neutrino or antineutrino, is taken from \cite{FM79}
and \cite{YL95}. Explicitly,
\be
q_\nu^{\rm Murca \; n} = 8.55 \times 10^{21} \; 
     \left(\frac{m_n^*}{m_n}\right)^3
     \left(\frac{m_p^*}{m_p}\right)    
     \nonumber \\
     \left[\left(\frac{k_{F e }}{k_{F0}}\right) +
           \left(\frac{k_{F\mu}}{k_{F0}}\right) \right]
     \alpha_n \beta_n \left(\frac{T}{10^9 \; \rm K}\right)^8 \,,
\label{Eq:Murca_n}
\ee
where $k_{F0}$ = 1.68 fm$^{-1}$ is a fiducial normalization factor.
The emissivity from the proton branch of the modified Urca process
\be
\begin{array}{rcl}
n + p'     & \rightarrow & p + p' + l + \bar{\nu}_l \\
p + p' + l & \rightarrow & n + p' + \nu_l \;\;\;\;\;\;\; ,
\end{array}
\ee
is taken from \cite{YL95} in the form 
%
\be
q_\nu^{\rm Murca \; p} = 8.55 \times 10^{21} \; 
     \left(\frac{m_n^*}{m_n}\right) \!\!
     \left(\frac{m_p^*}{m_p}\right)^3    
   \nonumber \\
     \left[\left(\frac{k_{F e }}{k_{F0}}\right) \!\!
           \left(\! 1 \! - \! \frac{k_{Fe}}{4 k_{Fp}}\right) +
           \left(\frac{k_{F\mu}}{k_{F0}}\right) \!\!
           \left(\! 1 \! - \! \frac{k_{Fe}}{4 k_{Fp}}\right) \right] 
   \nonumber \\
     \alpha_p \beta_p \left(\frac{T}{10^9 \; \rm K}\right)^8 \,.
\label{Eq:Murca_p}
\ee
In equations~(\ref{Eq:Murca_n}) and (\ref{Eq:Murca_p}), the coefficients
$\alpha_n$, $\alpha_p$, $\beta_n$, and $\beta_p$ are of order
unity and describe corrections due to the momentum transfer dependence of the
matrix element in the Born approximation ($\alpha_{n,p}$) and due to non-Born
corrections and strong interaction effects beyond the one pion exchange plus
Landau coefficients ($\beta_{n,p}$) \citep{FM79,YL95}.
To be specific, following \cite{YL95} we use
\be
\begin{array}{rcl}
\alpha_n = \alpha_p & = &  
    1.76 - 0.63 \left(\frac{k_{F0}}{k_{Fn}}\right)^2
\\
\beta_n  = \beta_p  & = & 0.68 \,.
\end{array}
\ee
In addition to the above two charged current modified Urca 
processes,  three neutral current bremsstrahlung processes
\be
\begin{array}{ccc}
n + n' & \rightarrow & n + n' + \nu_l + \bar{\nu}_l \; \\
n + p' & \rightarrow & n + p' + \nu_l + \bar{\nu}_l \; \\
p + p' & \rightarrow & p + p' + \nu_l + \bar{\nu}_l \; ,
\end{array}
\ee
where the pairs $\nu_l \bar{\nu}_l$ can be an 
$e$, $\mu$ or $\tau$ neutrino pair, also contribute. Their
emissivities are \citep{FM79,YL95}
\be
q_\nu^{\rm Brem \; nn} = 3 \times 7.4 \times 10^{19} \; 
     \left(\frac{m_n^*}{m_n}\right)^4
     \nonumber \\
     \left(\frac{k_{F n }}{k_{F0}}\right)
     \alpha_{nn} \beta_{nn} \left(\frac{T}{10^9 \; \rm K}\right)^8 \,,
\label{Eq:Bem_nn}
\ee
\be
q_\nu^{\rm Brem \; np} = 3 \times 1.5 \times 10^{20} \; 
     \left(\frac{m_n^*}{m_n}\right)^2
     \left(\frac{m_p^*}{m_p}\right)^2
     \nonumber \\
     \left(\frac{k_{F p }}{k_{F0}}\right)
     \alpha_{np} \beta_{np} \left(\frac{T}{10^9 \; \rm K}\right)^8 \,,
\label{Eq:Bem_np}
\ee
and
\be
q_\nu^{\rm Brem \; pp} = 3 \times 7.4 \times 10^{19} \; 
     \left(\frac{m_p^*}{m_p}\right)^4
     \nonumber \\
     \left(\frac{k_{F p }}{k_{F0}}\right)
     \alpha_{pp} \beta_{pp} \left(\frac{T}{10^9 \; \rm K}\right)^8 \,,
\label{Eq:Bem_pp}
\ee
where the $\alpha$'s and $\beta$'s are corrections of order unity for which we use
\citep{YL95}
\be
\alpha_{nn} = 0.59 \;\;\;\;\; \beta_{nn} = 0.56
\nonumber \\
\alpha_{np} = 1.06 \;\;\;\;\; \beta_{np} = 0.66
\nonumber \\
\alpha_{pp} = 0.11 \;\;\;\;\; \beta_{pp} = 0.70 \,.
\nonumber
\ee
It is important to note that the emissivities of the modified Urca and
bremsstrahlung processes have not been accurately calculated,
particularly at the highest densities reached in the center of the
stars we are considering.  \cite{VS86} have proposed that when the
density approaches the critical density for the onset of charged pion
condensation, the softening of the pion mode induces a strong increase
in the above emissivities. Since this approach assumes the occurrence
of charged pion condensation, we will not consider it here as part of
the minimal scenario.  Nevertheless, less dramatic medium effects are
certainly at work.  Recently, \cite{HPR01,VDDT04,SJG04} have revisited
the bremsstrahlung processes, including hadronic polarization up to
the two loop level, and found a reduction of the rates by a factor of
about 4 at saturation density.  In view of this, we will at first take 
equations~(\ref{Eq:Murca_n}), (\ref{Eq:Murca_p}),
(\ref{Eq:Bem_nn}), (\ref{Eq:Bem_np}), and (\ref{Eq:Bem_pp}), with the
quoted  $\alpha$'s and $\beta$'s at face value, but
will, in addition, consider the effects of ``cranking up'' or ``down'' all
modified Urca and bremsstrahlung emissivities by a significant factor
in \S~\ref{Sec:IncreasedMurca}.

\begin{figure}
\plotone{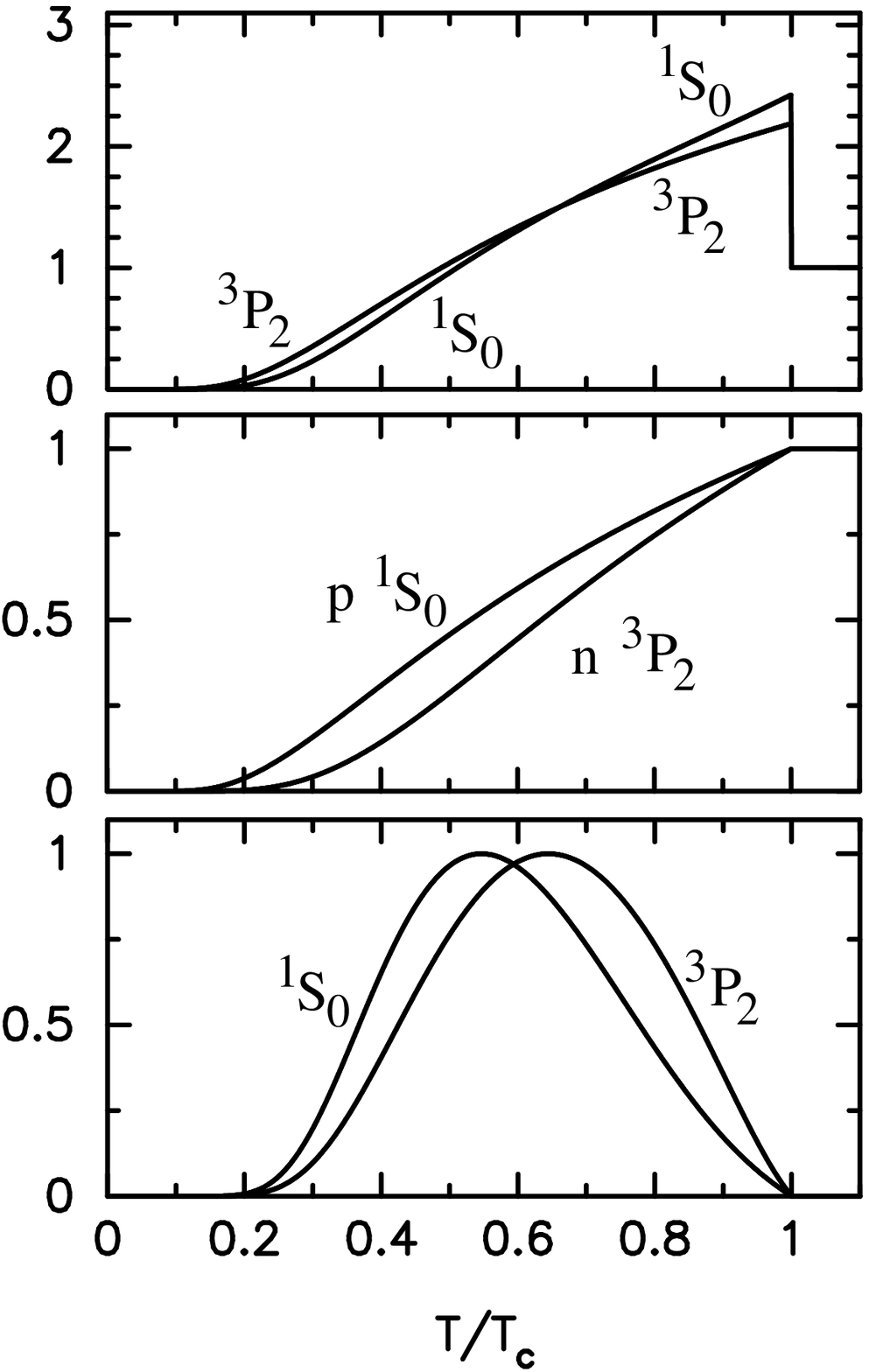}
\caption{Control functions for pairing effects in $^1S_0$ (for neutrons
         and/or protons) and $^3P_2$ (neutrons) channels. The 
         top panel shows the function relavant 
	 for the specific heat (\S~\protect\ref{Sec:Cv}), the
         central panel that for the neutron branch of the modified Urca process
         with either $p$~$^1S_0$ or $n$~$^3P_2$ pairing
	(\S~\protect\ref{Sec:nu}) and the 
         bottom panel is for the PBF process (\S~\protect\ref{Sec:nu}).
         \label{Fig:All_coeff}}
\end{figure}

Once the temperature $T$ reaches the pairing critical temperature
$T_c$ of either the neutrons or protons in a given layer of the star,
the corresponding neutrino emission process becomes suppressed by the
development of an energy gap $\Delta(T)$ in the single particle
excitation spectrum (see equation~(\ref{Eq:egap0})).  Similarly to
what happens for the specific heat, the neutrino emissivities are
suppressed by factors
%
%
which vary approximately like $\exp(-\Delta(T)/k_B T)$.
In our calculations, we employ the accurate calculations of these
various various control functions including pre-exponential factors (see
\cite{YL95} for details) for the two modified Urca processes and the three
bremsstrahlung processes, in the presence of $n$ $^1S_0$ or $^3P_2$
pairing and/or $p$ $^1S_0$ pairing. 
Two representative examples, for the neutron branch of the modified Urca
process, are plotted in the central panel of 
Figure~\ref{Fig:All_coeff}: for this specific case neutron pairing has
a much stronger effect than proton pairing since three neutrons, but
only one proton, participate in the reaction.

As the temperature begins to approach $T_c$, new channels for neutrino
emission through the continuous formation and breaking of Cooper pairs
\citep{FRS76,VS87} begin to become operative.  We take 
emissivities from these ``Pair Breaking and Formation'' or PBF processes
as \citep{YL95}
\be
q_\nu^{\rm p ^1S_0} =
   2.6 \times 10^{21} \left(\frac{n_b}{n_0}\right)^{1/3}
   \left(\frac{m_p^*}{m_p}\right)  \times
   \nonumber \\
   \tilde{F}_{\rm ^1S_0}(T/T_c) 
   \left(\frac{T}{10^9 \rm K}\right)^7
\ee
for $p$ $^1S_0$ pairing,
\be
q_\nu^{\rm n ^1S_0} =
   1.0 \times 10^{22} \left(\frac{n_b}{n_0}\right)^{1/3}
   \left(\frac{m_n^*}{m_n}\right)  \times
   \nonumber \\
   \tilde{F}_{\rm ^1S_0}(T/T_c) 
   \left(\frac{T}{10^9 \rm K}\right)^7
\ee
for $n$ $^1S_0$ pairing, and
\be
q_\nu^{\rm n ^3P_2} =
   8.6 \times 10^{21} \left(\frac{n_b}{n_0}\right)^{1/3}
   \left(\frac{m_n^*}{m_n}\right)  \times
   \nonumber \\
   \tilde{F}_{\rm ^3P_2}(T/T_c) 
   \left(\frac{T}{10^9 \rm K}\right)^7
\ee
for $n$ $^3P_2$ pairing, assuming again that pairing in this last
case occurs in the $m_J = 0$ phase.  
The control functions $\tilde{F}_{\rm ^1S_0}(T/T_c)$ and
$\tilde{F}_{\rm ^3P_2}(T/T_c)$ are shown in the lower panel of
Figure~\ref{Fig:All_coeff} and give the dependence on $T_c$.
These PBF processes can be
regarded as $nn$ or $pp$ bremsstrahlung processes with a strong
correlation in the initial state in the case of the breaking of a
Cooper pair, or in the final state in the case of the formation of a
Cooper pair, and exemplify an extreme case of medium correction to the
bremsstrahlung processes.  Their efficiencies are similar to those of
the bremsstrahlung processes of equations~(\ref{Eq:Bem_nn}),
(\ref{Eq:Bem_np}), and (\ref{Eq:Bem_pp}). However, they are less
sensitive to the values of the nucleon effective masses because they are
proportional to $m^*$ instead of $m^{*4}$.  Furthermore, the $T^7$
dependence of the PBF processes, compared to the $T^8$ dependence of
the bremsstrahlung processes, allows the PBF processes to eventually
dominate the total neutrino luminosity.  The precise value of $T_c$
for the neutron and/or proton pairing determines when this
dominance occurs, as will become apparent in \S~\ref{Sec:Coop}.  
The form of the control functions (Figure~~\ref{Fig:All_coeff})
show clearly that the PBF
process turns on when $T$ reaches $T_c$, increases its efficiency
as $T$ decreases, and becomes exponentially suppressed when the gap
approaches its maximum size $\Delta(0)$ when $T \simless 0.2 T_c$.


\section{THE NEUTRON STAR ENVELOPE}
         \label{Sec:Envelope}

It is customary to separate cooling models into the {\em interior}
and the {\em envelope}, the latter being the upper layer in which a strong
temperature gradient exists whereas the interior designates
everything inside which becomes isothermal within a few years after
the birth of the neutron star.

Precisely, the envelope can be defined as the layer extending from the
{\em photosphere}, the uppermost layer where the emitted spectrum is
determined, down to a boundary density $\rho_b$ such that
the luminosity in the envelope is equal to the total
surface luminosity of the star, $L(r) = L(R)$.
The thermal relaxation time scale of the envelope is
much shorter than the cooling time scale of the interior so that it
can be treated separately as a layer constantly in a stationary state.
Equation~(\ref{Eq:dLdr}) therefore implies that the
neutrino emission is negligible in the envelope.  Since the thickness
of the envelope is of the order one hundred meters or less, the
envelope can be treated in the plane parallel approximation.  Within
these approximations, integration of the heat transport and hydrostatic
equilibrium equations gives
a relationship between the temperature at the bottom of the envelope,
$T_b$, and the flux $F$ going through it, or, equivalently, a relationship
between the effective
temperature $T_e$ and $F$:  
$F \equiv \sigma_{\scriptscriptstyle SB} T_e^4$.  This
relationship is commonly termed as the ``$T_e - T_b$ relationship''.

Detailed numerical calculations along this line were presented by
\citet{GPE82,GPE83} and an analytical approximation to these results
was provided by \citet{HA84}, who assumed that the chemical composition
of matter corresponds to that in beta-equilibrium.  \citet{GPE82}
found a simple analytical relationship
\be
T_e = 0.87 \times 10^6 \; 
(g_{s \, \scriptscriptstyle 14})^{1/4} \; (T_b/10^8 {\rm K})^{0.55} \,,
\label{Eq:Tb_TeGPE}
\ee
where $g_{s \, \scriptscriptstyle 14}$ is surface gravity $g_s$
measured in $10^{14}$ cgs units.  (As a rule of thumb, this gives $T_e
\propto T_b^{1/2}$ and $T_e \sim 10^6$ K when $T_b \sim 10^8$ K.)
This equation illustrates the fact that the dependence of the envelope
structure on $M$ and $R$ is entirely contained in $g_s$ and that
$T_e/g_{s \, \scriptscriptstyle 14}^{1/4}$ is independent of $M$ and
$R$.  This allows us to use ``generic'' envelope models and glue them
to the upper layer of any stellar model.

\subsection*{The Sensitivity Strip: Effects of Chemical Composition
             and Magnetic Fields}

The most important finding of \citet{GPE82,GPE83} is that the $T_e -
T_b$ relationship is mostly determined by the value of the thermal
conductivity $\lambda$ in a thin layer in which ions are in the liquid
phase and where $\lambda$ is dominated by electron conduction. 
This layer was thus called the ``sensitivity strip'' in the $\rho-T$
plane.  The sensitivity strip is located at lower densities for lower
temperatures and spans about one and a half order of magnitude in
density depth.

\begin{figure} 
   \plotone{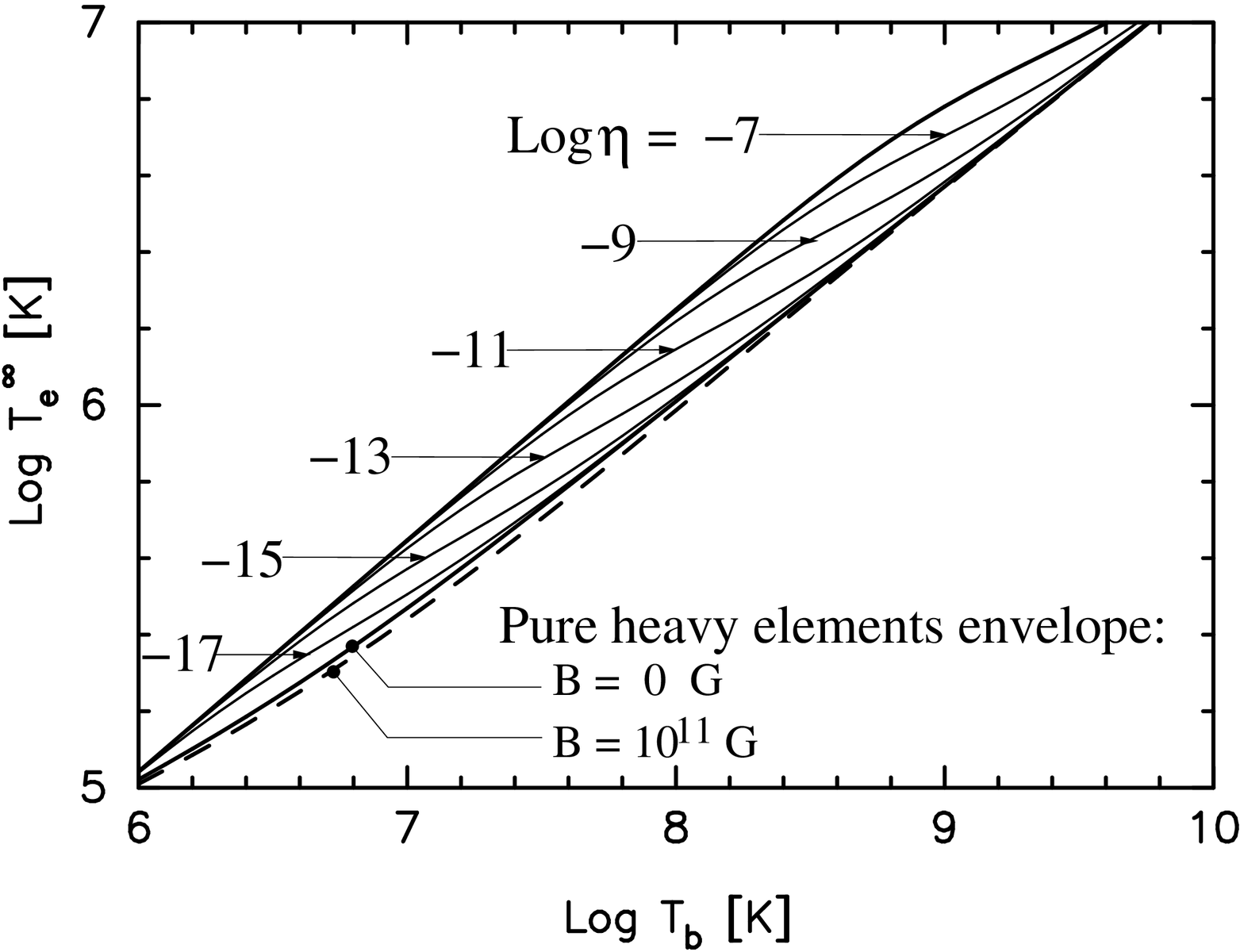}
   \caption{Relationship between the effective temperature $T_e^\infty$ and
            the interior temperature $T_b$ at the bottom of the
            envelope assuming various amounts of light elements
            parametrized by 
            $\eta \equiv g_{s \, 14}^2 \Delta M_{\rm L}/M$ 
            ($\Delta M_{\rm L}$ is the mass in light elements in the
            envelope, $g_{s \, 14}$ the surface gravity  in units of
            $10^{14}$ cm s$^{-1}$, and $M$ is the star's mass), in the
            absence of a magnetic field \protect\citep{PCY97}.  Also
            shown are the $T_e^\infty - T_b$ relationships for an envelope of
            heavy elements with and without the presence of a dipolar
            field of strength of $10^{11}$ G following \cite{PY01}.}
   \label{Fig:Tb_Te_accreted} 
\end{figure}

The presence of light elements (e.g., H, He, C or O) in the envelope
can significantly affect the $T_e - T_b$ relationship if the sensitivity
strip is populated by these elements \citep{CPY97}. Lighter elements
will burn into heavier ones in the thermonuclear regime at high enough
$T$ and in the pycnonuclear regime at high enough $\rho$, but
conditions in the envelope are usually such that H may be present up
to densities $\sim 10^7$ gm cm$^{-3}$, He up to $\sim 10^9$ gm
cm$^{-3}$ and C up to $\sim 10^{10}$ gm cm$^{-3}$.  The critical
temperatures for thermonuclear burning and densities for pycnonuclear
burning are well within the sensitivity strip and one can thus expect
a strong effect of light element presence on the $T_b - T_e$
relationship.  This problem was studied in detail by \citet{CPY97} and
\citet{PCY97}, who performed numerical calculations of envelope
structure with a pure iron plus catalyzed matter chemical composition
and with various amounts of light elements.  These authors found that
the presence of light elements can significantly raise the surface
temperature $T_s$ for a given $T_b$ if they are present in sufficient
amounts.  The larger the amount of light elements present, the higher
the temperature at which their effect will be felt due to the
temperature dependence of the location of the sensitivity strip.   But at
very high temperatures, the light elements have practically no effect
because they cannot penetrate deep enough.  The resulting $T_e^\infty - T_b$
relationships for various amounts of light elements are shown in
Figure~\ref{Fig:Tb_Te_accreted}.

The presence of a magnetic field can also affect the structure of the
envelope \citep{GH83}.  The effect is to enhance heat transport along
the field and inhibit transport along directions perpendicular to the
field.  This results in a nonuniform surface temperature distribution,
with a very cold region in which the field is almost tangential to the
surface as, e.g., around the magnetic equator for a dipolar field, and
a corresponding modification of the $T_e^\infty - T_b$ relationship
\citep{P95}.  However, the overall effect is not very large, but is
somewhat sensitive to the presence of strongly nondipolar surface fields
\citep{PS96}.  For a field of the order of $10^{11}$ - $10^{12}$ G,
one obtains a slight reduction of $T_e^\infty$ compared to the field-free
case, whereas for a higher field $T_e^\infty$ begins to be enhanced.  The
enhancement of $T_e^\infty$ is, however, much smaller than what is obtained
by the presence of light elements \citep{PYCG03}.  
Moreover, there are possible instabilities due to the non-uniformity
of the temperature \citep{U04} which have not yet been taken into account
in magnetized envelope calculations and may somewhat affect these
results, but we do not expect significant changes.
Hence, the important case for our purpose would be the maximal reduction of 
$T_e^\infty$ obtained for a pure heavy element envelope at $B = 10^{11}$ G, 
which is illustrated in Figure~\ref{Fig:Tb_Te_accreted}.

One must finally mention that our calculations  are based on the assumption
of spherical symmetry in the interior and that the only asymmetries 
considered, due to the presence of a magnetic field, are within the envelope
and hence included into this outer boundary condition.
However, this assumption is questionable in some  magnetic
field configurations where the field is confined to the stellar crust.
As shown by \cite{GKP04}, the crust is highly
non-isothermal in such cases and this can affect the thermal evolution
because the resulting photon luminosity is lowered compared to the
isothermal crust case.

\section{A GENERAL STUDY OF NEUTRON STAR COOLING
         WITHIN THE ``MINIMAL SCENARIO''}
\label{Sec:General}

In this section, we will consider the individual effects of the chief
physical ingredients which enter into the modeling of the cooling of
an isolated neutron star.  Our purpose here is twofold: \\

\noindent (1) to determine the sensitivity of results to uncertainties in
input physics in order to obtain a broad range of predictions which,
we hope, encompasses all possible variations within the minimal cooling
scenario; and, \\

\noindent (2) to provide us with the means to identify the types of
models that will result in the coldest possible neutron stars
within this paradigm.

Theoretical refutations of the critical physical ingredients needed
for these coldest models could allow us to raise the temperature
predictions and possibly provide more, or stronger, evidence for
``enhanced cooling''.  The task of identifying the minimally cooling
coldest star will be taken up in \$~\ref{Sec:Try-the best}. An object
colder than such a star could be considered as evidence for the
presence of physics beyond the minimal paradigm.

All results in this section use stars built using the APR EOS, except
for \S~\ref{Sec:EOS-effects} where the effects of the EOS are studied
for a star of 1.4 \Msol and for \S~\ref{Sec:Mass-effects} where 
effects of the stellar mass are studied.

\subsection{Neutrino vs. Photon Cooling Eras and the Effect of the Envelope
            \label{Sec:Nu-Phot-Env}}

The basic features of the thermal evolution of a neutron star can be easily
understood by considering the global thermal energy balance of the star
\be
\frac{d E_{\rm th}}{dt} \equiv C_V \frac{dT}{dt} = 
 -L_\nu - L_\gamma \,,
\label{Equ:cool_simple}
\ee
where $E_{\rm th}$ is the total thermal energy content of the star and
$C_V$ its total specific heat.  This equation is accurate when the
star is isothermal, which is the case for ages larger than a few
decades.  Since the dominant neutrino processes all have a $T^8$
temperature dependence, the neutrino luminosity can be expressed as 
\be
L_\nu = N T^8 \,.
\ee
Furthermore, most of the specific heat comes from the degenerate
fermions in the core for which 
\be
C_V = CT 
\ee
in the absence of pairing interactions.
The photon luminosity can be written as
\be
L_\gamma \equiv 4 \pi R^2 \sigma_{\scriptscriptstyle SB} T_e^4 = S T^{2+4\alpha}
\ee
where $T_e$, the effective temperature, is converted into the internal
temperature $T$ through an envelope model with a power-law dependence:
$T_e \propto T^{0.5 + \alpha}$ with $\alpha \ll 1$
(see equation~(\ref{Eq:Tb_TeGPE}) and Figure~\ref{Fig:Tb_Te_accreted} ).
Equation (\ref{Equ:cool_simple}) is easily integrated in the
dominantly neutrino and photon cooling eras. 

\noindent (1) The {\em neutrino cooling era} $(L_\nu \gg L_\gamma$):
In this case, 
\be
t = \frac{C}{6 N} \left( \frac{1}{T^6}-\frac{1}{T_0^6} \right)
\label{Eq:neutrino-cooling1}
\ee
where $T_0$ is the initial temperature at time $t_0 \equiv 0$.
For $T \ll T_0$, this gives
\be
T = \left(\frac{C}{6N}\right)^{\frac{1}{6}} t^{-\frac{1}{6}} 
\;\;\; {\rm and} \;\;\;\; 
T_e \appropto t^{-\frac{1}{12}} \,.
\label{Eq:neutrino-cooling2}
\ee
The very small exponent in the $T_e$ evolution during neutrino cooling is a 
direct consequence of the strong temperature dependence of $L_\nu$.

\noindent (2) The {\em photon cooling era} $(L_\gamma \gg L_\nu$): In
this case,
\be
t = t_1 + \frac{C}{4 \alpha \, S} \left( \frac{1}{T^{4\alpha}} - \frac{1}{T_1^{4\alpha}} \right)
\label{Eq:photon-cooling1}
\ee
where $T_1$ is the temperature at time $t_1$.
When $t \gg t_1$ and $T \ll T_1$, we have
\be
T = \left(\frac{C}{4 \alpha S}\right)^{\frac{1}{4\alpha}} t^{-\frac{1}{4\alpha}}
\;\;\;\; {\rm and} \;\;\;\; 
T_e \appropto t^{-\frac{1}{8\alpha}} \,.
\label{Eq:photon-cooling2}
\ee
Since $\alpha \ll 1$,  we see that, during the photon cooing era, 
the evolution is
very sensitive to the nature of the envelope, i.e., $\alpha$ and $S$, and to
changes in the specific heat, as induced by nucleon pairing.

\begin{figure}
\plotone{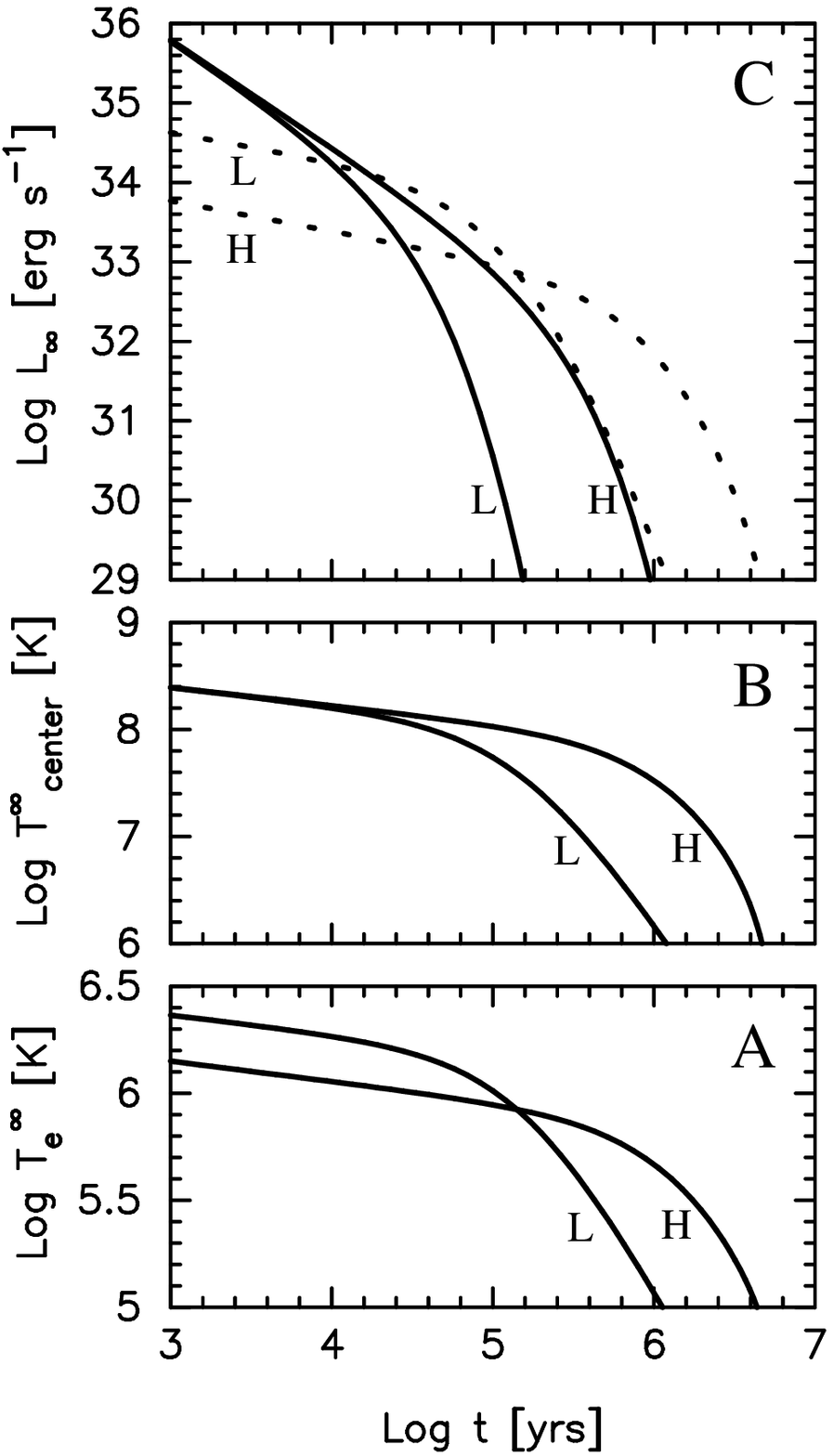}
\caption{Neutrino and photon cooling eras for two models of non
magnetized envelopes formed by heavy iron-like elements (labeled
``H'') and a maximum amount of light elements (labeled ``L'').  The
effective temperature (panel A), the central temperature (panel B) and
neutrino (continuous lines) and photon (dotted lines) luminosities
(panel C), all redshifted to infinity, are shown as a function of
time.  Pairing effects are not included in these
calculations. \label{Fig:Gen_Cooling_1}}
\end{figure}

\begin{figure}
\plotone{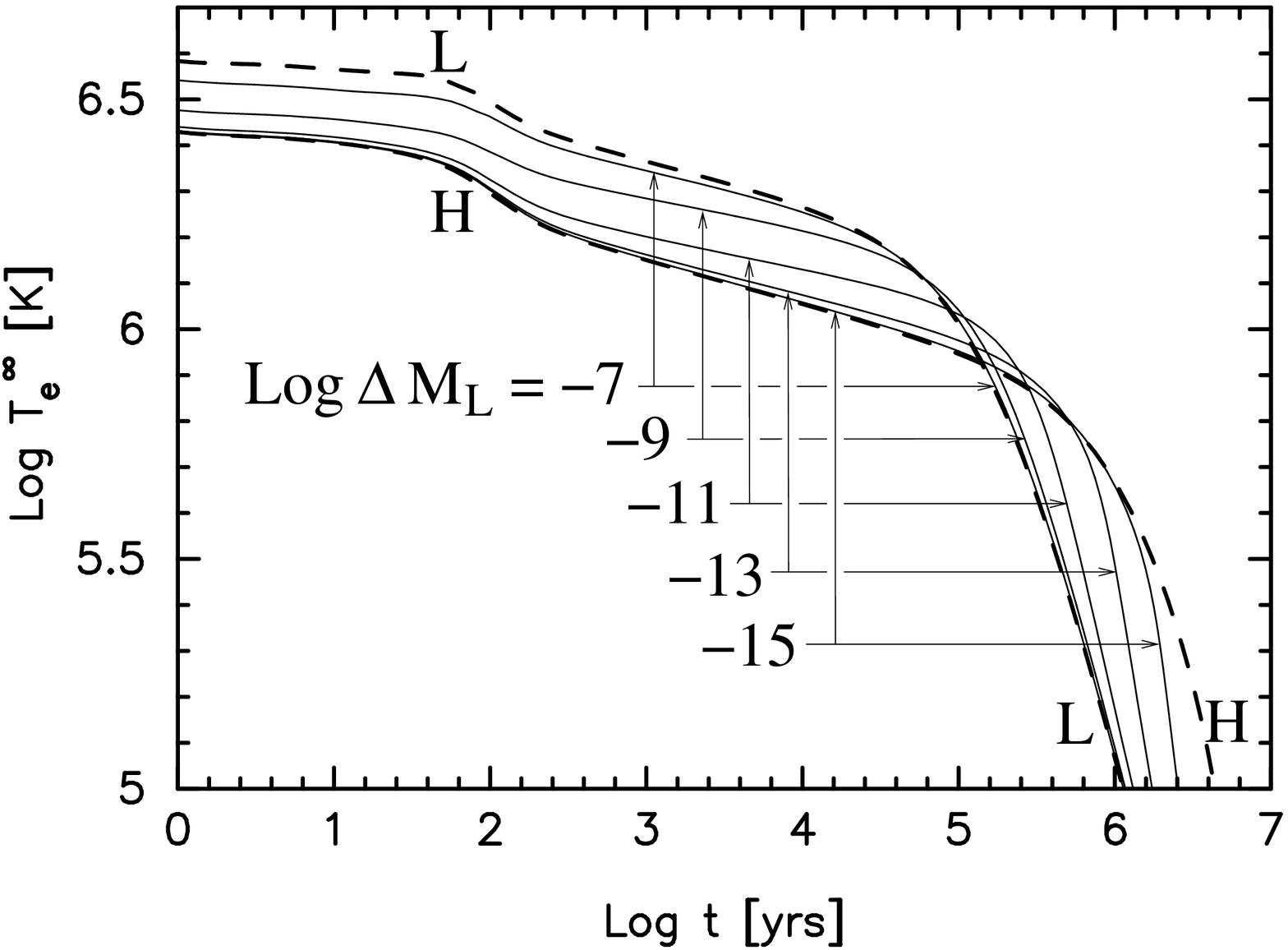}
\caption{Effect on the cooling of various amounts $\Delta M_{\rm L}$
(in solar masses) of light elements in the envelope.  The two
dashed curves, ``H'' and ``L'', are the same as in
Figure~\protect\ref{Fig:Gen_Cooling_1}.  Pairing effects are not
included in these calculations.
\label{Fig:Cooling-eta}}
\end{figure}

Figure~\ref{Fig:Gen_Cooling_1} shows the evolution of $T_e^\infty$,
$T_{\rm center}^\infty$, $L_\gamma^\infty$, and $L_\nu^\infty$ in a
simplified model in which no pairing has been included, but two
extreme cases of envelope chemical composition, iron-like elements and
light elements, are considered.  The $L_\gamma^\infty$ curves of panel
C are analogous to the $T_e^\infty$ curves of panel A, since
$L_\gamma^\infty = 4 \pi R_\infty^2 \sigma_{\scriptscriptstyle SB}
T_e^{\infty4}$.  For both envelope models the $T_e^\infty$ vs. $t$ and
$T_{\rm center}^\infty$ vs $t$ curves follow power laws (i.e.,
straight lines on a $\log - \log$ plot) in both the neutrino cooling
and photon cooling eras.  For $t \simless 10^4$ yrs, both
models have the same $T_{\rm center}^\infty$ because the
envelopes do not contribute to energy loss, neither through neutrino
emission (due to their low density and very small mass) nor through
photon emission (since $L_\gamma^\infty \ll L_\nu^\infty$).  At these
times, the model with a light element envelope, however, has a higher
$T_e^\infty$, and thus $L_\gamma^\infty$, due to the more efficient
transport of heat in this envelope and will consequently shift from
neutrino to photon cooling at a much earlier time.  This trend will
not be modified by the inclusion of more realistic physics.  
During the neutrino cooling era, $T_e^\infty$ simply follows the
evolution of the interior temperature and models with light element
envelopes appear hotter to an external observer than models with an
iron-like envelope, but they enter the photon cooling era sooner and
subsequently cool much faster.  Neutron stars with lesser amounts of
light elements in the envelope will cool intermediately between the
extremes of light-element and heavy-element dominated atmospheres, as
displayed in Figure~\ref{Fig:Cooling-eta}.

\subsection{Time Evolving Envelopes
            \label{Sec:Decay-Env}}

\begin{figure}
\plotone{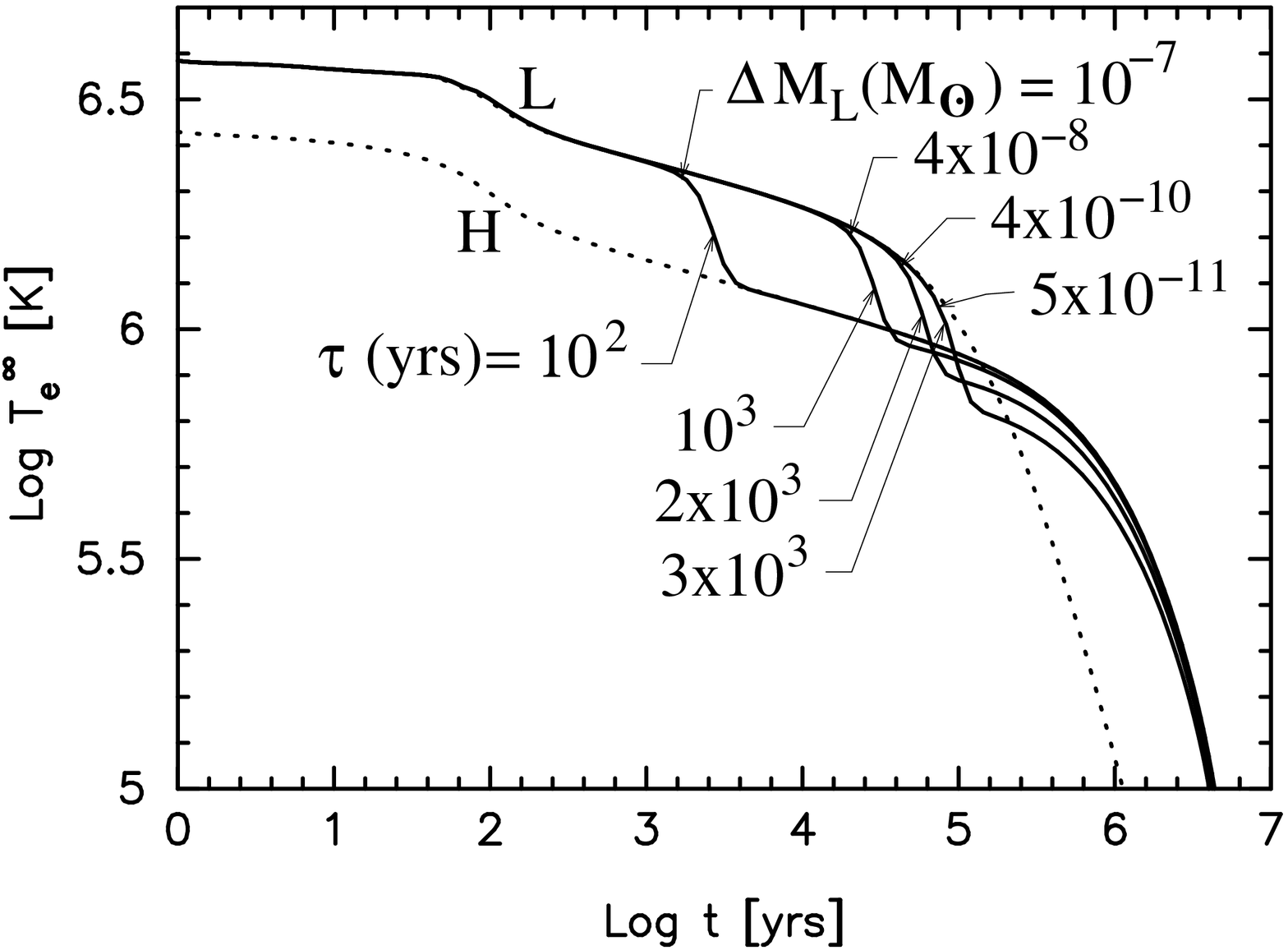}
\caption{Transition of cooling trajectories between a model with a
heavy element envelope (doted curve labeled ``H'') and a light element
envelope of maximum mass (dotted curve labeled ``L'').  Continuous
curves show evolution of models with ``decaying'' envelopes (see
equation~(\protect\ref{Eq:Decay-Env})) with various decay times $\tau$
as indicated.  Also indicated are masses of the light element
envelopes at the moment the star begins its shift toward the heavy
element envelope trajectory.  \label{Fig:Decay-Env}}
\end{figure}

We consider here the possibility of time evolution of the
chemical composition of the envelope.
We assume that the mass of the envelope consisting of light elements 
decays with time as
\be
\Delta M_{\rm L}(t) = e^{-t/\tau} \Delta M_{\rm L}(0)
\label{Eq:Decay-Env}
\ee
where $\Delta M_{\rm L}(0)$ is the initial mass in light elements.
This decay could be due to the pulsar mechanism which injects
light elements into the magnetosphere or due to nuclear
reactions which convert these elements into heavy ones
\citep{CB03a,CB03b}.
One can expect from this that the star will shift from a cooling
trajectory corresponding to a light element envelope toward a
trajectory with heavy elements envelope as $\Delta M_{\rm L}$ decreases.
Figure~\ref{Fig:Decay-Env} illustrates this evolution and shows that this shift
happens in a short time in the case of an exponential mass reduction.
This fast evolution is in agreement with the $T_e - T_b$ relationships
shown in Figure~\ref{Fig:Tb_Te_accreted} where one sees $T_e$ changing from
the light element case to the heavy element one within a small range of
variation of $\Delta M_{\rm L}$, at a value of $T_b$ depending strongly 
on $\Delta M_{\rm L}$.

\subsection{Basic Effects of Pairing 
            \label{Sec:Nu-Phot-Pairing}}

\begin{figure}
\plotone{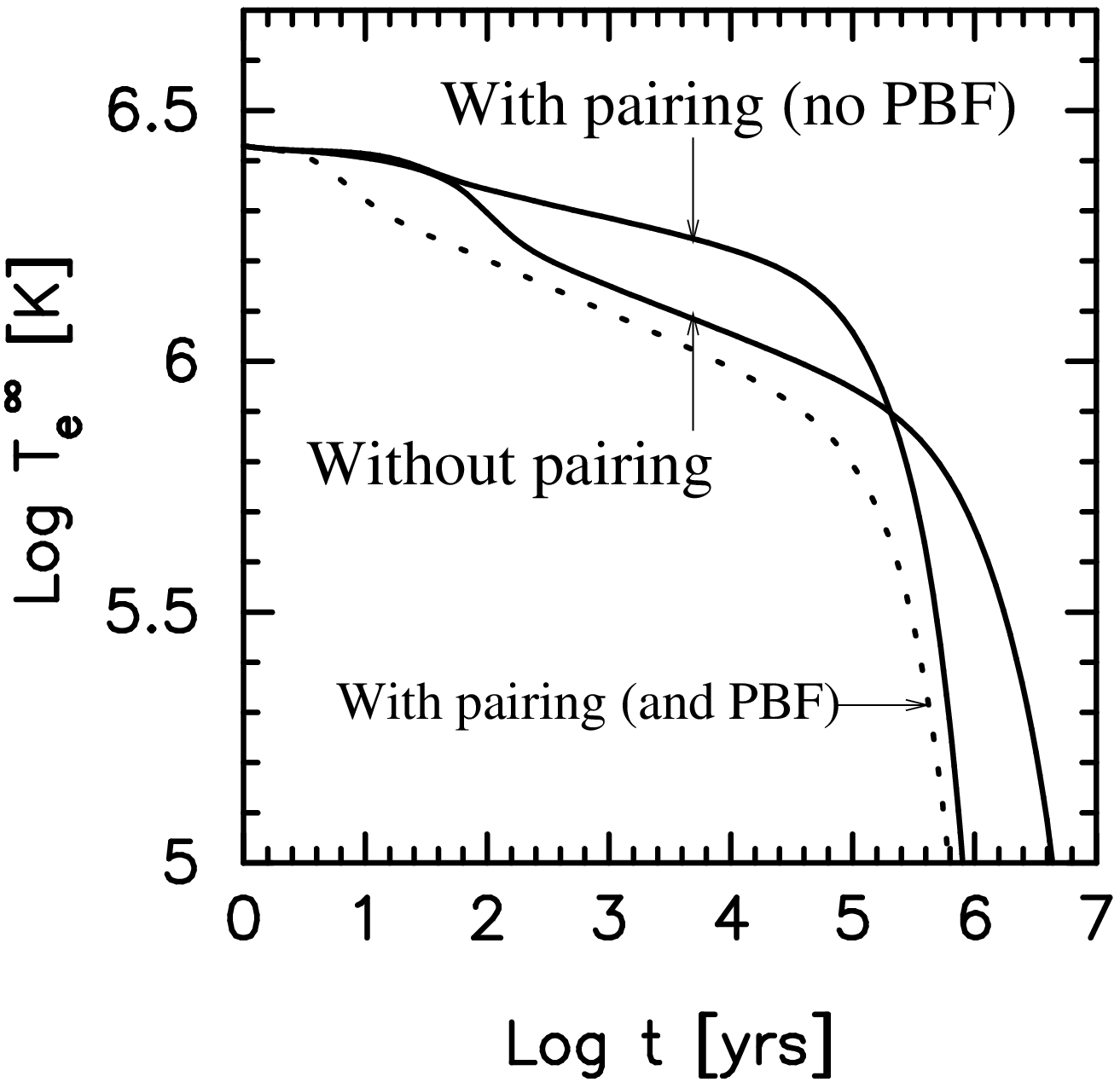}
\caption{Comparison of the cooling of a 1.4 \Msun star, built using
the EOS of APR, without and with nucleon pairing.  In the model with
pairing neutrino emission by the PBF process has been either
artificially turned off (continuous line) or allowed (dotted line).
Neutron $^1S_0$ pairing is from AWP, $^3P_2$ pairing from our case
``c'' and $p$ $^1S_0$ pairing from AO, as labeled in
Figures~\ref{Fig:n1S0}, \ref{Fig:n3P2}, and \ref{Fig:p1S0},
respectively.  The envelope is assumed to be composed of heavy
elements.  \label{Fig:Trivial}}
\end{figure}

In this subsection, we briefly illustrate the first two significant
effects of pairing, suppressions of $q_\nu$ and $c_v$.
The third effect, neutrino emission through the PBF process is
studied in the next section.
The continuous lines in Figure~\ref{Fig:Trivial} compare
thermal evolutions of the same neutron
star with and without nucleon pairing (the 
gaps have been chosen so as to maximize effects of suppression).
The results are very natural:
during the neutrino cooling era the paired star cools more slowly than
the unpaired one since its neutrino luminosity $L_\nu$ is severely suppressed
whereas during the photon cooling era it cools faster due to its
much reduced specific heat.
During the neutrino cooling era, the suppression of $c_v$ is 
present, but its effect is not so dramatic for three reasons: \\

\noindent (1) the lepton contribution to $C_V$ is not suppressed, whereas
$L_\nu$ is reduced by many orders of magnitude as only the very
inefficient electron-ion bremsstrahlung process in the crust is not
suppressed, \\

\noindent (2) when $T$ is not very much less than $T_c$, as is
partially the case in this example during the neutrino cooling era,
the suppression of $c_v$ is preceded by a phase of enhancement (see,
e.g., Figure~\ref{Fig:Cv-nnn}), and \\ 

\begin{figure}
\plotone{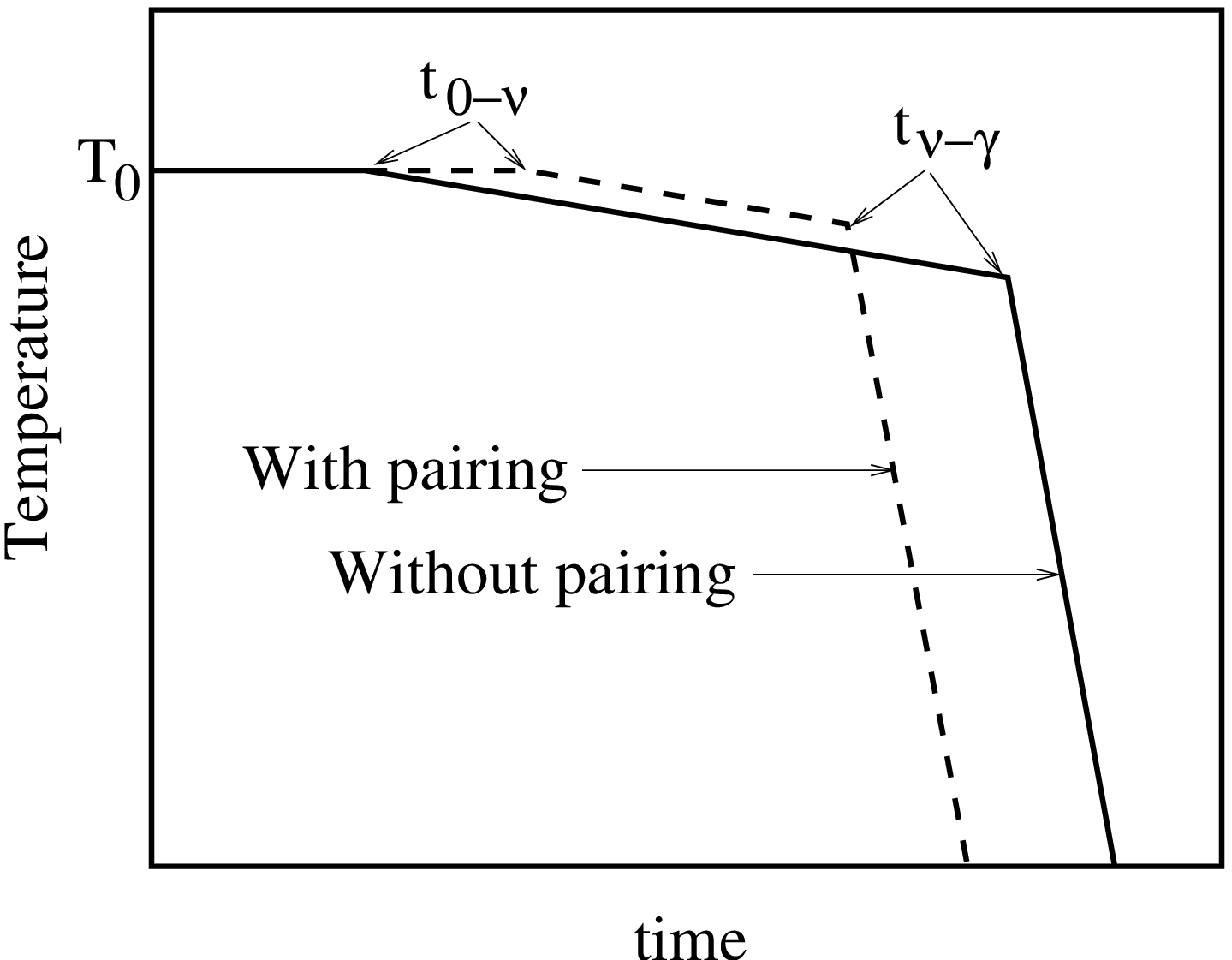}
\caption{Schematic representation of the power-law cooling
behaviors and the effect of pairing.  The
time $t_{0-\nu}$ denotes when the central temperature falls
to a value small enough that $T\propto t^{-1/6}$ becomes valid.
The time $T_{\nu-\gamma}$ denotes the transition from neutrino
to photon cooling eras.
 \label{Fig:Power-Laws}}
\end{figure}

\noindent (3) the cooling curve has a relatively small slope when $T\sim T_0$.
From equation~(\ref{Eq:neutrino-cooling1}), and as
represented schematically in Figure~\ref{Fig:Power-Laws}, one sees
that the shift to the $T \propto
t^{-1/6}$ power law occurs at a time $t_{0-\nu}$ determined by the
ratio $C/6N$.   This ratio is increased by pairing and results in a delayed
shift, but this amounts to an horizontal translation of the cooling
curve and hence shows no spectacular effect.  
In contradistinction, during the photon cooling era, the shift in the
transition time $t_{\nu - \gamma}$ from neutrino to photon cooling (this occurs
earlier with pairing than without pairing due to the smaller value of
$C/4 \alpha S$ in equation~(\ref{Eq:neutrino-cooling2})) acts on a
power law evolution with large slope. \\

However, the neutrino emission by the PBF process (artificially turned
off in this example) alters significantly these simple results as can
be seen from the dotted line in Figure~\ref{Fig:Trivial}, and is
described below.

\subsection{The PBF Neutrino Process
            \label{Sec:Coop}}


We consider here in detail the effect of the PBF neutrino process.
Given the strong $T^7 F(T/\Delta)$ temperature dependence and the
density dependence of $\Delta$, the overall effect
can only be assessed by complete calculations presented here and in
\S~\ref{Sec:Minimal-Data}. 

\begin{figure}
\plotone{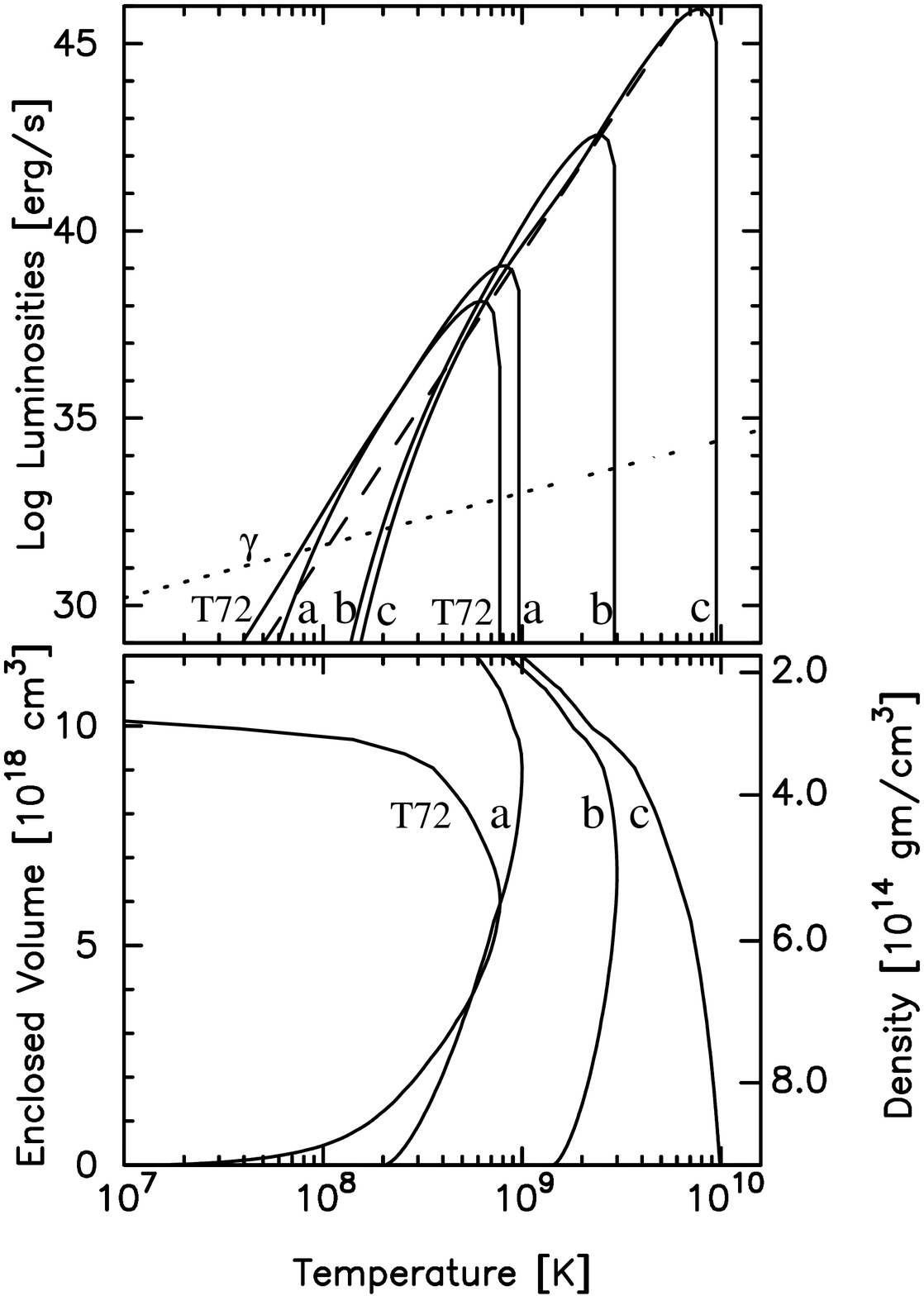}
\caption{Upper panel: neutrino luminosities vs temperature from the
PBF process for four different $n$ $^3P_2$ gaps
labeled as in Figure~\protect\ref{Fig:n3P2}.
Lower panel: $T_c$ for the four neutron
$^3P_2$ gaps vs density and enclosed volume.  
\label{Fig:Coop_n3P2}}
\end{figure}

As a first step, we consider separately the temperature dependence of
the luminosities
due to the $n$ $^3P_2$ and  $p$ $^1S_0$ gaps in the
core of a 1.4 \Msol star built with the EOS of APR.  Results for four
different $n$ $^3P_2$ gaps are shown in Figure~\ref{Fig:Coop_n3P2}.
The lower panel shows the $T_c$ profiles of these four gaps as a
function of the volume of the core (left hand scale) and the density
(right hand scale). A vertical line in this panel, which corresponds
to an isothermal core, allows one to visualize the amount of the
core's volume which is paired. The upper panel shows the corresponding
PBF neutrino luminosity $L_\nu^{PBF}$.  Also plotted are
the surface photon luminosity corresponding to an
iron-like envelope (dotted line) and the total neutrino luminosity
from the modified Urca and bremsstrahlung processes {\em without}
pairing suppression.  Notice that when $T \simgreater 10^9$ K, the
star is usually not isothermal: the crust is warmer than the core and
thus $L_\gamma$ is larger than indicated in this figure.

When $T$ decreases, $L_\nu^{PBF}$ grows very sharply for each gap
when $T$ reaches the maximum $T_c$ of the gap reached in the core (PBF
neutrino emission turns on) and then decreases with a $T$ dependence
which is between a $T^8$ and a $T^7$ power law. This results from the
overall $T^7 F(T/\Delta)$ temperature dependence of the PBF neutrino
emissivity combined with the density dependence of $T_c$ which
determines how much of the core, at this given $T$, is contributing to
$L_\nu^{PBF}$.  Once $T$ is much below the minimum value of $T_c$
reached in the core, $L_\nu^{\rm PBF}$ becomes exponentially
suppressed.  In the cases of gaps ``b'' and ``c'', this suppression
appears at $T \ll 10^9$ K, for gap ``a'' when $T \ll 2\times 10^8$ K,
whereas for the gaps ``T72'' this suppression does not appear since
$T_c$ reaches very low values and for any $T$ there is always a
significant volume of the core where $T \sim T_c$.

When $T \simless 10^8$ K, $L_\gamma$ dominates over $L_\nu$ so that
the important range of $T$ to consider is $10^8 - 10^9$ K and in this
range the figure shows clearly that the relatively small $n~^3P_2$
gaps as ``T73'' and ``a'' generate a $L_\nu^{PBF}$ which is higher, by
about one order of magnitude, than the combined $L_\nu$ that the
modified Urca and bremsstrahlung processes would produce when no
pairing is present (dashed line in the figure).  The modified Urca and
bremsstrahlung processes are of course strongly suppressed in the
presence of pairing, but these results show that, compared to the no
pairing case, pairing can actually {\em increase} the total neutrino
luminosity through the PBF neutrino emission if the gaps have the
appropriate size.

\begin{figure}
\plotone{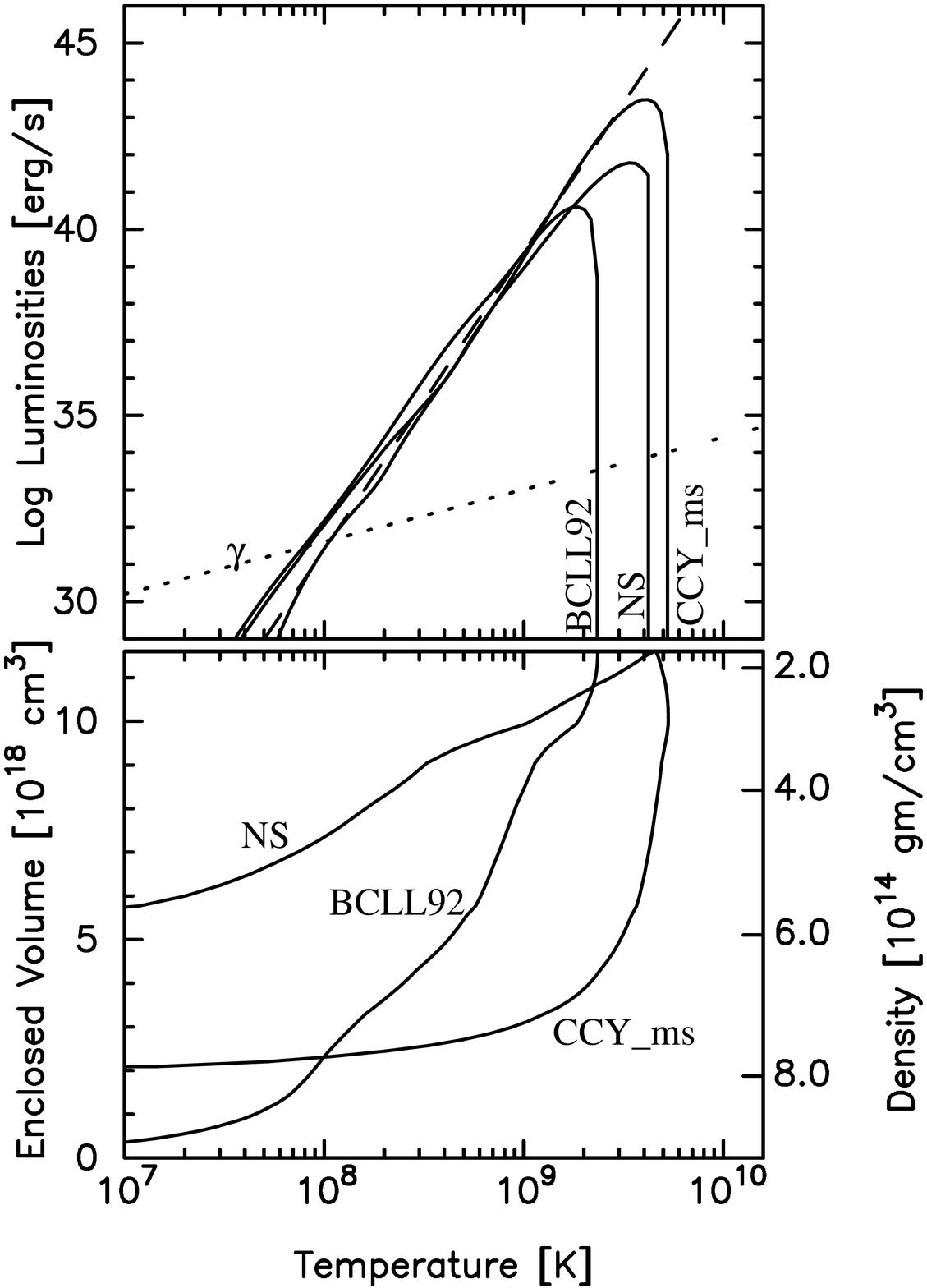}
\caption{Upper panel: neutrino luminosities vs temperature from
the PBF process for 3 different $p$ $^1S_0$ gaps labeled
as in Figure~\protect\ref{Fig:n1S0}.  
Lower panel: critical temperature $T_c$ for the 3 $p$
$^1S_0$ gaps vs density and enclosed volume.  
\label{Fig:Coop_p1S0}}
\end{figure}

Very similar results are obtained when considering the PBF neutrino
emission due to the $p$ $^1S_0$ gaps as shown in
Figure~\ref{Fig:Coop_p1S0}.  The three differences with respect to the
$n~^3P_2$ case are, first, that no calculation of the $p$ $^1S_0$ gap
reaches a $T_c$ as high as our case ``c'' for the $n$ $^3$P$_2$, second,
all $p~^1S_0$ gaps vanish in the inner part of the core which
implies that the suppression of $L_\nu^{PBF}$ at low T does not show
up and, third, the volume of the core in the superconducting state is
generally smaller than the volume in the $n$ superfluid state, resulting
in lower values of $L_\nu^{PBF}$ for $p$ than for $n$.

\subsection{Comparison of Various Neutrino Luminosities}

\begin{figure}
\plotone{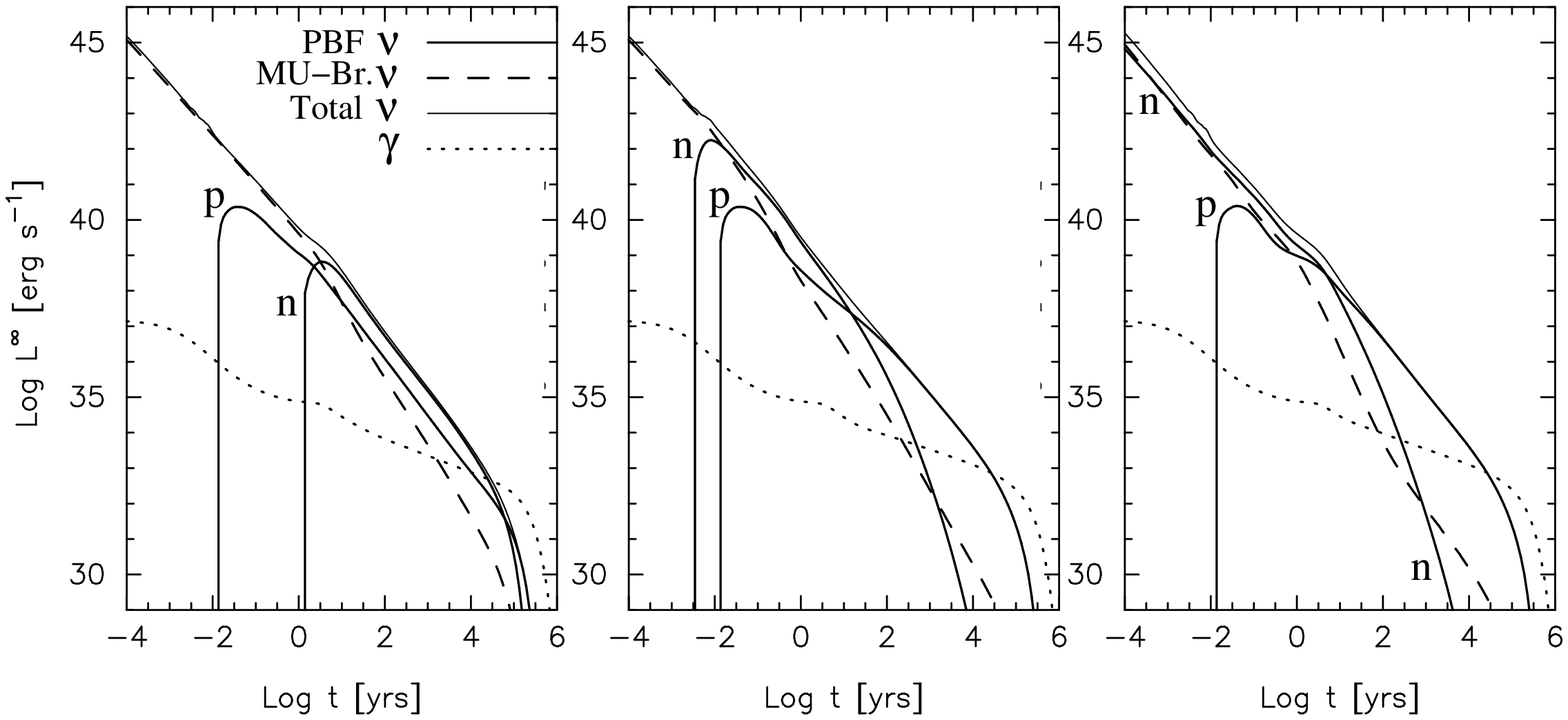}
\caption{Comparison of luminosities from various processes during
three realistic cooling histories: photon (``$\gamma$''), all
$\nu$-processes (``Total $\nu$''), modified Urca and nucleon
bremsstrahlung (``MU-Br. $\nu$''), and PBF
(``PBF $\nu$'') from $n$ $^3P_2$ and $p$ $^1S_0$ pairing marked
by ``$n$'' and ``$p$'', respectively.  PBF neutrinos from the $n$
$^1S_0$ gap are not shown explicitly, since their contribution is
always dominated by other processes, but they are
included in the total $\nu$ luminosity.  In all three cases, the $p$
$^1S_0$ gap is from AO, the $n$ $^1S_0$ gap from AWP, whereas the $n$
$^3P_2$ gap is our model ``a'' (left panel), ``b'' (central panel)
and ``c'' (right panel).  \label{Fig:Cool_Lum}}
\end{figure}

Having compared, in the previous subsection, the neutrino luminosities
from PBF assuming an isothermal interior, we now proceed to analyze
them, together with other processes, in realistic cooling calculations
which take into account the exact temperature profile within the star.
The results are shown in Figure~\ref{Fig:Cool_Lum}.  We use a 1.4
\Msol star built with the EOS of APR and we fix the $n$ and $p$ $^1S_0$ 
gaps considering that the major uncertainty in the
neutrino luminosity is due to uncertainty in the size of the $n$
$^3P_2$ gap (see the results of \S \ref{Sec:Coop}).  We
consider the three cases ``a'', ``b'', and ``c'' for the magnitude of
the $n~^3P_2$ gap.

The three panels of Figure~\ref{Fig:Cool_Lum} show clearly that at very
early times the cooling is driven by the modified Urca and nucleon
bremsstrahlung processes, but that once pairing occurs the neutrino
emission from the PBF process takes over because of its
efficiency and because the other processes are suppressed.

At ages relevant for the presently available data, $t \simgreater
10^2$ yrs, we find in all cases that the PBF neutrinos are the main
cooling agent, until photon emission takes over at $t \sim 10^5$
years.  There is a competition between the neutrino emission from $n$
$^3P_2$ and $p$ $^1S_0$ pairing: the smallest gap drives 
the cooling between $10^2$ and $10^5$ years.  With the assumed
$p$ $^1S_0$ gap from AO, we see that in case ``a'' the $n$ $^3P_2$
gap drives the cooling (Figure~\ref{Fig:Cool_Lum} left panel), whereas
in case ``b'' or ``c'' (Figure~\ref{Fig:Cool_Lum} central and right
panels, respectively) the proton pairing drives the cooling.

\begin{figure}
\plotone{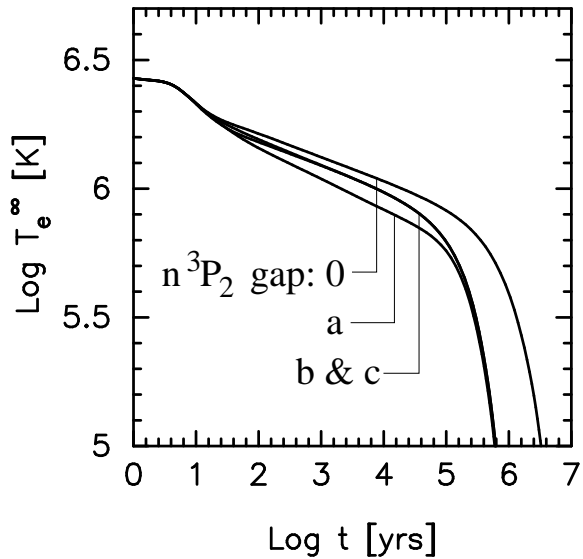}
\caption{
Comparison of cooling trajectories with vanishing $n$ $^3P_2$
gaps, labeled ``0'', and our three model gaps ``a'', ``b'', and ``c''
(see Figure~\protect\ref{Fig:n3P2}).
The $n$ $^1S_0$ gap is from AWPIII and the $p$ $^1S_0$ gap from AO
(see Figures~\protect\ref{Fig:n1S0} and
\protect\ref{Fig:p1S0}). Results are for a
1.4 \Msun star built using the EOS of APR with a heavy element envelope.
\label{Fig:cool-three-3P2}}
\end{figure}

This is in agreement with the results of Figure~\ref{Fig:Coop_n3P2}
which show that the case ``a'' $n$ $^3P_2$ gap is the most efficient and
that even smaller gaps, as in T72, do not result in a significant
enhancement of the PBF neutrino luminosity or do it too late, i.e., at a time
when photon emission dominates the cooling. Comparison of
Figure~\ref{Fig:Coop_n3P2} with Figure~\ref{Fig:Coop_p1S0} shows that in
the case of a larger $n$ $^3P_2$ gap, one can expect that the $p$
$^1S_0$ PBF neutrinos will dominate the cooling, in
agreement with our findings of this subsection.
Finally we compare cooling trajectories with our three $n$ $^3P_2$ gaps
and a vanishing gap in Figure~\ref{Fig:cool-three-3P2} explicitly confirming
that $n$ $^3P_2$ gaps with $T_c$ of the order of $10^9$ K are the most
efficient gaps with regard to neutrino cooling through the PBF process.

\onecolumn
\twocolumn

\subsection{More-Modified Urca Cooling }
           \label{Sec:IncreasedMurca}

The modified Urca and bremsstrahlung processes all involve 
four nucleons and are processes in which energy-momentum transfer
occurs via strong interactions in the medium.  
The associated emissivities are sensitive to one's assumptions about
in-medium strong interactions and their efficiencies are difficult to
assess with certainty.  
Given this, we consider it important to study the
effect of this uncertainty in a simple, but drastic way: 
we simply multiply the $q_\nu^{\rm MUrca}$ and $q_\nu^{\rm Brem}$
emissivities by a constant factor $F$, taking $F$ to be
$1/10$, $10$ or $100$.  
A factor $1/10$ or $10$ could be acceptable, whereas a factor $100$ is
probably exaggerated.

\begin{figure}
\epsscale{0.8}
\plotone{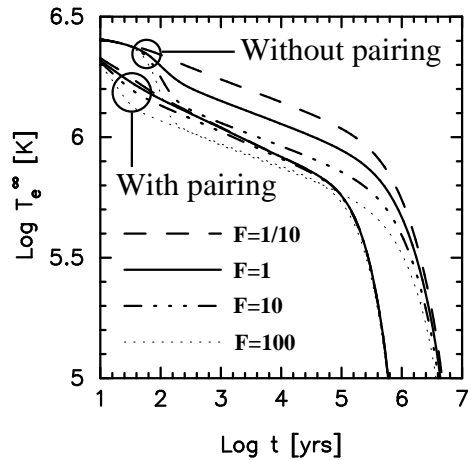}
%
\caption{Cooling with adjusted modified Urca, for $F=1/10$, $1$, $10$ and $100$
as indicated, with and without nucleon pairing.
Assumed pairing gaps are from our model ``a'' for $n$ $^3P_2$ and 
from AO for $p$ $^1S_0$  (and $n$ $^1S_0$ pairing
from AWP for which  the effect is very small).  
The envelope is assumed to be composed of heavy elements.
\label{Fig:Increased-MURCA}}
\end{figure}

The results of Figures~\ref{Fig:Coop_n3P2} and \ref{Fig:Coop_p1S0} showed
that in the presence of $n$ $^3P_2$ and $p$ $^1S_0$ pairing, most
reasonable gaps produce a neutrino emission by the PBF process
which is much more intense than the modified Urca and
bremsstrahlung in the absence of pairing by at least one order of magnitude
in the important temperature range $T \sim 10^8 - 10^9$ K.  Thus, in a
realistic calculation a factor $F=10$ is not expected to lead to a
significant change in the cooling.  
This is confirmed by our results shown in Figure~\ref{Fig:Increased-MURCA}. 
The models with no pairing clearly show enhanced cooling when $F=10$ or $100$,
and reduce cooling when $F = 1/10$, whereas when
pairing is included the models with $F=10$ and $1/10$ are practically
indistinguishable from the unenhanced case and only the, probably
unrealistic, case $F=100$ leads to a faster cooling.

These results are important and fortunate, since they show that the
uncertainty in the actual efficiency of the modified Urca rate has no
significant effect on the predictions of the minimal scenario when
pairing, and the corresponding neutrino emission from the PBF
process, is included in a realistic way.

\subsection{Effects of Neutron Star Mass
            \label{Sec:Mass-effects}}

In the case that neutrino cooling occurs only through the modified
Urca and bremsstrahlung processes, as required by the tenets of the
minimal cooling scenario, one can expect that the cooling curves in 
the neutrino cooling era will show practically no variation with neutron 
star mass, because there are no energy or density thresholds for these
processes. This situation will change drastically for the case in
which enhanced cooling through direct Urca processes becomes possible
either through nucleons or due to the presence of exotica.

\begin{figure}
\plotone{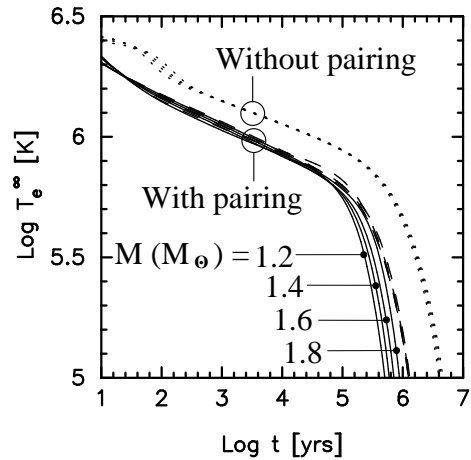}
\caption{Effects of the stellar mass: cooling of stars of various
masses built using the EOS of APR, with and without pairing.  Models
with pairing have $n$~$^1S_0$ gap from AWP and $n$~$^3P_2$ gaps from
our model ``a' and either no $n$~$^3P_2$ gap (dashed curves) or
$p$~$^1S_0$ gap from AO (continuous curves).  Stellar masses are
indicated in the cases with the three types of pairing, whereas for 
similar cases without proton pairing or with no pairing at all the
trajectories are too similar to be separately labeled.  The envelope is
assumed to be composed of heavy elements.  \label{Fig:Stellar-Mass}}
\end{figure}

Figure~\ref{Fig:Stellar-Mass} confirms that in the absence of pairing, there 
is almost no mass effect, both during the neutrino and the photon cooling era.
Similarly, when $n$, but not $p$, pairing is included, the mass dependence
is also small, though larger than with no pairing at all.
When the $p$ gap is included, the main variation with mass occurs in the photon
cooling era in which more massive stars cool more slowly. This is a
direct consequence of the lesser suppression of the proton specific
heat with increasing mass, since the $p$ $^1S_0$ gap vanishes at
high density and there is an increasingly larger unpaired region when
$M$ increases ($C_V(p)$ is larger for larger $M$).  
The chosen $n$ $^3P_2$ gap reaches the center of the star in all cases 
and thus $C_V(n)$ is strongly reduced for all masses, which explains the small
mass dependence when only $n$ gaps are taken into account.
In the case that the $n$ $^3P_2$ gap would also vanish at high density, 
we would obtain an additional mass dependence.

\subsection{Effects of the Equation of State
            \label{Sec:EOS-effects}}

In exploring the high density EOS, one can expect three sources of effects: \\

\noindent (1) general relativistic effects due to change in the star's
compactness, \\
\noindent (2) differences in the $n$ and $p$ effective masses, and \\
\noindent (3) differences in the volume of the star in the various
paired states. \\

\begin{figure}
\plotone{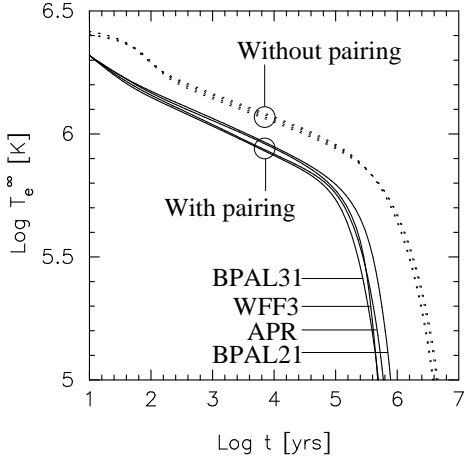}
\caption{Effects of the EOS: cooling of 1.4 \protect\Msun stars built
using the four chosen EOS's.  EOS's are labeled in the cases with
pairing, whereas for similar cases without pairing the trajectories are
too similar to be separately labeled.  Pairing gaps: $n$~$^1S_0$ from
AWP, $n$~$^3P_2$ from our model ``a'' and $p$~$^1S_0$ from AO.  The
envelope is assumed to be composed of heavy elements.
\label{Fig:Cool_EOS}}
\end{figure}

Figure~\ref{Fig:Cool_EOS} shows results for the four EOS's 
selected in \S~\ref{Sec:EOS}.  When no pairing is included, there is
little variation with the EOS, and slight variations exist when
pairing is considered.  The reasons are essentially the same as those
discussed in conjunction with the stellar mass (see the previous
subsection) and are due to the density dependence of the $p$
$^1S_0$ gap and, to a much lesser degree, that of the $n$ $^3P_2$
gap.  The very small differences in the unpaired models simply reflect
that the four chosen EOS's are rather similar because of constraints
imposed by the minimal cooling scenario: the differences in the
stars' compactness and nucleon effective masses are very small.

\section{MINIMALLY COOLING COLDEST NEUTRON STARS
            \label{Sec:Try-the best}}

One of the main goals in this work is to determine how cold an
observed neutron star should become to be incompatible with the
predictions of the minimal scenario.  Armed with the results of the
previous section, we can now identify the fastest cooling models within
this scenario.

\subsection{Neutrino cooling era}

During the neutrino cooling era, Figure~\ref{Fig:cool-three-3P2} shows
that the lowest $T_e$'s are obtained due to the PBF process when the
$n$ $^3P_2$ gaps are of the size of model ``a'', i.e., with $T_c$'s of
order at most $10^9$ K in most of the stellar core.  The $p$ $^1S_0$
gaps cannot compete with the most efficient $n$ $^3P_2$ gaps, because
proton gaps are restricted to a smaller volume; compare
Figures~\ref{Fig:Coop_n3P2}, \ref{Fig:Coop_p1S0}, and
\ref{Fig:Cool_Lum}.  These fastest neutrino cooling models have a very
weak dependence on the mass of the star
(Figure~\ref{Fig:Stellar-Mass}).  These models also require that the
envelope be made of heavy elements or, if it contains light elements,
their amounts should be much smaller than $10^{-11}$ \Msun (see
Figure~\ref{Fig:Cooling-eta}).

\subsection{Photon cooling era}

The physical processes that control cooling in the photon cooling era
are quite different from those in the neutrino cooling era.  Neutrino
emission from any of the possible processes make only a small
contribution in the photon cooling era.  

\begin{figure}
\plotone{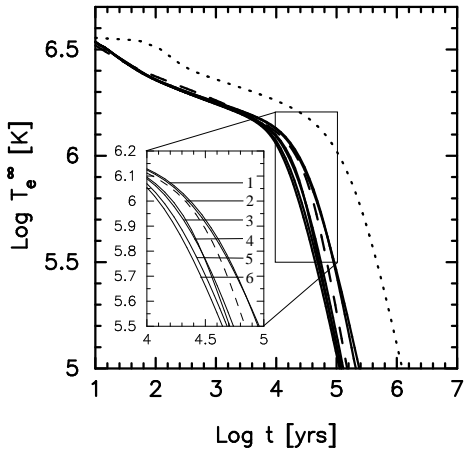}
\caption{Cooling of a 1.1 \Msun star (continuous lines) and with 
	various $p$~$^1S_0$ gaps (as labeled in the inset):
         1 - NS,
         2 - T,
         3 - AO,
         4 - BCLL,
         5 - CCY\_ms, and
         6 - CCDK
         (see Figure~\protect\ref{Fig:p1S0} for notation).
         The $n$~$^1S_0$ gap is from AWP, and the $n$~$^3P_2$ gap from 
	our model ``a''.
         The dotted line is the same 1.1 \Msun star,
	 but without any pairing  and
         the dashed line is for a 1.4 \Msun star, 
	with the same pairing gaps as in 
         Figure~\protect\ref{Fig:Stellar-Mass}.
         Envelopes are all assumed to be composed of light elements.
         \label{Fig:Mass-low-L}}
\end{figure}

The two crucial ingredients are the envelope, which determines the
photon luminosity and the specific heat, which controls pairing (see
\$~\ref{Sec:Nu-Phot-Env}).  A light element envelope, producing a
higher $T_e$ and hence a higher $L$ for a given core temperature,
leads to fast cooling; an amount above $10^{-9}$ \Msun of these
elements is necessary (see Figure~\ref{Fig:Cooling-eta}).  Concerning
the total specific heat, the strongest reduction can be achieved by
pushing baryon pairing to the extreme, and this means considering low
mass neutron stars so that the $p$ $^1S_0$ gap is more likely to reach
the center of the star.  Pursuing the trend indicated in
Figure~\ref{Fig:Stellar-Mass}, we consider effects of the various $p$
$^1S_0$ gaps of Figure~\ref{Fig:p1S0} for a low mass, 1.1 \Msun,
neutron star.  Results are shown in Figure~\ref{Fig:Mass-low-L}.  The
proton $k_F$ at the center of this star has a value of $1.1$ fm$^{-1}$
(see Figures~\ref{Fig:M-R} and \ref{Fig:kf-n}); the inset of
Figure~\ref{Fig:Mass-low-L} shows a direct mapping of the density at
which the $p$~$^1S_0$ gap vanishes with $T_e$ at these times.  The
fastest cooling model corresponds to the $p$~$^1S_0$ gap ``CCDK,''
which has a $T_c$ of $1.44 \times 10^9$ K at the center of the star
and hence produces a complete suppression of the proton specific heat
in the photon cooling era.  A $p$~$^1S_0$ gap with a higher $T_c$ at
the center of the star, or a gap that vanishes at higher densities
(not reached in this star), would lead to the same cooling curve.
Similar considerations apply to the $n$~$^3P_2$ gap. To illustrate
this, we used our model gap ``a'' in Figure~\ref{Fig:Mass-low-L}. 
Any other gap with a $T_c$ higher than a few times $10^8$ K would
result in the same total suppression of the neutron specific heat and,
therefore, to exactly the same cooling curve.

These results will be important when comparing our predictions with
data in the next section, particularly for young stars with ages of the
order of a few times $10^4$ years, such as the Vela pulsar and PSR 1706-44.

\section{COMPARISON OF THE MINIMAL COOLING SCENARIO WITH DATA}
\label{Sec:Minimal-Data}

In \S~\ref{Sec:General}, we analyzed in some detail the effect of each
physical ingredient that shapes the cooling history of a neutron star
within the minimal scenario.  In \S~\ref{Sec:Try-the best}, we
identified the fastest cooling neutron star models in this scenario.
The combined effects of these ingredients in realistic models,
together with comparisons to the presently
available temperature and luminosity measurements, are presented
below.

Our task is greatly simplified by the fact that the EOS is
considerably constrained by the tenets of the minimal scenario (see
results of \S~\ref{Sec:EOS-effects}).  Moreover, as shown in
\S~\ref{Sec:Mass-effects}, the precise mass of the neutron star also
has little effect, with the possible exception of low mass stars at
ages around a few times $10^4$ years.  (This, of course, is changed
drastically once we go beyond the minimal scenario and allow for
enhanced neutrino emission processes to occur at high density.)  We
can therefore restrict our attention mostly to the thermal evolution
of a ``canonical'' 1.4 \Msun neutron star built with the EOS of APR.
In contrast, the chemical composition of the envelope and the extent
of pairing of both neutrons and protons will play significant roles.

As shown in \S~\ref{Sec:Nu-Phot-Env}, the presence of light elements
in the envelopes of young neutron stars leads to effective
temperatures that are larger than those without any light elements
during the neutrino cooling era, whereas it implies a faster cooling
during the later photon cooling era. Thus, for an assumed high density
structure of the star, there exists a whole family of models limited
by the two extreme cases of envelopes: those with only heavy elements,
and those with a maximum amount of light elements.  Stars with
envelopes containing only a small amount of light elements will evolve
on intermediate tracks shifting from a track close to the former one
toward the latter one as illustrated in Figure~\ref{Fig:Cooling-eta}.
Conversely, stars can evolve in the opposite direction if the envelope 
composition changes with time from light elements
toward heavy elements; such an evolution is very abrupt as shown in
Figure~\ref{Fig:Decay-Env}.

On the other hand, the occurrence of pairing also accelerates 
cooling during the photon cooling era through the reduction of the
specific heat (see \S~\ref{Sec:Nu-Phot-Pairing}), whereas
pairing effects during the neutrino cooling era are more delicate.
Neutrino emission from the modified Urca and bremsstrahlung processes
is suppressed, but the breaking and formation of Cooper pairs can
easily, with appropriate gaps, become vastly more efficient than the
former processes.  As a result, depending on its size and
density dependence, pairing can lead to faster or slower cooling during the
neutrino cooling era; its effect has to be considered carefully.

Our main results are presented in Figure~\ref{Fig:plot_cool_lum_data}
and compared with data.  For the reasons discussed in
\S~\ref{sec:Data-DL}, we present them in two forms: effective
temperature $T_e^\infty$ versus time and luminosity $L^\infty$ versus
time.  We divide our results into three subclasses depending on the
size of the $n$ $^3P_2$ gap, given that this gap is the most uncertain
one: a vanishingly small gap and our gap models ``a'' and ``b''.
Figures~\ref{Fig:Cool_Lum} showed that case ``c'' results in neutrino
emissions very similar to that of model ``b'' and
Figure~\ref{Fig:cool-three-3P2} confirmed that the resulting cooling
trajectories are practically identical for these two large $n$ $^3P_2$
gaps and we therefore do not need to include results for the gap ``c''
here.  For each assumed neutron $^3P_2$ gap, it is still necessary to
consider uncertainties in the $n$ and $p$ $^1S_0$ gaps. Varying these
gaps is less dramatic than varying the $n~^3P_2$ gap and we consider
15 different combinations (see the caption of
Figure~\ref{Fig:plot_cool_lum_data}) which we plot together.  We
obtain, for each assumed $n~^3P_2$ gap, two sets of very closely
packed curves, one for each envelope composition.  The size and extent
of the $n~^1S_0$ gap has very little effect, since it is mostly
restricted to the crust and encompasses only a small part of the star's
volume. This leads to small differences in the early cooling when the
star has not yet reached isothermality, at ages $\sim 3 - 100$ years,
and the surface temperature is still controlled by the evolution of
the crust.  Among the $p~^1S_0$ gaps, the ones which can reach higher
densities will lead to slightly faster cooling both during the
neutrino cooling era, because of the enhanced neutrino emission from
the PBF process, and during the photon cooling era, due to the
resulting smaller specific heat.

Figure~\ref{Fig:cool-three-3P2} demonstrated that models with the
$n~^3P_2$ gap ``a'' yield the coldest stars and
Figure~\ref{Fig:plot_cool_lum_data} shows that the spread in results
due to the variation of the other two $^1S_0$ gaps is much smaller
that in the other two scenarios.  This is because the
$n~^3P_2$ gap ``a'' maximizes neutrino emission by the PBF process (see
Figure~\ref{Fig:Coop_n3P2}) and because the neutrino luminosity due to
the proton PBF process is lower (compare Figure~\ref{Fig:Coop_p1S0}
with Figure~\ref{Fig:Coop_n3P2}).  We have studied many other models
with a $n~^3P_2$ gap similar to our case ``a'', but with slightly
different maximum values of $T_c$ and different density dependences,
and discerned negligible differences.  We are thus confident that the
results presented here reflect the smallest temperatures possible
within the constraints of the minimal scenario.  We can obtain a
slightly faster cooling in the photon cooling era for low mass neutron
stars (near 1 to 1.2 M$_\odot$) as discussed in \S~\ref{Sec:Try-the
best}, and this case will be presented separately at the end of this
section.

We now compare observational data for specific neutron
stars with the suite of models encompassing the minimal cooling scenario.

\subsection{RX J0822-4247 and 1E 1207-52}

These two are young and are the hottest known stars.
Their inferred temperatures are higher than the predictions of all our
models with heavy element envelopes, but are compatible with all
models with light element envelopes. This may be considered as
possible evidence for the presence of a significant amount of light
elements in the upper layers of these stars.   However,
when considering luminosities, for select values of the $n~^3P_2$
gap, 1E 1207-52 is compatible with having an heavy elements envelope.
RX J0822-4247, however, remains more luminous than any of the heavy-element
envelope models but is compatible with light-element envelope models.

\subsection{PSR 0656+14, PSR 1055-52, Geminga, RX~J1856.5-3754, and 
RX J0720.4-3125}

These are the five oldest observed stars.  Fits of their spectrum to
light-element atmospheres result in radii too large to be compatible
with the neutron star hypothesis.  Blackbody spectral fits result in
too-small radii, but it is possible for heavy-element atmospheres 
or two-temperature black bodies
to
be constructed that produce compatible radii.  For consistency, we
have restricted the data appearing in these plots to be a result of
either light-element atmosphere fits or single-temperature 
blackbody fits.  Inferred
temperatures are more sensitive to the assumed  atmospheric
composition than are luminosities, and for these five objects, the $L$
versus age plots are probably more reliable and representative of the
observational uncertainties.  Consequently, blackbody fits result
in relative positions for temperatures that are quite
different than those of the luminosities.

Except for RX~J1856.5-3754, the large uncertainties on the age and the
luminosity of these objects preclude definite conclusions.  If we
consider the upper limits to their agea and/or luminosities, we find
them too bright and must invoke the presence of some strong heating
process.  On the other hand, considering the lower limits to ages
and/or luminosities they appear compatible with the minimal
scenario independently of assumptions about pairing.

\subsection{Vela, and PSR 1706-44}

Very intriguing objects are the pulsars PSR 0833-45 (``Vela'') and PSR
1706-44.  Vela has been repeatedly proposed as a candidate for
enhanced cooling or exotic matter, but our results are inconclusive
with respect to these claims.  For this star, the effective area is
compatible with emission from almost the entire surface of a neutron
star and both types of plots, $T$ or $L$ vs age, are equivalent.  With
an $n$ $^3P_2$ gap chosen to maximize neutrino emission from the PBF
process, as in our case ``a'', the discrepancy of Vela with the
theoretical prediction is not significant whereas for a vanishing $n$
$^3P_2$ gap it is very large.  However, for any non-vanishing $n$
$^3P_2$ gap and a low assumed stellar mass (see
Figure~\ref{Fig:plot_cool_data_11}), several of the light element
envelope models reach the temperature of Vela at an age of 20,000 yrs,
i.e., well within its age uncertainty and even less than some
estimates of the supernova remnant age, $1.8 \pm 0.9 \times 10^4$ yrs
\citep{AET95}.  These models correspond to $p$ $^1S_0$ gaps which
extend to relatively high densities and hence result in strong
suppression of the proton specific heat in most of, if not all, the
core (see \$~\ref{Sec:Try-the best}).  An interesting feature of these
models is the very fast decrease of temperature
\be
\frac{\Delta T_e}{T_e} \sim -0.85 \frac{\Delta t}{t} \,,
\ee
which for an age of 20,000 yrs gives a decrease of the observable
x-ray flux of 0.17\% every decade. 
In this case one could interpret these results as indicating that the
Vela pulsar is a low mass neutron star with a thick light
element envelope and in which neutrons and protons are paired in the
entire core.  
On the other hand, they could favor a neutron star whose core
neutrons have a $T_c$ of the order of $10^9$ K, without
any constraint about the proton pairing and the stellar mass, but this
star must have 
a heavy element envelope with at most $10^{-13}$ \Msun of light
elements at the surface (see Figure~\ref{Fig:Cooling-eta}).

PSR 1706-44 is in a similar situation, but with larger uncertainties
both in $T$ and in $L_\infty$, due to the large distance
and age uncertainties.
Confirmation of its association with the supernova remnant G 343.1-2.3
\citep{MOP1993} would help constrain both its distance and age.

\subsection{Barely detected and undetected objects}

Sources that have negligible or no observed thermal emissions, listed
in Table~\ref{utable} and plotted in Figure~\ref{Fig:UL}, are compared
with our results in Figure~\ref{Fig:plot_lum_cold-data-theroy}.  The
upper luminosity limits for the two objects CXO J232327.8+584842 (in
Cas A) and PSR J0154+61 are well within the prediction of the minimal
scenario whereas the limit of PSR J1124-5916 is on the lower side, but
still compatible.

Most interesting are the two stars PSR J0205+6449 (in 3C58) and RX
J0007.0+7302 (in CTA 1) whose upper limits are clearly below our
predictions.  The remaining four objects, with no point-like emissions
of any kind observed to date, would provide definitive evidence for
enhanced cooling if it could be shown that neutron stars in fact exist
in any of them.

\onecolumn
\begin{figure}
\epsscale{1.0}
\plotone{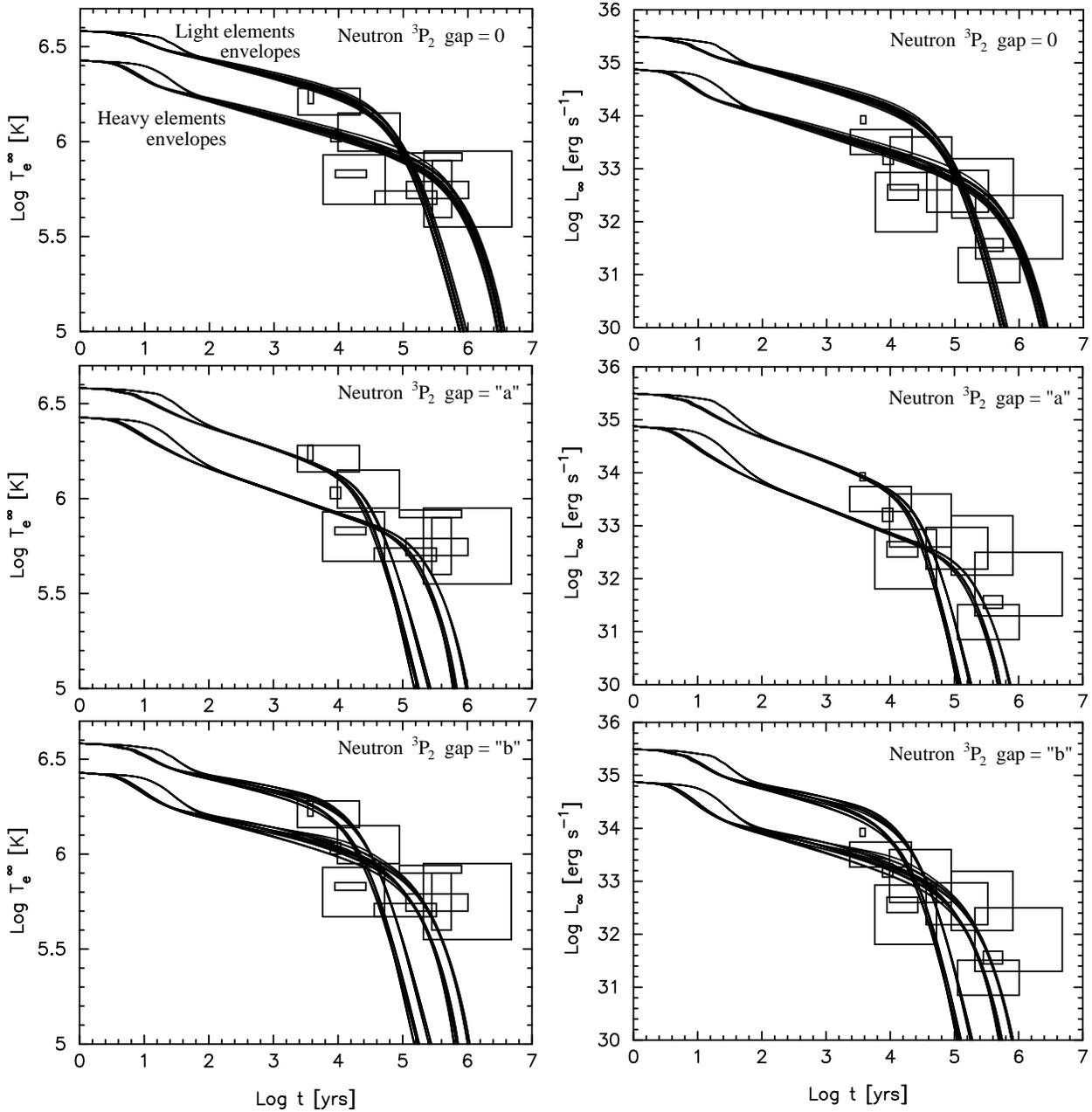}
\caption{Comparison of predictions of the minimal cooling scenario with data,
all models being 1.4 \Msun stars built using the EOS of APR.  Left
panels: effective temperature $T_e^\infty$ vs. age.  Right panels:
luminosity $L_\infty$ vs age.  The upper, middle, and lower panels
correspond to three different assumptions about the size of the 
$n$~$^3P_2$ gap as indicated in the panels.  In each panel, the two sets
of curves correspond to the two extreme models of envelope chemical
composition: light elements or heavy elements, as labeled in the
upper left panel.  For each set of curves, the 15 different curves
correspond to three different choices of the $n$ $^1S_0$ gap
(``AWPII'', ``AWPIII'', and ``SCLBL'' as labeled in
Figure~\protect\ref{Fig:n1S0}) and five different $p$ $^1S_0$
gaps (``CCYms'', ``T'', ``NS'', ``AO'', and ``BCLL'' as labeled in
Figure~\protect\ref{Fig:p1S0}).
         \label{Fig:plot_cool_lum_data}
}
\end{figure}
\twocolumn

\onecolumn
\begin{figure}
\plotone{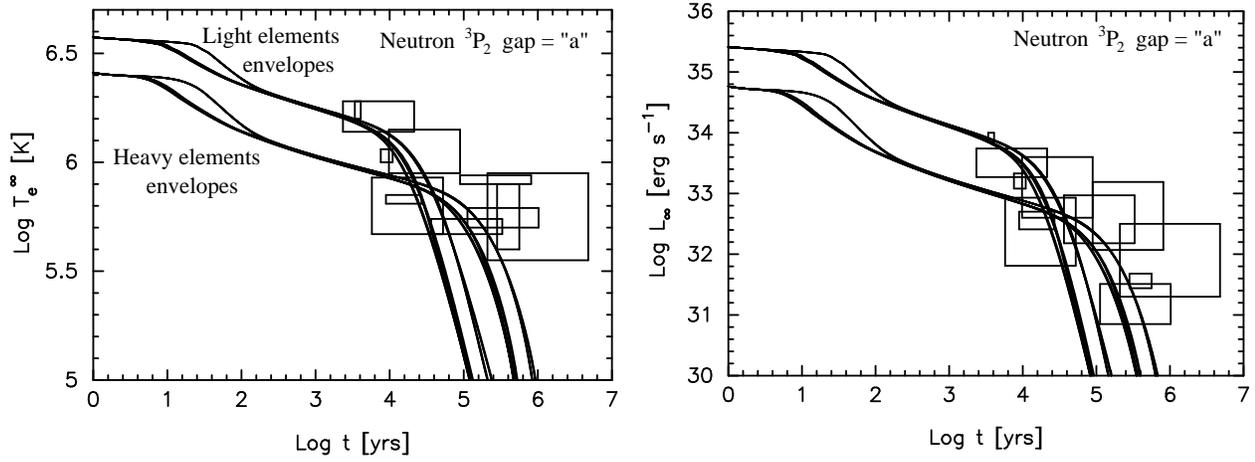}
\caption{Same as the central panels of 
	Figure~\protect\ref{Fig:plot_cool_lum_data}, 
         but for a 1.1 \Msun star built using the EOS of APR. 
         \label{Fig:plot_cool_data_11}
}
\end{figure}
\twocolumn

\begin{figure}
\plotone{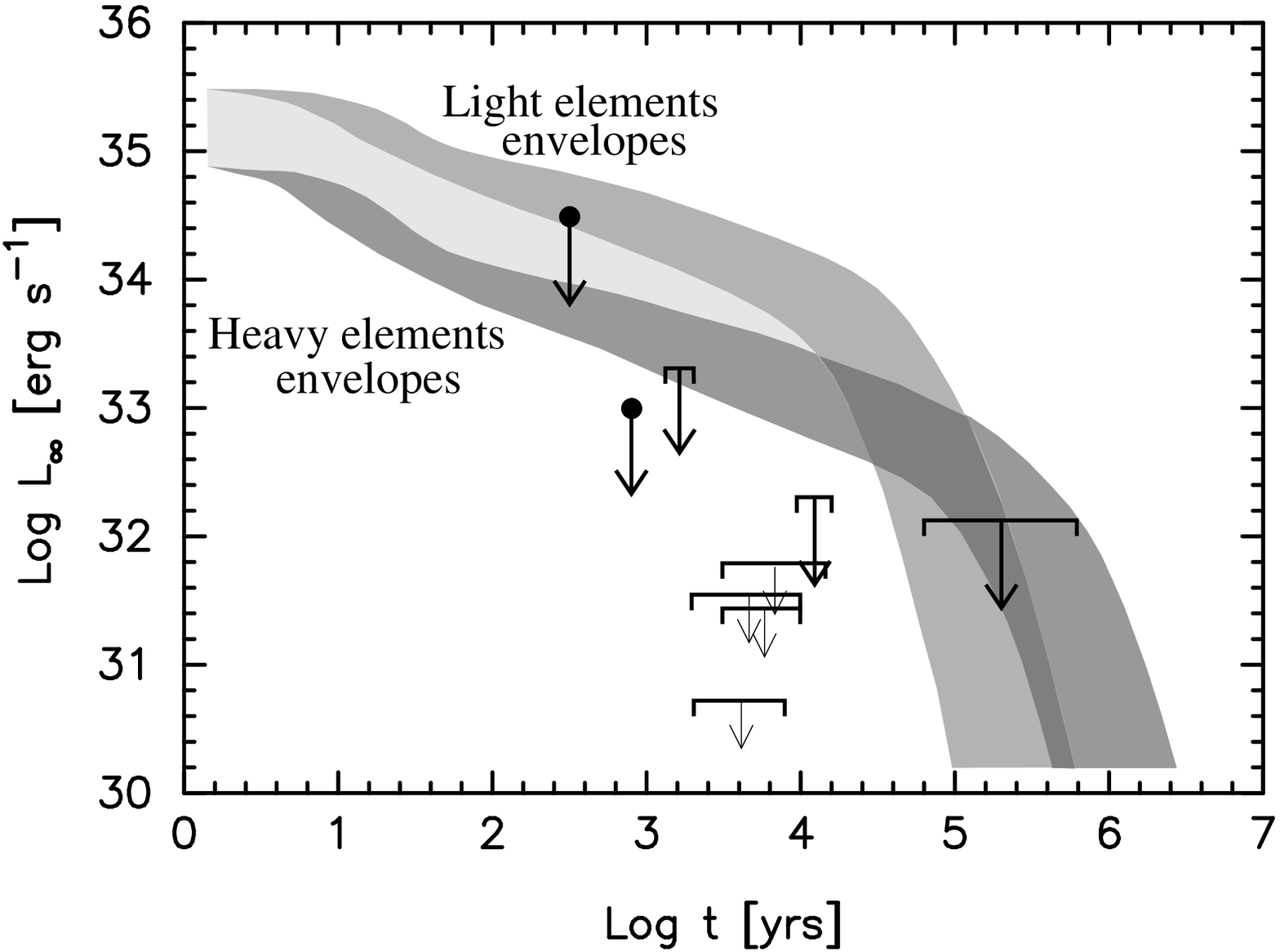}
\caption{Global comparison of the upper luminosity limits for sources
  lacking apparent thermal emissions with cooling trajectories,
  satisfying the minimal cooling scenario. The area with the 
  darkest gray shading
  contains models with heavy-element dominated envelopes, whereas that
  with the
  intermediate shading contains models with ligh-element envelopes. 
The region with the lightest shading contains models with 
intermediate compositions.
         \label{Fig:plot_lum_cold-data-theroy}
}
\end{figure}

\section{COMPARISON WITH OTHER WORKS}
         \label{Sec:Other}

The literature on neutron star cooling is extensive and dates
back to the early 60's.  Detailed studies of the ``standard'' cooling
of isolated neutron stars have been presented, e.g., by \cite{NT87},
\cite{vR88}, and \cite{SWWG96}, but none of these works had included
the neutrino emission from the PBF process.  Models incorporating the PBF
process were first presented by \cite{SVSWW97} and subsequently by
\cite{P98}.  These two works, however, did not explore effects of
various EOS's and/or various sets of pairing gaps within the framework
of the ``standard'' scenario.
Our present results are in agreement with the results of \cite{NT87},
\cite{vR88} and \cite{P98} when similar input physics are used and
allow a close comparison.

We find, however, large differences with the results of \cite{SWWG96}
and \cite{SVSWW97} during the photon cooling era in which 
the models of these authors cool much more slowly than our
models. This discrepancy is probably attributable to differences in
the specific heat.  For example, during the photon cooling era, the
Schaab et al.  models
in which pairing is considered cool in almost the same
way as models in which the effects of pairing are neglected.  The
models without pairing have luminosities about one order of magnitude
higher than our corresponding models at ages in the range $10^6 -
10^7$ years.

Extensive calculations of the effect of the PBF process have been
presented by \cite{LSY99} in a simplified model in which both the
$p$ $^1S_0$ and $n$ $^3P_2$ gaps were assumed to be constant in
the entire core of the star.  This simplification of uniform $T_c$
(which implies that protons are paired in the entire core of their
stars, whereas in our models the protons in the central part of the star
are unpaired for 1.4 \Msun stars) has the effect of 
overestimating the neutrino emission by the PBF process.  
Their results are consistent
with our findings in that they obtain somewhat cooler stars. For
example, they can reach $\log T_e^\infty = 5.8$ at age $t = 10^4$
years, whereas our coolest model with $n$ $^3P_2$ gap of our model ``a'' has
$\log T_e^\infty = 5.9$ at the same age.

Subsequent works of \cite{YKHG02} and \cite{YH03}, which explored more
realistic, density-dependent gaps, obtained results which are in 
good agreement with ours.  For models in which no enhanced neutrino
emission is at work, these two works obtain a minimal $\log
T_e^\infty$ of 5.9, as we do, at an age of $t =10^4$ years.

Several recent works, \cite{KHY01}, \cite{YKG01}, \cite{KYG02},
and \cite{TTTTT02}, studied the cooling of neutron stars within the
``standard'' scenario (including the PBF processes), but with enhanced
neutrino emission at high density.  In addition, they did not explore
the range of parameters that are considered here.  Consequently, these
studies cannot be compared to the minimal scenario presented here.

In a recent review, \cite{YP04} have examined standard and enhanced
cooling in an attempt to fit all the data within one model.  In this
work,  the
data on individual objects are fit by treating the mass of the star as
a free parameter.  Models in which the $n$~$^3P_2$ gap is taken to be
vanishingly small \citep{SF04} are favored, together with a
$p$~$^1S_0$ gap that persists up to very high densities and with a
$T_c$ of the order of $7\times 10^9$ K.  This latter feature 
is incompatible with all calculations of this gap we report in
Figure~\ref{Fig:p1S0}, but is not absolutely 
excluded.  Our models for 1.1 \Msun
with nucleon pairing in the entire core (see
Figure~\ref{Fig:plot_cool_data_11}) are similar to those of Yakovlev
\& Pethick with non-vanishing $n$~$^3P_2$ gaps, whereas
our models of heavier stars with extensive neutron pairing but a large
volume of unpaired protons are similar to their models with vanishing
$n$~$^3P_2$ gaps and extensive proton pairing (see
Figures~\ref{Fig:plot_cool_lum_data}).

\section{DISCUSSION AND CONCLUSIONS}
        \label{Sec:Conclusion}

We have presented a detailed study of the thermal evolution of an
isolated neutron star using what we term as the minimal cooling
scenario.  This scenario is an extension of the well-known ``standard
cooling'' scenario to include the effects of nucleon pairing and
complements neutrino emission by the modified Urca and nucleon
bremsstrahlung processes with the pair breaking and formation (PBF)
process.  We have confirmed the results of previous works by others 
that for many models of nucleon pairing, the PBF process
actually dominates the cooling of the star (see \S~\ref{Sec:Coop}) and
hence is an essential ingredient of the minimal cooling scenario.

Among the four parameters we proposed for an
overall classification of neutron star cooling models, we found that
the EOS at supranuclear densities is well constrained
by the requirements of the minimal cooling scenario. Moreover, we showed that
the stellar mass has little effect on the results.  We emphasize that
for scenarios beyond the minimal one, i.e., when new particles and
neutrino emission processes appear, these two parameters definitely
gain importance.  The other two parameters we considered, pairing
properties of the nucleons and chemical composition of the envelope, 
introduce the largest uncertainties in our theoretical predictions.

We singled out three subclasses of scenarios due to uncertainties in
the size and extent of the $n$ $^3P_2$ gap.  For this gap, we
considered three different cases: a vanishingly small gap, a
somewhat small gap (our model ``a'') and a relatively large gap (model
``b'').  Within each of these cases, variations of the $n$ and
$p$ $^1S_0$ gaps covering the published ranges of these gaps were also
considered.

With respect to the chemical composition of the envelope, we singled
out two extreme cases: an envelope consisting of heavy iron-like
elements and an envelope containing essentially only light elements.
For each choice of the $n$ $^3P_2$ gap, we obtained two families
of closely packed cooling curves representing each of the two extreme
envelope cases, with some spreads due to variations of the $n$ and
$p~^1S_0$ gaps.  Intermediate envelope chemical compositions, or its
possible temporal evolution, result in cooling trajectories intermediate
between the extremities (see Figures~\ref{Fig:Cooling-eta} and
\ref{Fig:Decay-Env}).

Comparing our results with observationally inferred temperatures
$T_\infty$ and luminosities $L_\infty$ of eleven isolated neutron stars
(Figures~\ref{Fig:plot_cool_lum_data} and \ref{Fig:plot_cool_data_11}), 
we found that the observations were in overall good agreement with the
minimal cooling scenario taking into account the uncertainties of
envelope and pairing properties as well as those of the ages and
inferred $T_\infty$ and $L_\infty$ for these stars.  It is probably
not possible to understand these data within a single model with a
unique envelope chemical composition.  Considering that the
compositions of the upper
layers are strongly dependent on poorly understood processes which
occurred during and soon after the birth of the star, and  possibly
later due to the dynamics, it appears likely that different
stars have envelopes with different chemical compositions.

The Vela pulsar 0833-45, and possibly, but with much larger
uncertainties, PSR 1706-44, are marginal candidates for
enhanced cooling as their inferred temperatures and luminosities are
lower than most of our models.  Nevertheless, we found that low mass
neutron stars, $\sim 1.1 - 1.2$ \Msun, with a light element envelope
and extensive nucleon pairing covering essentially the entire core, could
reach the inferred values, $T_\infty \simeq 10^{5.8}$ K at an age of $\sim
20,000$ years (Figure~\ref{Fig:plot_cool_data_11}) which is within the range
of estimated age of the associated supernova remnant.

An essential component of the minimal cooling scenario is neutrino
emission by the PBF process, which leads, in the presence of
appropriate nucleon gaps, to more rapid cooling than possible in the
standard cooling scenario.  The low observed temperatures of the two
pulsars, Vela and 1706-44, can be accomodated by the PBF process with
a $n$ $^3P_2$ gap of sufficient size.  However, if this gap were
vanishingly small \citep{SF04}, then the temperature and thermal
luminosity measurements of these two objects would be evidence for the
presence of processes beyond the minimal cooling paradigm.

The five older stars, PSR 0656+14, PSR 1055-52, PSR 0633+1748
(Geminga), and RX J0720.4-3125, unfortunately have such large
uncertainties on both their ages and thermal luminosities that their
interpretation is delicate.  They do not require the occurence of
enhanced neutrino emission and can be accomodated within the minimal
cooling scenario when the lowest values for $T$ and $L_\infty$ are
chosen.  But, should the upper limits for $T$ and $L_\infty$ prevail,
they would be good candidates for the occurrence of some ``heating
mechanism'', i.e., dissipative processes, which inject heat into the
star (see, e.g., \cite{USNT93} \& \cite{SSWW99}).  In the case of RX
J1856.5-3754, which has a much more tightly constrained age and
luminosity, the agreement with the minimal scenario is excellent.

The two objects standing apart from the other observed neutron stars
are J0205+6449 (in the supernova remnant 3C58) and RX J0007.0+7302 (in
CTA 1). Upper limits on their luminosities are well below the
predicted values for any of our models.  These two objects are the
best candidates, to date, for the necessity to go beyond the minimal
scenario.

Finally, the four upper luminosity limits on the undetected objects in
shell supernova remnants (G084.2-0.8, G093.3+6.9, G127.1+0.5, and
G315.4-2.3) recently found by \cite{Ketal2004} are so low that they
will constitute the strongest evidence for enhanced neutrino emission
well beyond the minimal scenario if it can be demonstrated that they
actually correspond to neutron stars and not quiescent black-holes.

\acknowledgements
The authors wish to acknowledge many discussions over the years with
D. G. Yakovlev and O. Y. Gnedin, including extensive exchanges of numerical 
results to compare our two, independently developed, cooling codes.
This work was supported by a binational (NSF-Conacyt) US-Mexico grant.
DP acknowledges extra support from grants UNAM-DGAPA (\#IN112502) and 
Conacyt (\#36632-E). JL, MP, and AS acknowledge research support from  
the U.S. DOE grants DE-AC02-87ER40317 and DE-FG02-88ER-40388, and the
NSF grant INT-9802680. 

\appendix

\section{Observational Data}
\label{App:data}

\subsection{Supernova Remants G084.2-0.8, G093.3+6.9, G127.1+0.5, \& G315.4-2.3}

These four SNR's are considered to be the product of core-collapse
supernovae and are hence expected to contain either a neutron star or a 
black-hole.
Nevertheless, the searches of \cite{Ketal2004} found no evidence
of any sort for the presence of compact objects.
In case a neutron star is present, these observations provide us with
upper limits on the thermal luminosity of the star which we take from the
Figure~37 of \cite{Ketal2004} and report in our Table~\ref{utable}.

\subsection{CXO J232327.8+584842 in Cas A}

Discovered in the first light of the {\em Chandra} observatory, this object
is still enigmatic but evidence points toward an isolated neutron star
\citep{MTI2002}.
We take the upper limit on $L_\infty$ from \cite{Petal2000}, 
which results from a composite model of a hot polar cap and a warm
(but barely detected) surface.
The age is from the association with the supernova SN 1680.

\subsection{PSR RX J0205+6449 in 3C58}

A first upper limit on $T_\infty$ had been obtained by \cite{SHM02} from the
non detection of a thermal component in the pulsar spectrum.
Analysis of a deeper {\em Chandra} observation \citep{Setal2004b} requires the
presence of a thermal component and leads to a lower upper limit on
$L_\infty$ reported in Table~\ref{utable}. 
This value is a conservative estimate since even lower values are possible 
when an atmosphere model is used in the spectral fits.
The association with the historical remnant of SN 1181 gives an age of
822 years.  The pulsar spindown age is about 5400 years \citep{Metal2002}.
The distance estimate is from 21 cm (HI) absorption \citep{Retal93}.

\subsection{PSR J1124-5916 in G292.0+1.8}

This x-ray and radio pulsar is seen as a point source in a composite SNR.
The {\em Chandra} spectrum of the pulsar \citep{Hetal2003} is adequately 
fit by a power law, with  
no evidence for thermal emission, and provides 
an upper limit on the thermal luminosity.
Distance and kinematic age estimates are taken from \cite{Cetal02}.

\subsection{PSR RX J0822-4247 in Puppis A}
The ages are taken from spin-down (8000 years) and from motions of
filaments in the Puppis A supernova remnant.   The distance is estimated
from 21 cm (H) absorption.  Both quantities are discussed in
\cite{ZTP1999} and references therein. 
Blackbody and H atmosphere spectral fitting, from ROSAT and ASCA data,
are also from \cite{ZTP1999}. 

\subsection{PSR 1E 1207-5209 in G296.9+10.0}
Estimates of the kinematic age are from \cite{RMK1988} and the distance
are from \cite{GDG2000}.  
The spindown age is from \cite{PZST2002}.
Blackbody spectral fitting is from \cite{MBC1996} from ROSAT and 
\cite{ZPT1998} from ROSAT + ASCA.

This is one of the very few isolated neutron stars which shows spectral lines
in its spectrum \citep{Setal02}, but the phase variation of these lines
\citep{Metal02} may indicate they are of magnotespheric origin.
Moreover, its peculiar spin-down behavior may be a sign
of accretion, making the interpretation of this star as an isolated cooling
neutron star questionable \citep{ZPS04}.

\subsection{RX J0002+6246}
Spectral fitting for both H atmosphere and blackbody surfaces are from
{\em Chandra} observations \citep{P2002}, and are consistent with ROSAT
observations reported by \cite{HC1995}.

\subsection{RX J0007.0+7302 in CTA 1}

An x-ray point source within a pulsar wind nebula in a composite SNR.
No pulsations are detected but the general morphology of the object
makes it very similar to Vela, including a very likely association
with the EGRET $\gamma$-ray source 3EG J0010+7309.  The x-ray spectrum of
the point source, from either {\em XMM-Newton} \citep{Setal2004} or
{\em Chandra} \citep{HGCH2004}, is fitted by a power law plus a
thermal component.  The latter could be either originating from a hot
polar cap or the cooler entire stellar surface.  The upper value for
the thermal luminosity we use is from \cite{HGCH2004}, because the
high angular resolution of {\em Chandra} allows a better estimate than
the {\em XMM-Newton} data.  Distance and kinematic age estimates are
taken from \cite{Setal2004}.  

\subsection{PSR 0833-45 (Vela)}
Spectral fitting from {\em Chandra} for both H atmosphere and
blackbody surfaces are from \cite{PZS2001}, and the spindown age is
also quoted there.  The kinematic age of the SNR is from \cite{AET95}
and the VLBI interferometric distance measurement of 250 pc is due to
\cite{Detal2003}.  This is probably the most reliable data point
available and the first isolated neutron star whose radius is well
determined because of a well known distance \citep{PSZ96}.

\subsection{PSR 1706+44}
Blackbody spectral fitting for the {\em Chandra} data is from \cite{GHD2002},
and spindown age is quoted in the same source whereas H atmosphere
spectral fitting for the {\em XMM-Newton} data is from
\cite{McGetal2004}.  Estimates of distance are from \cite{TC1993} and
\cite{KJW1995}.

\subsection{PSR 0538+2817}

We employ the results of spectral fitting from {\em Chandra} by 
\cite{MS2002} for both blackbody and magnetized hydrogen models.
We assume a typical error of 0.1 on $\log_{10} T$ and of
0.5 on $\log_{10} L$, since these authors do not report any
uncertainty estimates.

\subsection{PSR J0154+61}
We use the results of an {\em XMM-Newton} observation by
\cite{GKLP2004} for $L_\infty$.  The distance and $t_{sd}$ values are
taken from the same paper.  

\subsection{PSR 0656+14}
We employ the results of spectral fitting from {\em Chandra} \citep{MS2002},
which result in a lower temperature than the ROSAT value
$\log_{10}T_\infty=5.96^{+0.02}_{-.03}$ quoted by
\cite{PMC1996}. \cite{MS2002} also suggest there is a hard component
with a temperature of $2\times10^6$ K.  The spindown age is from
\cite{TML1993}.  As with other objects in this study, we employ the
softer component's temperature as being more characteristic of the
underlying surface temperature.  The distance to this object is
constrained by the VLBA parallax measurement of \cite{B2003}.

\subsection{PSR 0633+1748 (Geminga)}
Blackbody spectral fitting with ROSAT data is from \cite{HW1997}; later
analyses have not  changed these results significantly.  The distance is the
result of parallax measurements by \cite{CBM1996}; however, \cite{P2002} 
suggests these measurements may not be reliable.

\subsection{PSR 1055-52}
\cite{P2002} quotes results from {\em Chandra} observations of two thermal
components: a soft component with temperature $8.9\pm0.01\times10^5$ K
and emitting radius $13d_{1000}$ km and a hard component with
temperature $1.9\pm0.1\times10^6$ K and emitting radius of
$0.5\pm0.1d_{1000}$ km.  These temperatures are consistent with ASCA
results quoted in \cite{G1996} and ROSAT results in \cite{O1995}.
We employ the soft component temperature as being characteristic of
the average surface temperature.

\subsection{RX J1856-3754}
Absence of spectral lines in the high resolution {\em Chandra} LETGS
data rules out non-magnetic, non-rotating heavy element atmosphere
models \citep{BZN2001}.  Blackbodies provide the best fits to the
x-ray data but the optical data require the presence of a colder
blackbody component \citep{PWL2002}.  We take the blackbody spectral
fitting from \cite{DMD2002} and \cite{Betal2003} for the warm
component and fits for the colder component from \cite{PWL2002}.  The
range of $T_\infty$ listed in Table~\ref{btable} correspond to these
two, cold and warm, components.  The cold blackbody component gives
the lower bound on the radius but it makes a very small contribution
($\sim$ 5\%) to the luminosity.

\subsection{RX J0720.4-3125}
As for RX J1856-3754, fit of both the x-ray and optical data requires
a two blackbody model.  We take both warm and cold blackbody fits from
\cite{Ketal2003} which gives us the range of $T_\infty$ we report in
Table~\ref{btable}.  The luminosity has a significant contribution
from the cold component.  The distance is unknown and the values we report
are a guess based on the observed low absorption of the x-ray
spectrum.  This distance results in a large uncertainty in
$L_\infty$.  Notice also that the spectrum is known to vary on long
time scales \citep{dVVMF2004} and contains a phase-dependent absorption
line \citep{HZTB2004}.  Both $P$ and $\dot{P}$ are from \cite{CHZZ04}.



\section{The Equations of Structure and Evolution
         \label{Sec:Equ}}

We employ the standard structure equations derived from spherically
symmetric, general relativistic, considerations.
It should be mentioned that the stellar surface in our computation
is fixed by
\be
R = R_{\rm star} = r(\rho = \rho_b) \,,
\ee
where $\rho_b = 10^{10}$ g cm$^{-3}$.
This guarantees that the EOS is temperature independent.
The layers at densities below $\rho_b$, called the {\em envelope},
are treated separately (see \S~\ref{Sec:Envelope}).

At the temperatures of interest here,
neutrinos have a mean free path much larger than the radius of the
star (\cite{ST83}) and thus leave the star once they are produced.
Energy balance arguments (see for instance \citet{T66}) then imply
\be
\frac{d(l e^{2 \Phi})}{dr} = -\frac{4 \pi r^2 e^{\Phi}}{\sqrt{1 - 2Gm/c^2r}}
      \, \left( \frac{d\epsilon}{dt} + e^{\Phi} (q_{\nu}-q_h) \right) \,, 
\label{Eq:dLdr}
\ee
where $\Phi$ is the gravitational potential, $l$ is the internal
luminosity, $\epsilon$ is the internal energy per unit volume,
$q_{\nu}$ and $q_h$ are respectively the neutrino emissivity and
heating rate, both per unit volume.  The corresponding inner boundary
condition for $l$ is
\be
l(r=0) = 0
\ee
The time derivative of $\epsilon$ can be written in the form \be
\frac{d \epsilon}{dt} = \frac{d \epsilon}{dT} \cdot \frac{dT}{dt} =
c_v \cdot \frac{dT}{dt} \,,
\label{Eq:dEdt}
\ee where $T$ is the local temperature and $c_v$ is the specific heat
per unit volume at constant volume, which for degenerate matter is the
same as the specific heat at constant pressure $c_P$.

The energy transport equation is
\be
\frac{d(T e^{\Phi})}{dr} = - \frac{1}{\lambda} \;\;
   \frac{l e^{\Phi}}{4 \pi r^2 \sqrt{1 - 2Gm/c^2r}}\,,
\label{Eq:dTdr}
\ee where $\lambda$ is the thermal conductivity.  (Notice that within
the relativistic framework, an `isothermal' configuration is defined
by $e^{\phi} \cdot T = {\rm constant}$, instead of $T$ = constant.)
The associated boundary condition is \be T_b = T_b(l_b) \,,
\label{Eq:Tb-Te}
\ee which relates the temperature $T_b$ at the outer boundary (defined
more precisely further below) to the luminosity $l_b$ in this layer.
The location of this outer boundary layer is chosen such that $l_b$ is
equal to the total photon luminosity of the star, $l_b = l(r=R) \equiv
L$.  $L$ is commonly expressed through the ``effective'' temperature
$T_e$, which is {\em defined} by \be L \; \equiv \; 4 \pi R^2 \cdot
\sigma_{\scriptscriptstyle SB} T_e^4 \ee where
$\sigma_{\scriptscriptstyle SB}$ is the Stefan-Boltzmann constant.  We
emphasize that $L$ and $T_e$ are, modulo $R$, essentially equivalent
quantities.
We can thus write equation~(\ref{Eq:Tb-Te}) as $T_b = T_b(T_e)$; this
`$T_e - T_b$ relationship'' is discussed further in
\S~\ref{Sec:Envelope}.

We will present our results of thermal evolution by using the
``effective temperature at infinity''
\be
T_e^\infty \equiv T_e \cdot e^{\Phi(R)}
\label{eq:t_e_inf}
\ee
related to the ``luminosity at infinity'' $L^{\infty}$ through the
``radiation radius'' $R_{\infty} \equiv R \cdot e^{-\Phi(R)}$ by \be
L^{\infty} \; \equiv \; e^{2\Phi(R)} L(R) \; = \; 4 \pi R^{\infty \; 2}
\cdot \sigma_{\scriptscriptstyle SB} T_e^{\infty \; 4}.
\label{eq:L_inf}
\ee The three quantities $T_e^\infty, L^\infty$ and $R^\infty$, are,
in principle, measurable.   In particular, $R_{\infty}$ would be the
areal radius of the star that an observer `at infinity' would measure
with an extremely high angular resolution instrument \citep{P95}.




\end{document}